\documentclass[%
reprint, 
superscriptaddress,
preprintnumbers,
 amsmath,amssymb,
 aps,
 prd,
 nofootinbib
]{revtex4-1}

\usepackage{graphicx}
\usepackage{dcolumn}
\usepackage{bm}
\usepackage{color}
\usepackage{hyperref}
\usepackage{cancel}
\usepackage{multirow}
\usepackage{enumitem}
\usepackage{booktabs}
\usepackage{float}
\usepackage{subfigure}
\usepackage{lineno}
\usepackage[normalem]{ulem}

\def\met{\cancel{E}_T}
\newcommand{\neutralino}[1]{\tilde{\chi}_{#1}^0}
\newcommand{\chargino}  [1]{\tilde{\chi}_{#1}^\pm}

\newcommand{\slepton} {\tilde{\ell}}
\newcommand{\sleptonL}{\tilde{\ell}_L}

\newcommand{\sneutrino}{\tilde{\nu}}
\newcommand{\selectronL}{\tilde{e}_L}

\newcommand{\iab}{{\rm ab}^{-1}}
\newcommand{\ifb}{{\rm fb}^{-1}}
\newcommand{\GeV}{\,{\mathrm{GeV}}}
\newcommand{\TeV}{\,{\mathrm{TeV}}}
\newcommand\Order{\mathop{\mathcal O}}

\definecolor{Green}{rgb}{0.2,0.7,0}

\usepackage[utf8]{inputenc}

\begin{document}

\title{Search for the SUSY Electroweak Sector at $ep$ Colliders}

\author{Georges Azuelos}
\email{georges.azuelos@cern.ch}
\affiliation{Universit\'e de Montr\'eal, Montr\'eal,  Quebec H3C 3J7,  Canada}
\affiliation{TRIUMF, Vancouver V6T 2A3, Canada}

\author{Monica D'Onofrio}
\email{Monica.D'Onofrio@cern.ch}
\affiliation{ University of Liverpool, Oliver Lodge Laboratory, P.O.\ Box 147, Oxford Street, Liverpool L69 3BX, United Kingdom }%

\author{Sho Iwamoto}%
\email{sho.iwamoto@ttk.elte.hu}
\affiliation{Dipartimento di Fisica e Astronomia, Universit\`a di Padova, Via Marzolo~8, I-35131 Padua, Italy}
\affiliation{INFN, Sezione di Padova, Via Marzolo~8, I-35131 Padua, Italy}
\affiliation{ELTE E\"otv\"os Lor\'and University, P\'azm\'any P\'eter s\'et\'any 1/A, H-1117 Budapest, Hungary}

\author{Kechen Wang}
\email{kechen.wang@whut.edu.cn (Corresponding author)}
\affiliation{ Department of Physics, School of Science, Wuhan University of Technology, 430070 Wuhan, Hubei, China }


\begin{abstract}\noindent
The sensitivity of future electron-proton colliders, the LHeC and FCC-eh, to weakly-produced supersymmetric particles is evaluated in this article. Supersymmetric scenarios where charginos ($\tilde{\chi}_1^{\pm}$) and neutralinos ($\tilde{\chi}_1^{0}$ and $\tilde{\chi}_2^{0}$) are nearly degenerate in mass are considered. Two sets of models, which differ in the mass of sleptons ($\tilde{\ell}$), are studied.  Under the hypothesis that slepton masses are at the multi-TeV scale (``decoupled'' scenario), the production processes for charginos and neutralinos at $ep$ colliders, $p\, e^- \to j\, e^-\, \tilde{\chi} \tilde{\chi}$ with $\tilde{\chi}=\tilde{\chi}_1^{0}$, $\tilde{\chi}_1^{\pm}$ or $\tilde{\chi}_2^{0}$, are considered.
For the models where slepton masses are above but close to $\tilde{\chi}_1^{\pm}$, $\tilde{\chi}_2^{0}$ masses (``compressed'' scenario), contributions from the processes $p\, e^- \to j\, \tilde{\chi}\, \tilde{e}_L^-$ and $j\, \tilde{\chi}\, \tilde{\nu}$ followed by the decays $\tilde{e}_L^- \to \tilde{\chi}_{1,2}^{0} + e^-$ and $\tilde{\nu} \to \tilde{\chi}_1^+ + e^-$ are also taken into account.   These scenarios are analysed with realistic detector performance, using multivariate techniques. Effects of systematic uncertainties and electron beam polarization dependence are also discussed. The reach is found to be complementary to the one obtained at $pp$ colliders, in particular for the compressed-slepton scenario. 
\end{abstract}

\maketitle

\section{Introduction}
\label{sec:intro}

\noindent
Supersymmetry (SUSY) is one of the most promising new physics scenarios beyond the Standard Model (SM). It has been explored widely in various production and decay channels at the Large Hadron Collider (LHC).
Due to the large production cross sections in strong interactions at $pp$ colliders, the current LHC experiments have produced impressive constraints on the SUSY coloured sector (squarks and gluinos), excluding masses up to approximately 2 TeV at a centre-of-mass energy of $\sqrt{s} = $ 13 TeV by the CMS collaboration with integrated luminosity of 137 $\ifb$~\cite{Sirunyan:2019xwh,Sirunyan:2019ctn} and by the ATLAS collaboration with 139 $\ifb$ integrated luminosity~\cite{ATLAS:2019vcq}. 

Because of their much lower direct production cross sections, less stringent constraints have been placed on weakly-produced SUSY particles, namely neutralinos  $\neutralino{}$, charginos $\chargino{}$, and sleptons $\slepton^{\pm}$.
Here, $\neutralino{}$ and $\chargino{}$ (collectively referred to as electroweakinos) are the SUSY partners of the Higgs and gauge bosons described in the mass eigenstates.
If R-parity is conserved, the lightest neutralino $\neutralino1$ can be the lightest SUSY particle (LSP) and a candidate for dark matter.

A variety of searches for $\chargino{}$, $\neutralino{}$, and $\slepton^\pm$ at $\sqrt{s} = 13\TeV$ have been reported by the ATLAS~\cite{Aaboud:2017leg,Aaboud:2018jiw,Aaboud:2018sua,Aaboud:2018ngk} and CMS~\cite{Sirunyan:2017zss,Sirunyan:2017lae,Sirunyan:2017eie,Sirunyan:2017qaj,Sirunyan:2018iwl,Sirunyan:2018ubx,Sirunyan:2018nwe,Sirunyan:2019zfq} collaborations with 36 $\ifb$ of integrated luminosity.
Recently, the collaborations have updated the analyses with up to 139 $\ifb$ integrated luminosity~\cite{ATLAS:2019cfv,ATLAS:2019lov,ATLAS:2019lnq,ATLAS:2019nnv,Aad:2019vvf,Aad:2019vnb}.
Most of the studies target the processes $pp \to \chargino1\neutralino2 \to W^{\pm}Z\, \neutralino1\neutralino1$ and $pp \to \tilde{\ell}^+ \tilde{\ell}^- \to l^+ l^- \neutralino1\neutralino1$, exploiting events characterized by multiple charged leptons and large missing transverse momentum, $\met$. Dedicated searches for decays through the Higgs boson ($pp \to \chargino1\neutralino2 \to W^{\pm}h\, \neutralino1\neutralino1$) are also performed. Searches for electroweakinos decaying through sleptons exclude masses above 1 TeV (see for example ~\cite{Aaboud:2018jiw}).
Lower bounds of $\Order(100)\GeV$ on $\chargino1/\neutralino2$ masses are obtained, depending on the scenarios; for example, Ref.~\cite{Aaboud:2018sua} excludes $\chargino1/\neutralino2$ decaying via $W^{\pm}$ and $Z$ bosons up to approximately 600 GeV with 95\% confidence level (CL), while Ref.~\cite{Aad:2019vnb} reports that the left-handed selectron $\selectronL$ masses can be excluded up to approximately 550 GeV. 
These limits mostly apply to scenarios characterised by a large mass difference, $\Delta m$, between $\neutralino1$ and the next lightest supersymmetric particle (NLSP), either $\chargino1 / \neutralino2$ or $\slepton^\pm$.
In the case of small mass differences (e.g., $1\GeV< \Delta m < 50\GeV$, with $\Delta m$ still sufficiently large to guarantee prompt $\chargino1/\neutralino2$ decays), $W^{\pm}$ and $Z$ bosons are off-shell and scenarios are referred to as ``compressed'':  the visible decay products from $\slepton$ and $\chargino1 / \neutralino2$ have very soft transverse momenta ($p_T$) and the SM backgrounds are kinematically similar to the signal.  The analyses therefore become challenging and sensitivities decrease substantially.

Dedicated searches have been recently developed by the LHC experiments to target  low-$\Delta m$ scenarios. They exploit the recoil of the sparticles system against an initial state radiation (ISR) jet for efficient detection of soft decay products. Assuming $\Delta m \sim 5$ GeV, the ATLAS collaboration has excluded $\selectronL$ masses up to around 160 GeV and $\chargino1/ \neutralino2$ masses up to around 200 GeV~\cite{ATLAS:2019lov}. A similar reach has been reported by the CMS collaboration~\cite{Sirunyan:2018iwl}, as well as searches exploiting vector-boson fusion (VBF) production~\cite{Sirunyan:2019zfq}. Phenomenological studies  on weakly-produced SUSY particles in compressed scenarios have been also reported in Refs.~\cite{Barr:2015eva,Han:2014kaa,Baer:2014kya,Dutta:2014jda,Han:2014aea,Han:2014xoa,Delannoy:2013ata,Delannoy:2013dla,Schwaller:2013baa,Gori:2013ala,Han:2013usa,Giudice:2010wb}.
The studies on the Bino-Wino coannihilation scenario at the LHC with $\sqrt{s} = $ 13 TeV and the high-energy LHC (HE-LHC) with  $\sqrt{s} = $ 27 TeV can also be found in~\cite{Duan:2018rls, Abdughani:2019wss}.

Studies on the potential of the high-luminosity LHC (HL-LHC)~\cite{CidVidal:2018eel}, which foresees 3 $\iab$ of data taken at a centre-of-mass energy of 14 TeV, have shown that searches for low-momentum leptons and ISR-jet boost will be sensitive to chargino masses up to 400 (350)~GeV for $\Delta m (\chargino1 / \neutralino2, \neutralino1) \sim 5$ GeV, and to mass splittings between 1 and 50 GeV, 
assuming Wino-like (Higgsino-like) cross sections for electroweakino productions. Similar search techniques can also be used to target pair produced sleptons in compressed scenarios.  
It is worth noting that, to suppress SM backgrounds, analyses targeting very small $\Delta m$ require soft-momentum leptons from the decays of the sleptons, charginos or neutralinos and are thus complementary to searches targeting very large $\Delta m$ via multiple high-$p_T$ leptons. Regions of intermediate $\Delta m(\chargino1 / \neutralino2, \neutralino1)$ ($\sim 20$--$50$~GeV) may still be elusive after the HL-LHC. 

This article focuses on \emph{compressed} electroweakinos scenarios, produced assuming Wino-like cross sections and mass differences between $\chargino1, \neutralino2$ and $\neutralino1$ small, {$\mathcal{O}$}(GeV), but still allowing their prompt decay. Two SUSY scenarios with different hypotheses on the charged slepton masses are considered to evaluate the sensitivity of future $ep$ colliders, the Large Hadron electron Collider (LHeC) and the electron-hadron mode of the Future Circular Collider (FCC-eh), to discover SUSY electroweakinos and sleptons. 
The search uses the multivariate analysis approach (MVA) and Monte Carlo (MC) generated events passed through a realistic detector-level simulation. The electron and proton beam energies are assumed to be $60\GeV\times7\TeV$ ($60\GeV\times50\TeV$) at the LHeC (FCC-eh), which correspond to $\sqrt s = 1.3\TeV$ $(3.5\TeV)$. The maximal integrated luminosity at the LHeC is expected to be 1 $\iab$, while it could reach 2.5 $\iab$ at the FCC-eh after a 25-year running period. 
The centre-of-mass energy, $\sqrt{s}$, of $ep$ colliders are considerably lower than the HL-LHC. However, since there are no gluon-exchange diagrams, the SM QCD backgrounds, which are dominant at $pp$ colliders, are much smaller. Furthermore, the number of additional interactions in the same event (pile-up) is negligible at $ep$ colliders, whilst it is expected to be very large at the HL-LHC.
Previous studies on Higgsino-like $\neutralino{}/\chargino{}$ production at $ep$ colliders can be found in Ref.~\cite{Han:2018rkz, Curtin:2017bxr}, 
where Ref.~\cite{Han:2018rkz} focuses on decoupled-scenario with light Higgsinos, and assume a proton beam of 50 (7)~TeV energy, colliding with an electron beam of 60 (60–140)~GeV energy at the FCC-eh (LHeC), 
while Ref.~\cite{Curtin:2017bxr} explores the parameter space for long-lived Higgsinos.

The article is organised as follows. Sec.~\ref{sec:scenarios} presents the SUSY models considered. In Sec.~\ref{sec:strategy} data simulation, signal and background processes and search strategy are reported. In Sec.~\ref{sec:reuslts_compressedSlep}, the results of the compressed-slepton scenario are presented. The results of the decoupled-slepton scenario are reported in Sec.~\ref{sec:reuslts_decoupledSlep}. Conclusions and discussions on the effects of electron beam polarizations are presented in Sec.~\ref{sec:Summary}.

\section{SUSY scenarios}
\label{sec:scenarios}

\begin{figure}[h]
\includegraphics[width=4cm,height=3cm]{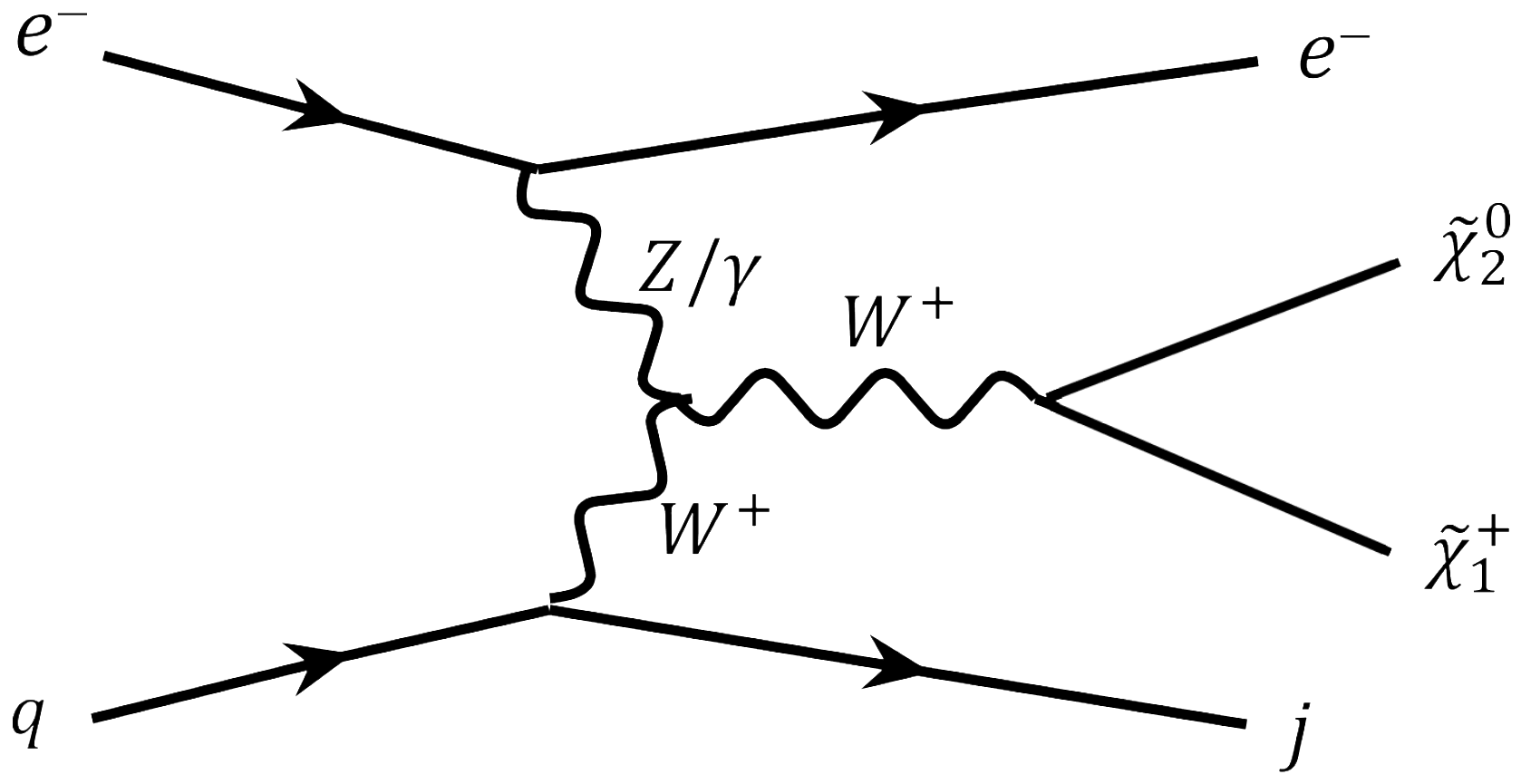}
\includegraphics[width=4cm,height=3cm]{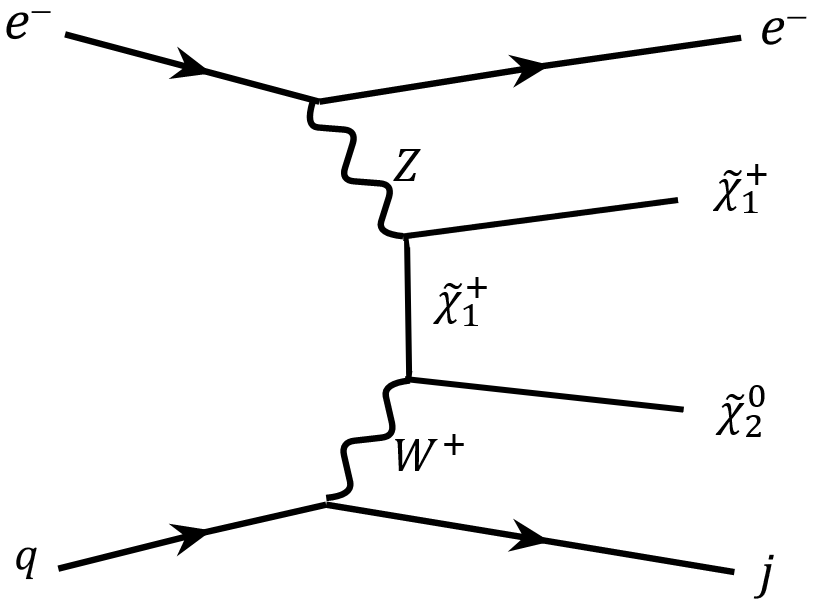}
\includegraphics[width=4cm,height=3cm]{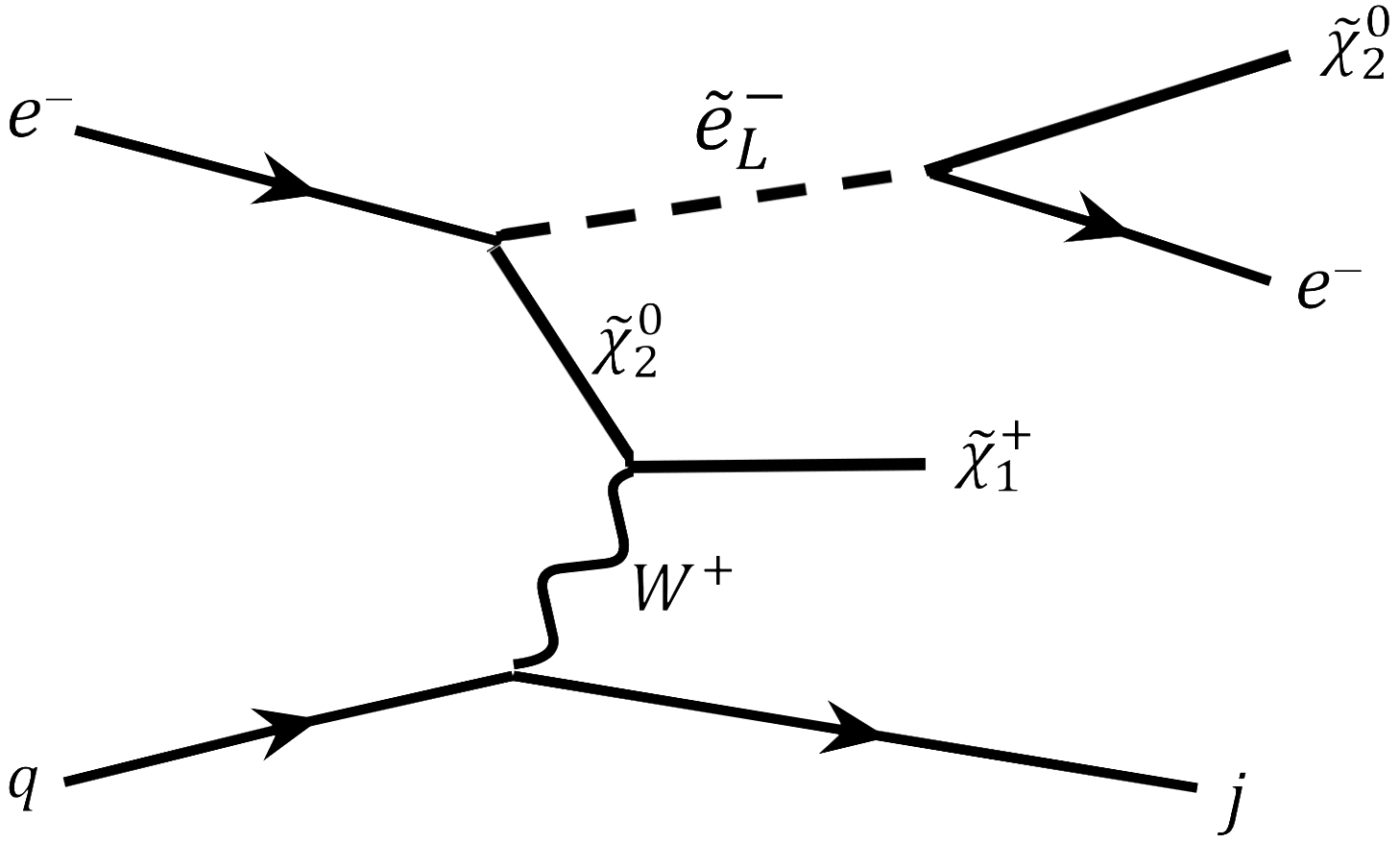}
\includegraphics[width=4cm,height=3cm]{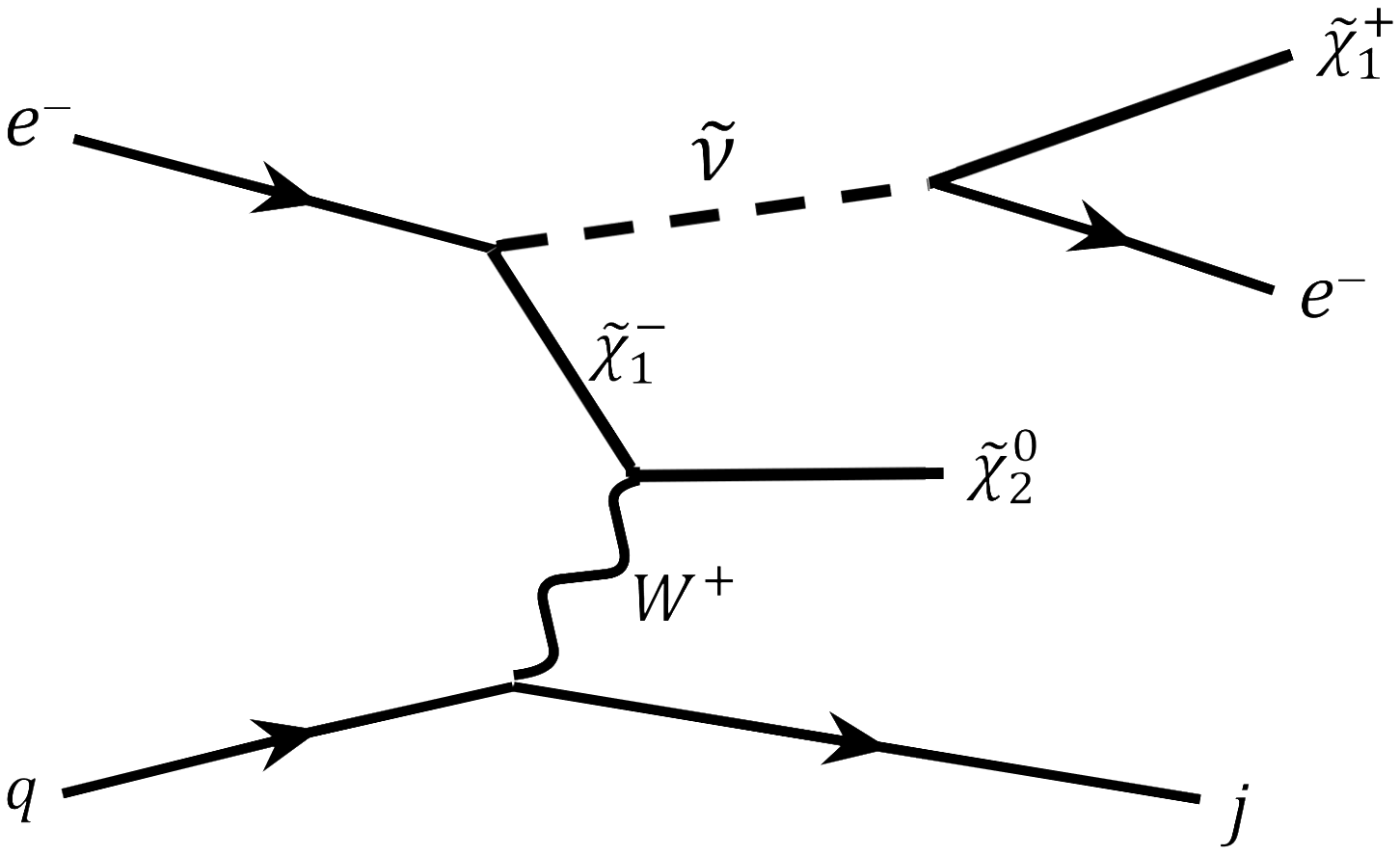}
\caption{
Some representative production diagrams for the signal processes considered in this article. The \emph{decoupled-slepton} scenario includes only the upper two diagrams, while the \emph{compressed-slepton} scenario includes both the upper and lower diagrams.
}
\label{fig:production}
\end{figure}

\noindent
Two SUSY compressed scenarios with different hypotheses on the charged slepton masses are considered.  In the first scenario,  referred to as the ``decoupled-slepton'' scenario, we assume that the only SUSY particles within kinematic reach are the electroweakinos $\neutralino{1,2}$ and $\chargino{1}$.
The LSP $\neutralino1$ is assumed to be Bino-like, the $\neutralino{2}$ and $\chargino1$ are Wino-like and degenerate in masses, and the mass difference between $\neutralino1$ and $\chargino1$ is small ($\Delta m \sim 1$~GeV). All other SUSY particles are at the multi-TeV scale and therefore decoupled.
The upper two diagrams in Fig.~\ref{fig:production} represent the typical production processes in this scenario; charginos and neutralinos are produced via VBF processes.
This scenario is motivated by dark matter coannihilation arguments~\cite{BirkedalHansen:2002am,Baer:2005jq,ArkaniHamed:2006mb,Harigaya:2014dwa}. For a Bino-like dark matter, the annihilation cross section is usually too low. The Wino-like $\chargino1/ \neutralino2$ with masses of order 1-10 of GeV larger than the LSP enhance the coannihilation processes, which may result in the dark matter relic density consistent with observations. 
In the second scenario, referred to as ``compressed-slepton'' scenario, the left-handed charged sleptons $\sleptonL^\pm$ and sneutrinos $\sneutrino$ are assumed to be kinematically accessible and slightly heavier than $\chargino1/\neutralino2$.  
The signal production in this case includes both the upper and lower diagrams in Fig.~\ref{fig:production}.
This scenario is also motivated by coannihilation arguments~\cite{Ellis:1998kh}.
Furthermore, since the main SUSY contributions to the muon anomalous magnetic moment, $(g-2)_\mu$, are given by the neutralino-smuon and chargino-sneutrino loop diagrams, SUSY mass spectra containing neutralinos and sleptons of mass ${\cal{O}} (100)$ GeV may explain the discrepancy of the measured $(g-2)_\mu$ with theoretical predictions (see Ref.~\cite{Endo:2017zrj} for a recent summary). 

In the decoupled-slepton scenario, since the charged sleptons and sneutrinos $\sneutrino$ are at the multi-TeV scale,  signal events  $pe^-\to je^- \tilde{\chi} \tilde{\chi}$ ($\tilde{\chi} = \neutralino{1}, \neutralino{2}, \chargino1$) are produced via VBF processes of charginos and neutralinos only (c.f. upper two diagrams in Fig.~\ref{fig:production}).
In the compressed-slepton scenario, however, since the left-handed slepton $\sleptonL$ and $\sneutrino$ are slightly heavier than the $\chargino1$ and $\neutralino2$,  the signal events $pe^-\to je^- \tilde{\chi} \tilde{\chi}$ include  not only VBF processes of $\chargino1$ and $\neutralino2$ but also the direct production of  $\sleptonL$ and $\sneutrino$ followed by the decays of $\tilde{e}_L^- \to \neutralino{2,1} + e^-$ and $\sneutrino \to \tilde{\chi}_1^+ + e^-$  (i.e. $pe^- \to j \tilde{\chi} \tilde{e}_L^-,\, j \tilde{\chi} \sneutrino \to   je^- \tilde{\chi} \tilde{\chi}$). Thus both the upper and lower diagrams in Fig.~\ref{fig:production} contribute to the signal rate in the compressed-slepton scenario.

\begin{figure}[h]
\includegraphics[width=7.5cm,height=4.5cm]{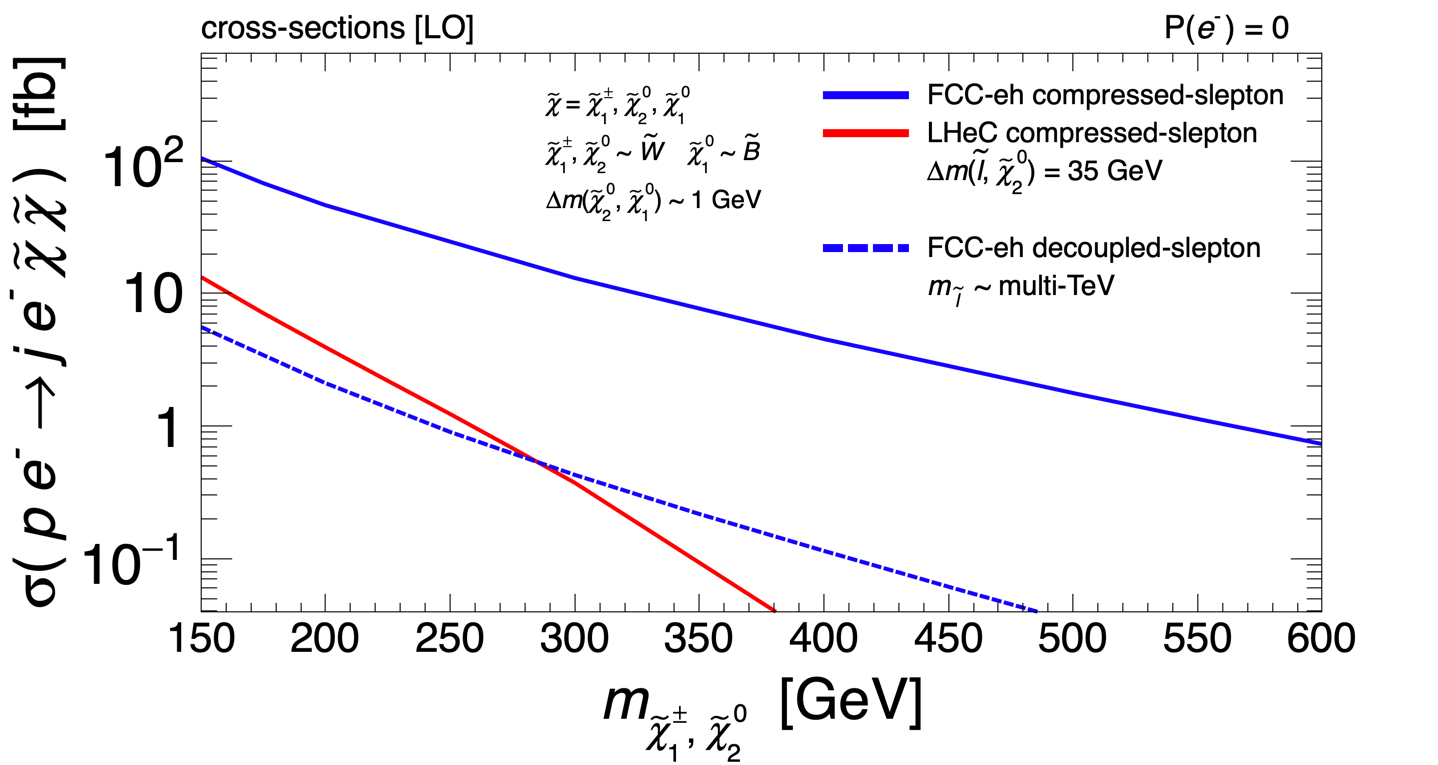}
\caption{
Production cross sections $\sigma(pe^- \to je^- \tilde{\chi} \tilde{\chi})$ for both the compressed- and decoupled-slepton scenarios at the FCC-eh and for the compressed-slepton scenario at the LHeC with unpolarized $e^-$ beam as varying the masses of $\chargino1$ and $\neutralino2$. 
Here $\neutralino1$ is Bino while $\chargino1$ and $\neutralino2$ are Wino with almost degenerate masses. 
Cross sections are calculated at leading order by MadGraph5\_aMC@NLO  version 2.4.3.
For the compressed-slepton scenario, the mass difference $\Delta m = m_{\slepton} - m_{\chargino1, \neutralino2} = 35\GeV$, while the decoupled-slepton scenario assumes sleptons are heavy and decoupled.
}
\label{fig:crs}
\end{figure}

Figure~\ref{fig:crs} shows the signal cross sections $\sigma(pe^- \to je^- \tilde{\chi} \tilde{\chi})$ at FCC-eh for both the compressed- and decoupled-slepton scenarios as a function of the masses of the $\chargino1$ and $\neutralino2$, and at the LHeC only for the compressed-slepton scenario since the decoupled-slepton scenario yields much smaller cross section. 
Cross sections are calculated at leading order (LO) by MadGraph5\_aMC@NLO version 2.4.3~\cite{Alwall:2014hca} .
Higher-order corrections are expected to be at the same level as those for electroweakino pair-productions at the LHC~\cite{Debove:2010kf}, in general up to 20\% with respect to LO calculations, but they are not taken into account here. This leads to conservative results in terms of signal sensitivity.
For both SUSY scenarios, the signal final state events are characterised by the presence of one jet, one electron, large $\met$, and undetected very soft-momenta particles arising from $\chargino1$ or $\neutralino2$ decays.
%

\section{Search Strategy}
\label{sec:strategy}

\noindent
Simulated data are generated at parton-level by MadGraph5\_aMC@NLO  2.4.3 by importing the general MSSM model with the model parameter file prepared by the SUSY-HIT package~\cite{Djouadi:2006bz}. Pythia performs parton showering and hadronization, and Delphes~\cite{deFavereau:2013fsa} is used for detector simulation.
The detectors at $ep$ colliders are assumed to have a cylindrical geometry comprising a central tracker followed by an electromagnetic and a hadronic calorimeter.
The forward and backward regions are also covered by a tracker, an electromagnetic and a hadronic calorimeter.
The angular acceptance of charged tracks and the detector performance in terms of momentum and energy resolution of electrons, muons, and jets are based on the LHeC (FCC-eh) detector design~\cite{AbelleiraFernandez:2012cc,LHeCDetector}. In particular,  electrons are selected in the pseudorapidity range of $-4.3 < \eta < 4.9$ ($-5.0 < \eta < 5.2$) at the LHeC (FCC-eh), while jets are selected within an $\eta$ range between $\pm~5$ for the LHeC and $\pm~6$ for the FCC-eh. 
For the simulation, a modified Pythia version tuned for the $ep$ colliders
\footnote{
The modification concerns the treatment of beam remnants, that in some cases might cause generation failure. The parameter setting preserves the correct flow of the generation in these instances. 
As a cross check, the hadronic showering is not expected to change the kinematics of the DIS scattered lepton. This has been shown, see page 11 of Ref.~\cite{Py_mod2014}: a very good level of agreement of the NC DIS electron kinematics with and without the ep-customized Pythia showering is found.
}
and the Delphes card files for the LHeC and FCC-eh detector configurations~\cite{Delphes_cards} are used.

Due to the presence of large missing transverse momentum in the final state, processes with production of neutrinos will contribute to the SM background. We separate the background into two categories: the 2-neutrino process $p\, e^- \to j\, e^-\, \nu \nu$ and the 1-neutrino process $p\, e^- \to j\, e^-\, \ell \nu$,
where $\nu$ stands for the neutrino and anti-neutrino with all flavors of $e, \mu, \tau$. 
The 2-neutrino process has the same final state as the signal and is an irreducible background. The 1-neutrino process has a large cross section and will therefore have a non-negligible contribution to the background if one of the two leptons is undetected.

To make our simulation more realistic, we set following cuts at the parton level in the MadGraph: $p_T(j) > 5 \GeV, p_T(\ell) > 1 \GeV, |\eta(j)| < 10, |\eta(\ell)| < 10$, etc., which are looser than the default cuts.
The following pre-selections in the final state after detector simulation are applied for LHeC (FCC-eh, reported if differing) studies:
\begin{enumerate}
\item At least one jet with $p_T > 20$ GeV, $|\eta|\leq 5.0 (6.0)$; 
\item Exactly one electron with $p_T > 10$ GeV, $-4.3 < \eta < 4.9$ ($-5.0 < \eta < 5.2$); 
\item No b-jet with $p_T > 20$ GeV;
\item No muon or tau with $p_T > 10$ GeV; 
\item Missing transverse momentum $\met > 50$ GeV.
\end{enumerate}
After the pre-selections, 17 observables listed in the following are used as inputs  to the TMVA package~\cite{Hocker:2007ht} to perform a multi-variate analysis through a boost decision tree (BDT) approach.
\begin{enumerate}[label*=\arabic*.]

\item Global observables:
\begin{enumerate}[label*=\arabic*.]
\item the missing transverse momentum $\met$;
\item the scalar sum of $p_T$ of all jets, $H_T$.
\end{enumerate}

\item Observables for the visible objects:
\begin{enumerate}[label*=\arabic*.]
\item $p_T$ and the pseudorapidity $\eta$ of the first leading jet $j_1$ and the first leading electron $e_1$: $p_T(j_1)$, $\eta(j_1)$, $p_T(e_1)$, $\eta(e_1)$;
\item the pseudorapidity difference $\Delta\eta$ and the azimuthal angle difference $\Delta\phi$ between $j_1$ and $e_1$: $\Delta\eta(j_1, e_1)$, $\Delta\phi(j_1, e_1)$;
\item the invariant mass $M$, $p_T$ and $\eta$ reconstructed from the 4-vector of the system of $j_1$ and $e_1$: $M(j_1+e_1)$, $p_T(j_1+e_1)$, $\eta(j_1+e_1)$.
\end{enumerate}

\item Angular correlations between the missing transverse momentum and visible objects:
\begin{enumerate}[label*=\arabic*.]
\item $\Delta\phi$ between $\met$ and $j_1$, $e_1$, or $j_1+e_1$: $\Delta\phi(\met,j_1)$, $\Delta\phi(\met,e_1)$, $\Delta\phi(\met, j_1+e_1)$;
\item the transverse mass $M_T$ of the system of $\met$ and $j_1$, $e_1$, or $j_1+e_1$: $M_T(\met,j_1)$, $M_T(\met,e_1)$, $M_T(\met,j_1+e_1)$.
\end{enumerate}
\end{enumerate}

The BDT score for each event is used to separate the signal events from the SM background. The BDT threshold value is optimized for each SUSY scenario as well as for each collider, since kinematical distributions might vary as a function of $m_{\chargino1, \neutralino2}$ and the beam energies.
The number of signal and background events passing the BDT selection, $N_s$ and $N_b$, are used to calculate the significance. Assuming no systematic uncertainty, the statistical significance,
$\sigma_{\rm stat}$, of the potential signal is evaluated as
\begin{equation}
\sigma_{\rm stat} =
\sqrt{2 [(N_s+N_b) {\rm ln}(1+\frac{N_s}{N_b}) - N_s ] }.
\label{eqn:sgf1}
\end{equation}
Including a systematic uncertainty $\sigma_b$ in the evaluation of the number of background events, the significance is given by
\begin{equation}\begin{split}
\sigma_{\rm stat+syst} &=
\Bigg[ 2 \bigg( (N_s + N_b) \ln \frac{(N_s+N_b)(N_b+\sigma_b^2)}{N_b^2+(N_s+N_b)\sigma_b^2}
\\ &\qquad
 - \frac{N_b^2}{\sigma_b^2} \ln\left[1+ \frac{\sigma_b^2 N_s}{N_b(N_b+\sigma_b^2) } \right]\bigg) \Bigg]^{1/2}.
\end{split}\label{eqn:sgf2}
\end{equation}

\section{Compressed-Slepton Scenario}
\label{sec:reuslts_compressedSlep}

\noindent
The compressed-slepton scenario is characterised by three sets of masses, which are
\begin{equation}
    m_{\tilde\ell, \sneutrino},\, m_{\chargino1,\neutralino2},\, m_{\neutralino1}
\end{equation}
in the descending order, where $m_{\tilde\ell, \sneutrino}$ represents the mass of charged sleptons, here assumed dominated by the left-handed component, or sneutrinos. In this article, we assume 
nearly degenerate electroweakinos, with $m_{\chargino1,\neutralino2} = m_{\neutralino1} + $ 1 GeV, and prompt $\chargino1,\neutralino2$ decays. The mass splitting between $\sleptonL$ and $\sneutrino$ is generally below 10 GeV for $\tan\beta$ between 3 and 40. In this study, the benchmark $\tan\beta = $ 30 is assumed, which results in the mass difference between $\sleptonL$ and $\sneutrino$ of around 9 GeV.

In this section, the $\Delta m$ symbol is used to denote the mass difference between the $\slepton$ and $\chargino1, \neutralino2$, i.e.
\begin{equation}
    \Delta m = m_{\slepton} - m_{\chargino1, \neutralino2} \,\,.
\end{equation}

\subsection[Results with fixed Delta-m]{Results with fixed $\Delta m$}
\label{subsec:massDiff}

We first discuss the case with  $\Delta m = 35\GeV$, 
where 35~GeV is used as a proxy for intermediate-low-$\Delta m$ scenarios. The branching ratios for sleptons are ${\rm BR}(\tilde{e}_L^- \to \neutralino{2,1} + e^-) \approx 40\%$ and ${\rm BR}(\sneutrino \to \chargino1 + e^-) \approx 60\%$. 
The other decay channels, $\tilde{e}_L^- \to \tilde{\chi}_1^- + \nu $ and $ \sneutrino \to \neutralino{2,1} + \nu $, have no electrons in the final state (except for very soft leptons), and therefore will not contribute to the signal. 

\begin{figure}[h]
\includegraphics[width=7.5cm,height=4.5cm]{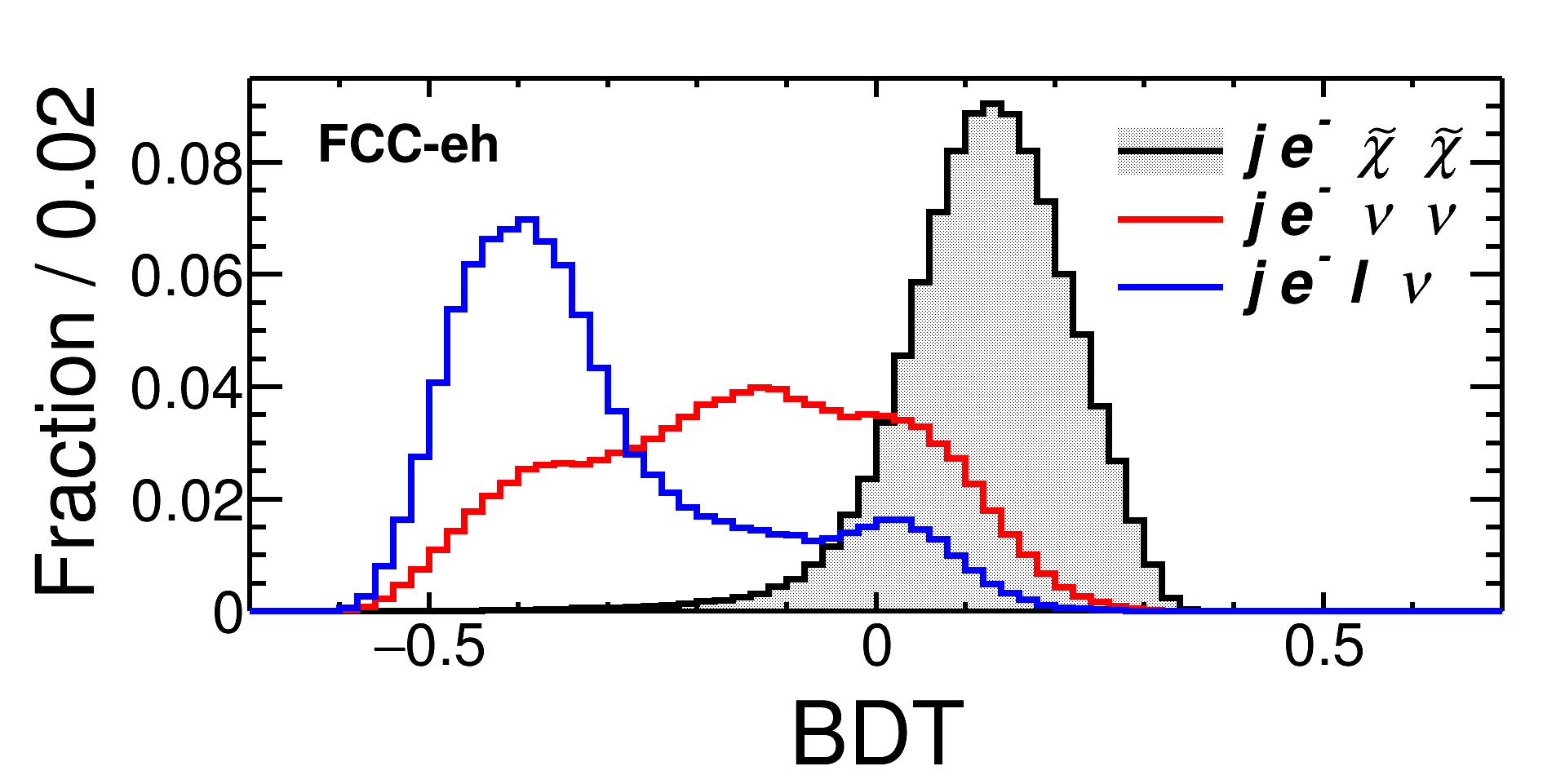}
\includegraphics[width=7.5cm,height=4.5cm]{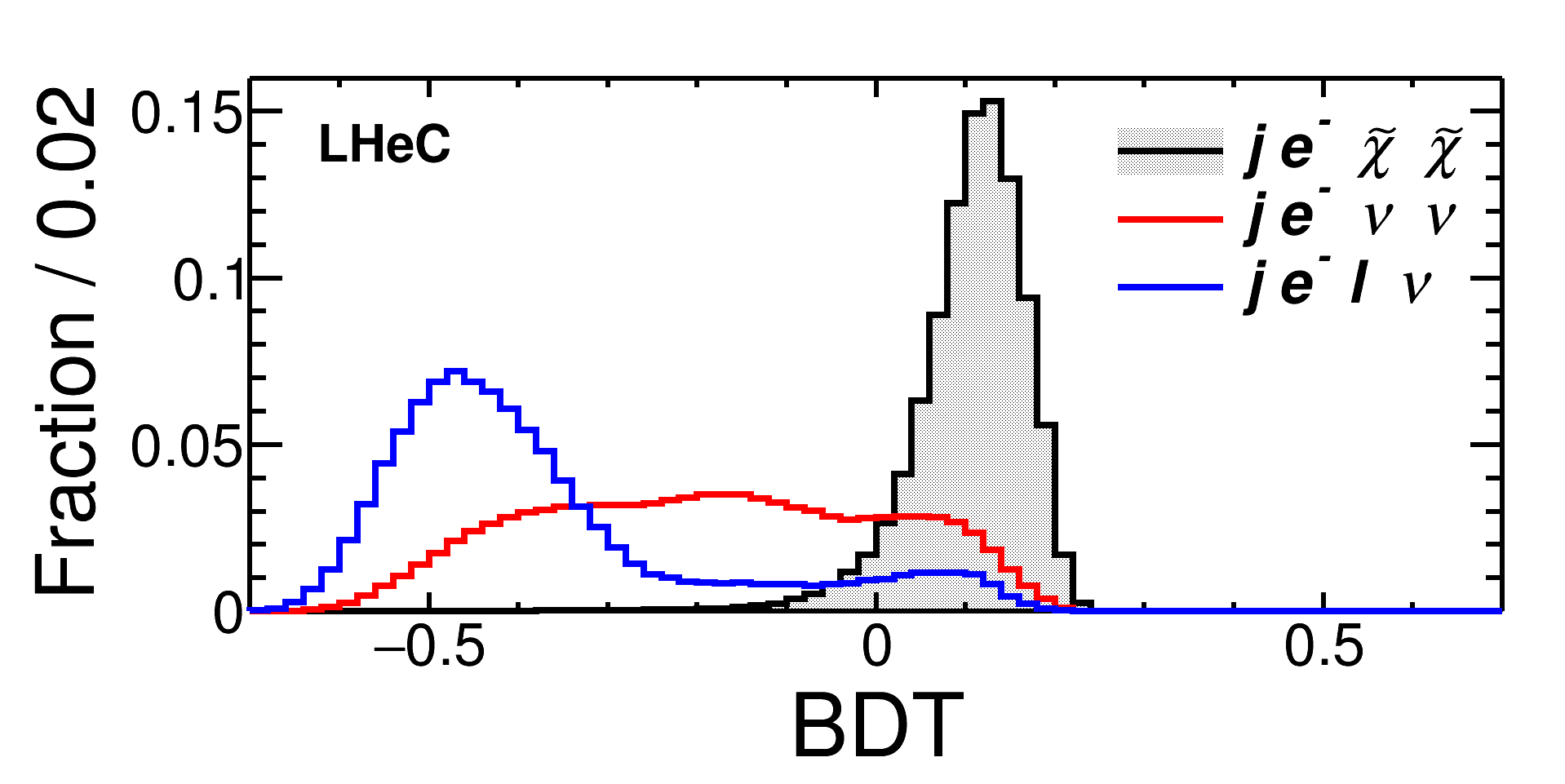}
\caption{Distributions of BDT response at the FCC-eh (upper) and the LHeC (lower) with the unpolarized electron beam for signal $j\, e^-\, \tilde{\chi} \tilde{\chi}$ (black with filled area) in the compressed-slepton scenario, and the SM background of the $j\, e^-\, \nu \nu$ (red) and $j\, e^-\, \ell \nu$ (blue) processes after applying the pre-selection cuts.
For signals, $m_{\chargino1, \neutralino2}$ is 400 GeV and 250 GeV at the FCC-eh and LHeC, respectively.}
\label{fig:BDT_compressedSlep_fixdm}
\end{figure}

Figure~\ref{fig:obs_FCCeh_compressedSlep_m400} of the Appendix shows kinematic distributions of the input observables, both for signal and background, for the FCC-eh case assuming an unpolarized electron beam (P(e$^{-}$)=0).
A SUSY model with $m_{\chargino1, \neutralino2}=400\GeV$ and production $e^-p\to j\, e^-\, \tilde{\chi} \tilde{\chi}$ is assumed, while the two sources of SM backgrounds, $e^-p\to j\, e^-\, \nu \nu$  and $j\, e^-\, \ell \nu$, are shown separately. Distributions of the corresponding BDT response are presented in the upper panel of Fig.~\ref{fig:BDT_compressedSlep_fixdm}. 

\begin{table}[h]
\begin{tabular}{c|c|cc}
\hline
\hline
FCC-eh [1 $\iab$] & Signal & \multicolumn{2}{c}{Background} \\
\hline
$m_{\chargino1, \neutralino2}$ [GeV]&  400 &\multirow{2}{*}{$j\, e^-\, \nu \nu$} &\multirow{2}{*}{$j\, e^-\, \ell \nu$} \\
\,\,\,\,\,\,\,\,\,\,\,$m_{\slepton}\,$   [GeV]&  435 &  & \\
\hline
initial            & 4564 & $1.08 \times 10^6$ & $7.96 \times 10^6$ \\
Pre-selection      & 3000 & $3.87 \times 10^5$ & $5.71 \times 10^5$ \\
${\rm BDT}>0.262$  &  149 &                600 &                86 \\
\hline
$\sigma_{\rm stat+syst}$ & 3.3 & &  \\
\hline
\hline
\end{tabular}
\caption{Number of events after selections applied sequentially on SUSY signal $j\, e^-\, \tilde{\chi} \tilde{\chi}$ with $m_{\chargino1, \neutralino2}$ = 400 GeV in the compressed-slepton scenario, and for the SM background processes of $j\, e^-\, \nu \nu$ and $j\, e^-\, \ell \nu$. The numbers of events correspond to an integrated luminosity of $1~\mathrm{ab}^{-1}$ at the FCC-eh with unpolarized electron beam.
The significances including 5\% systematic uncertainties on the background are presented in the last row.}
\label{tab:FCCeh_compressedSlep_m400}
\end{table}

\begin{table}[h]
\begin{tabular}{c|c|cc}
\hline
\hline
LHeC [1 $\iab$] & Signal & \multicolumn{2}{c}{Background} \\
\hline
$m_{\chargino1, \neutralino2}$ [GeV]&  250 &\multirow{2}{*}{$j\, e^-\, \nu \nu$} &\multirow{2}{*}{$j\, e^-\, \ell \nu$} \\
\,\,\,\,\,\,\,\,\,\,\,$m_{\slepton}\,$ [GeV]  &  285 &  & \\
\hline
initial            & 1231 & $2.80 \times 10^5$ & $2.01 \times 10^6$ \\
Pre-selection      &  453 & $6.60 \times 10^4$ & $1.66 \times 10^5$ \\
${\rm BDT}>0.172$ & 49 &                486 &                278 \\
\hline
$\sigma_{\rm stat+syst}$ & 1.0 & &  \\
\hline
\hline
\end{tabular}
\caption{Number of events after selections applied sequentially on SUSY signal $j\, e^-\, \tilde{\chi} \tilde{\chi}$ with $m_{\chargino1, \neutralino2}$ = 250 GeV in the compressed-slepton scenario, and for the SM background processes of $j\, e^-\, \nu \nu$ and $j\, e^-\, \ell \nu$. The numbers of events correspond to an integrated luminosity of $1~\mathrm{ab}^{-1}$ at the LHeC with unpolarized electron beam.
At each stage, the significance including 5\% systematic uncertainties on the background are presented in the last row.}
\label{tab:LHeC_compressedSlep_m250}
\end{table}

The number of events at each selection stage is shown in Table~\ref{tab:FCCeh_compressedSlep_m400}, which corresponds to an integrated luminosity of $1~\mathrm{ab}^{-1}$.
The optimized value for the BDT is found to be 0.262 for this SUSY model.
Assuming a systematic uncertainty of 5\%, i.e., $\sigma_b=0.05N_b$, the number of events after the optimized BDT cut, $N_s=149$ and $N_b=686$, results in a significance of $\sigma_{\rm stat+syst}=3.3$
\footnote{In case the $\eta$ range of the jet is restricted to $\pm~ 5.5$, equivalent to a less-than-ideal calorimeter coverage at the FCC-eh, both the signal and background rates are reduced by less than 5\%, and the significance is only slightly reduced.}. The 5\% systematic uncertainties on SM backgrounds are assumed to include experimental sources (e.g. jet energy scale and resolution, lepton identification) and modelling uncertainties (renormalisation and factorisation scale choices, PDF). They are expected to be constrained by semi data-driven techniques and improved Monte Carlo predictions and the 5\% value is assumed on the basis of a conservative extrapolation of uncertainties estimated in $pp$ monojet prospect studies at HL-LHC~\cite{ATL-PHYS-PUB-2018-043}.

The same analysis is performed for the LHeC case, where we assume $m_{\chargino1, \neutralino2}=250\GeV$ as the benchmark masses. The distributions of input observables are shown in Fig.~\ref{fig:obs_LHeC_compressedSlep_m250} and the BDT-score distribution is reported in the lower panel of Fig.~\ref{fig:BDT_compressedSlep_fixdm}.
The cut flow is given in Table~\ref{tab:LHeC_compressedSlep_m250}, where the optimized BDT cut of 0.172 is used.
Considering an integrated luminosity of $1~\mathrm{ab}^{-1}$ and 5\% systematic uncertainty on the background, the significance of about 1.0 is obtained for this benchmark point.

\begin{figure}[h]
\includegraphics[width=7.5cm,height=4.5cm]{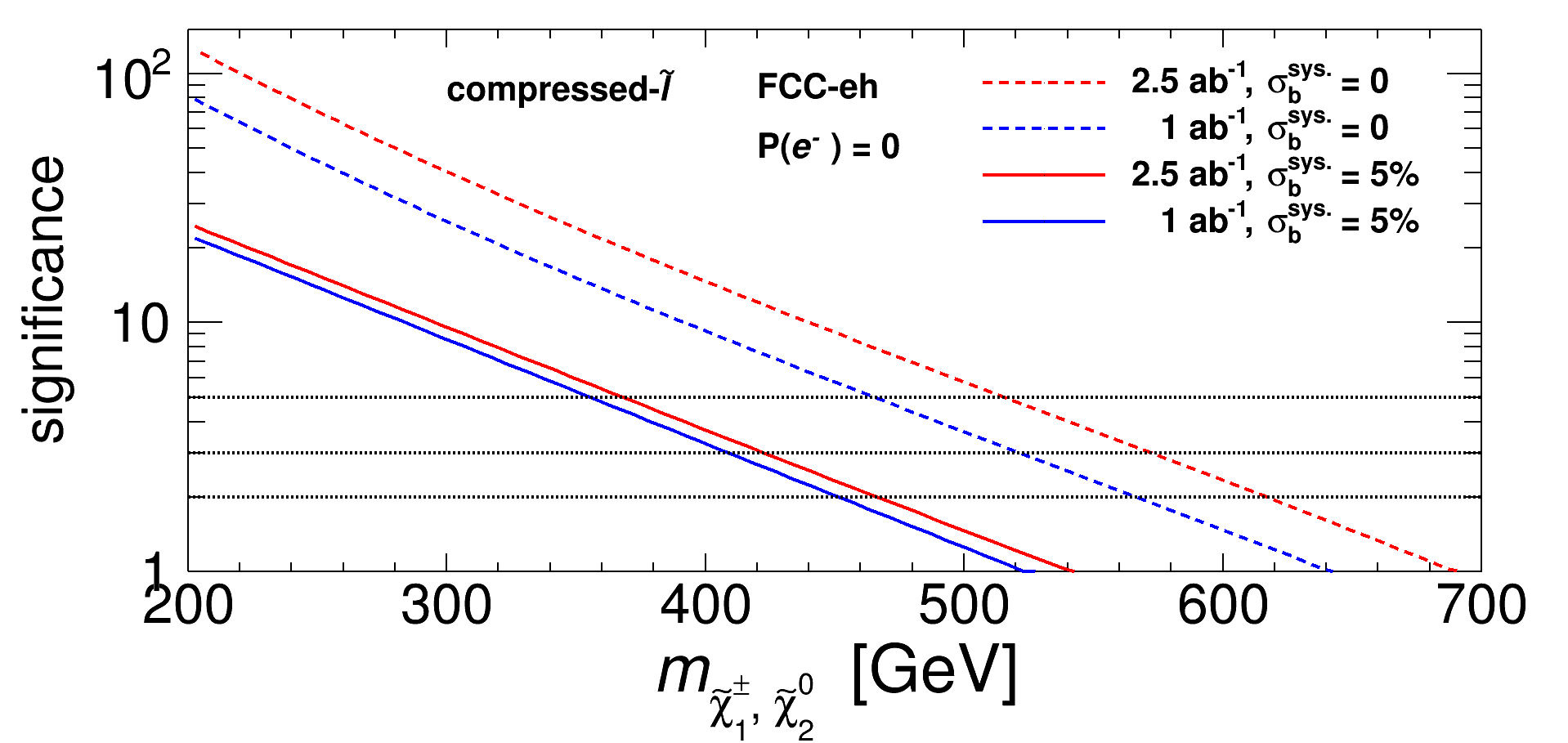}
\includegraphics[width=7.5cm,height=4.5cm]{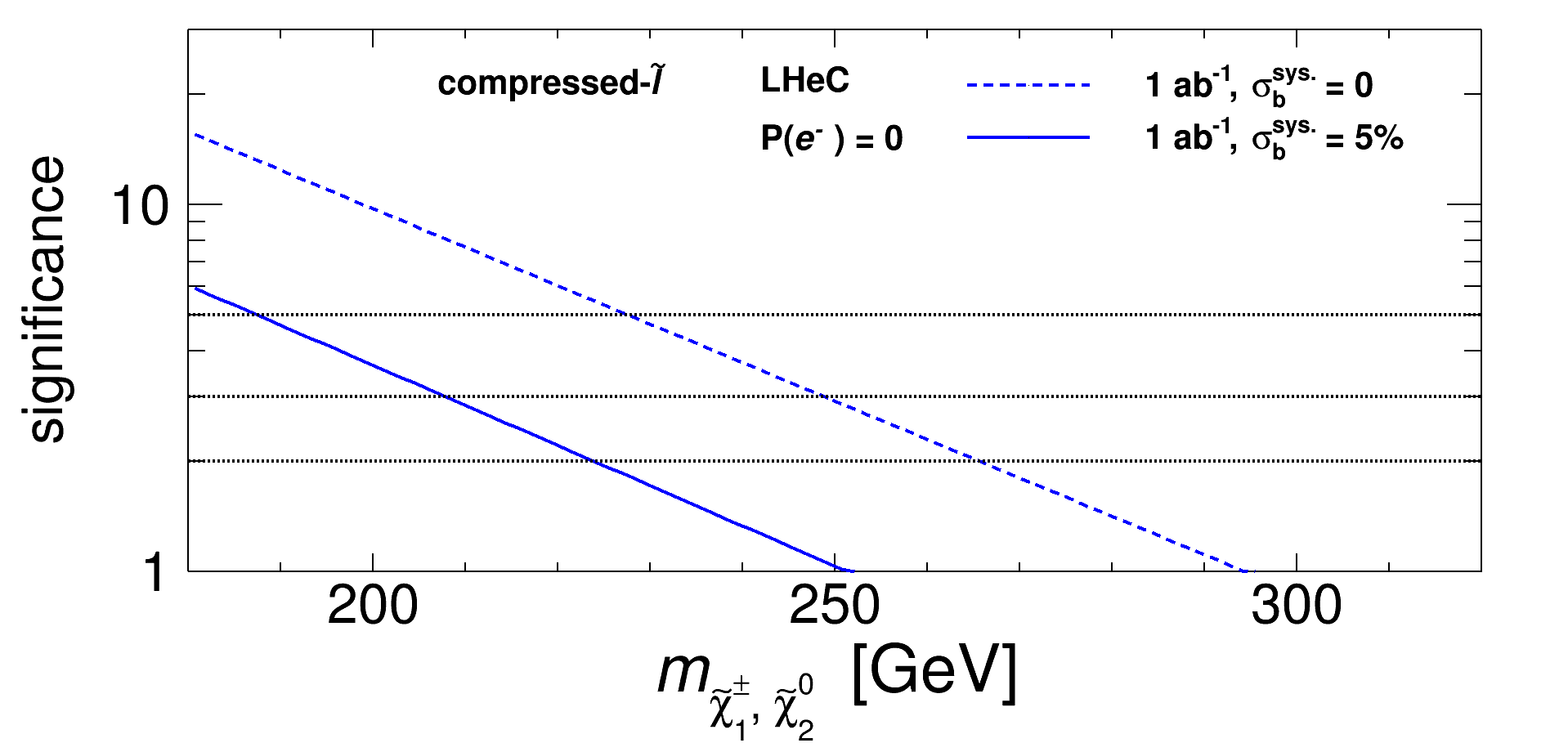}
\caption{Significances as varying the masses of $\chargino1$ and $\neutralino2$ for the compressed-slepton scenario. Upper plot: at the FCC-eh with unpolarized beams and integrated luminosities of 1 $\iab$ and 2.5 $\iab$; Lower plot: at the LHeC with unpolarized beams and 1 $\iab$ luminosity. 
For dashed (solid) curve, a systematic uncertainty of 0\% (5\%) on the background is considered.
}
\label{fig:sgf_compressedSlep}
\end{figure}
In the upper panel of Fig.~\ref{fig:sgf_compressedSlep}, the significance curves as a function of $m_{\chargino1,\neutralino2}$ for the FCC-eh with unpolarized electron beam and integrated luminosities of 1 $\iab$  and 2.5 $\iab$ are presented. 
With 0\% (5\%) systematic uncertainty on the background and 2.5 $\iab$ integrated luminosity, the 2-$\sigma$ limits on the $\chargino1$, $\neutralino2$ mass are 616 (466) GeV, while the 5-$\sigma$ discovery limits are 517 (367) GeV.

The lower panel of Fig.~\ref{fig:sgf_compressedSlep} shows the significance curves at the LHeC with unpolarized electron beam and an integrated luminosity of 1 $\iab$.
With 0\% (5\%) systematic uncertainty on the background, the limits on the mass are 266 (224) GeV and 227 (187) GeV corresponding to the 2 and 5-$\sigma$ significances, respectively.

\subsection[Effects of Varying Delta-m]{Effects of Varying $\Delta m$ }
\label{subsec:varySlep}

\begin{figure}[H] 
\includegraphics[width=4cm,height=3cm]{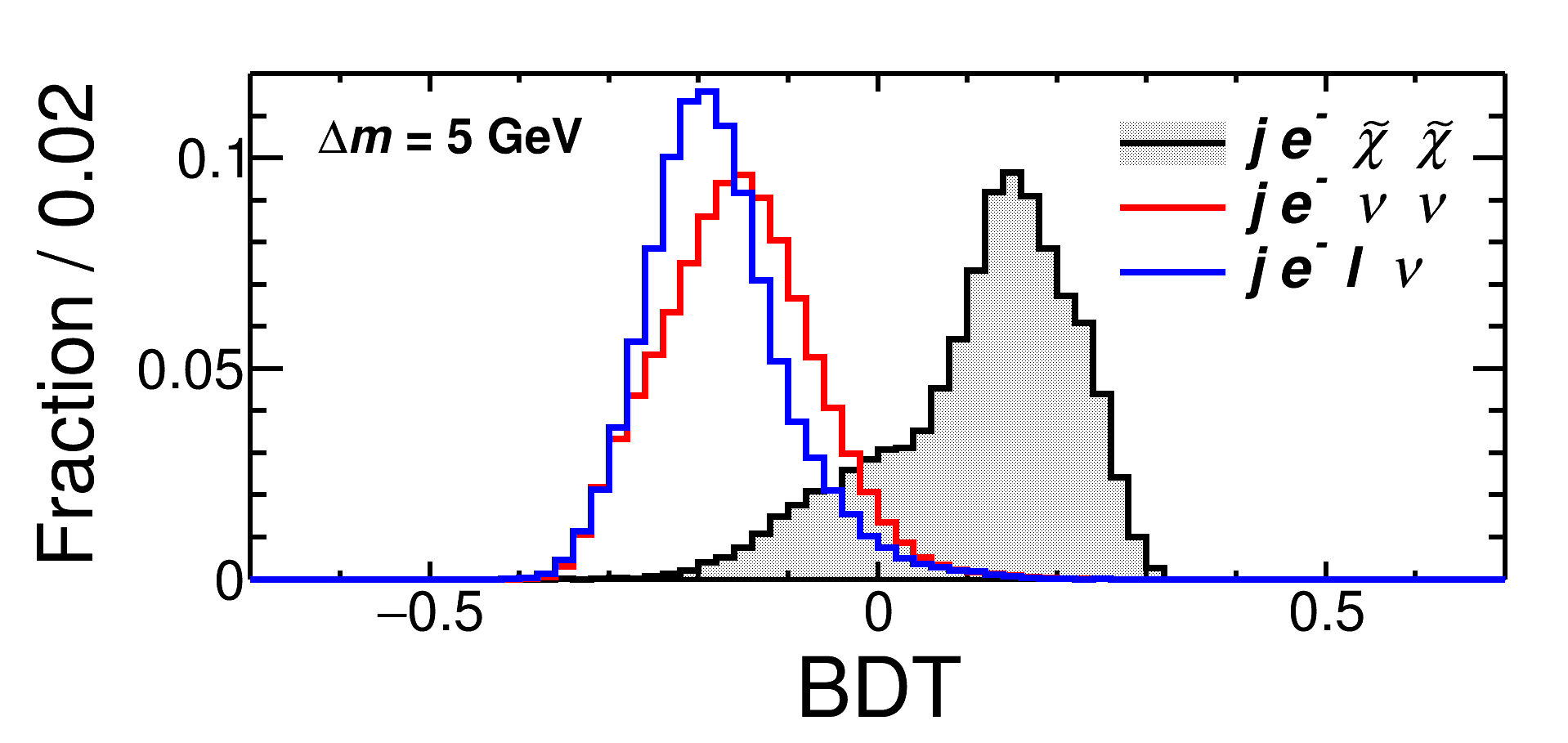}
\includegraphics[width=4cm,height=3cm]{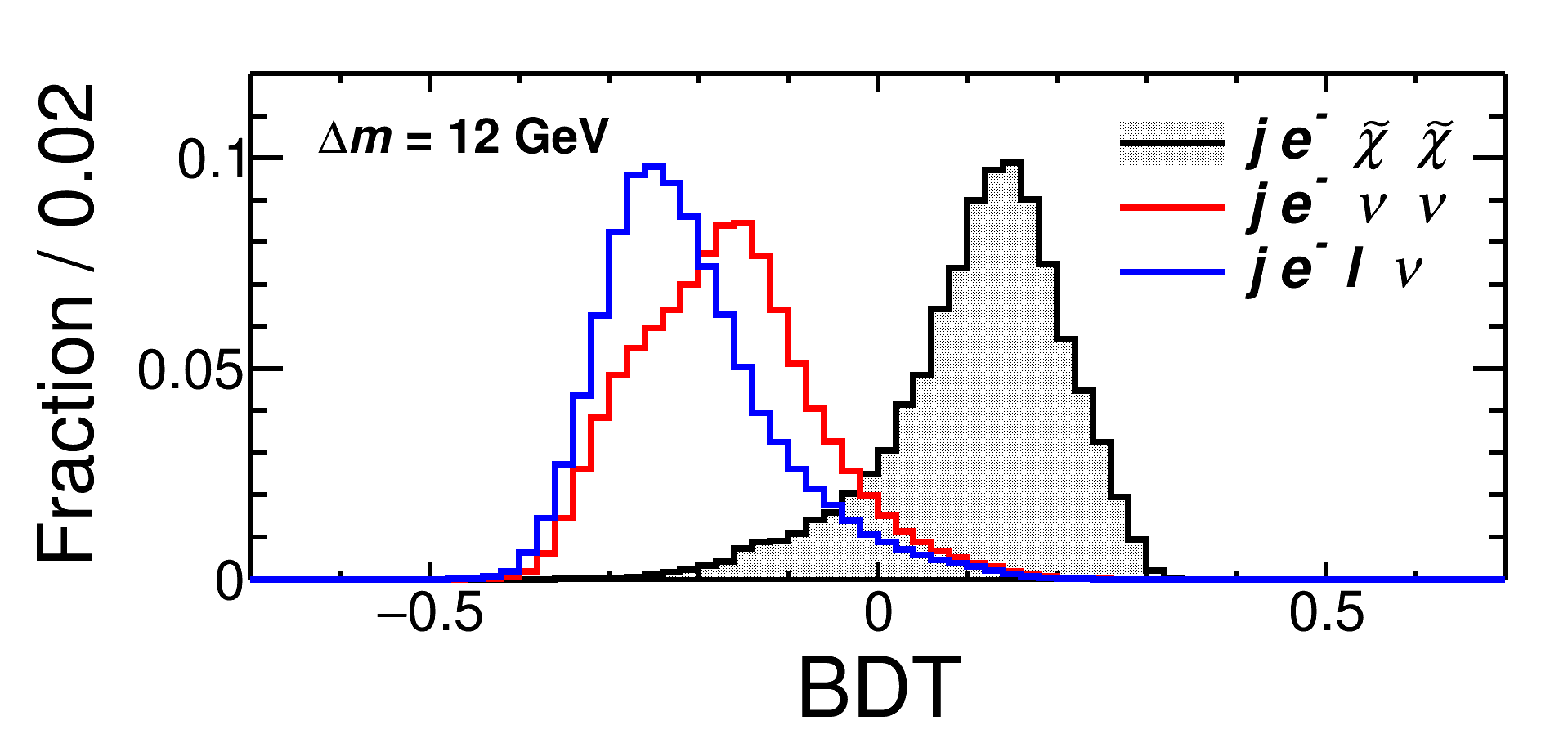}
\includegraphics[width=4cm,height=3cm]{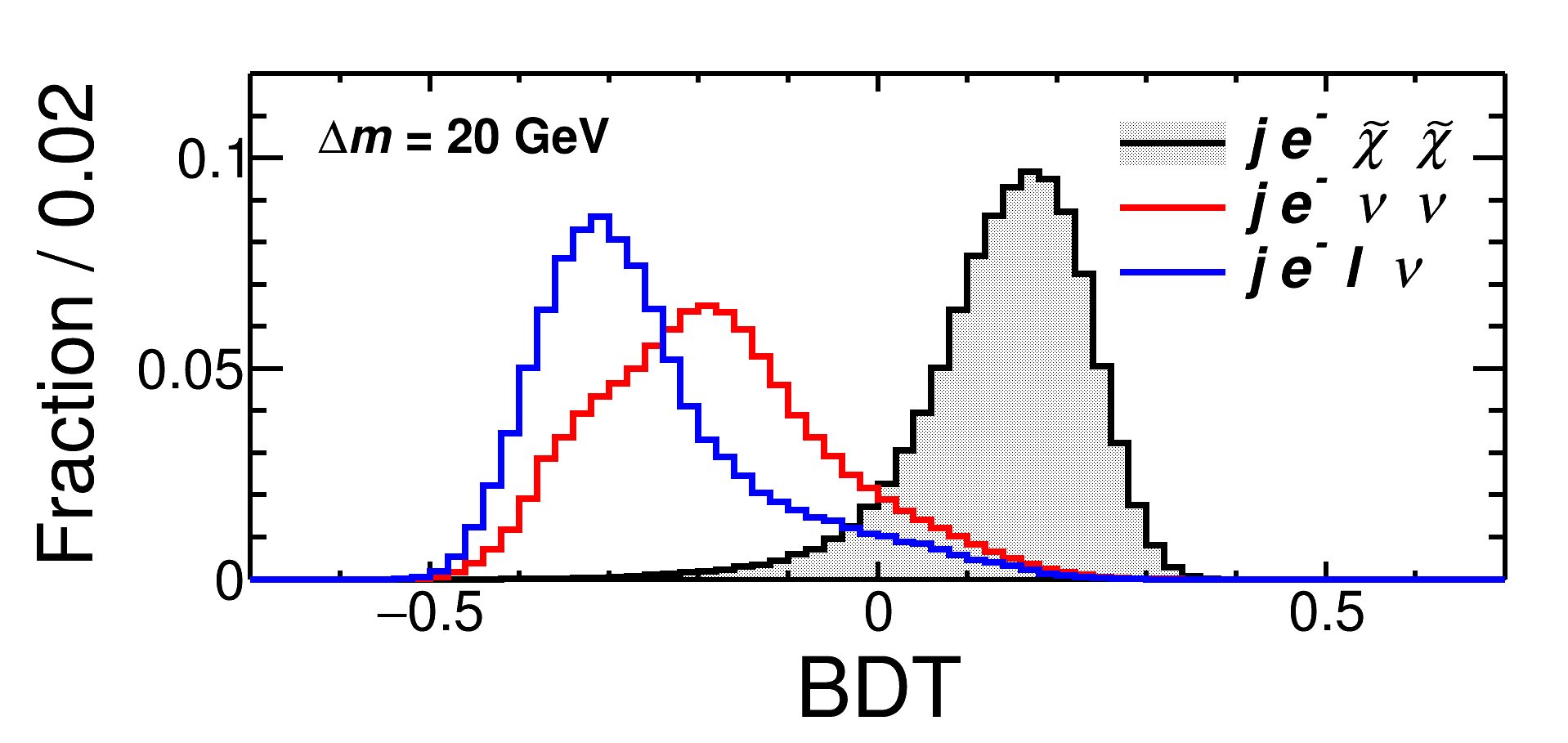}
\includegraphics[width=4cm,height=3cm]{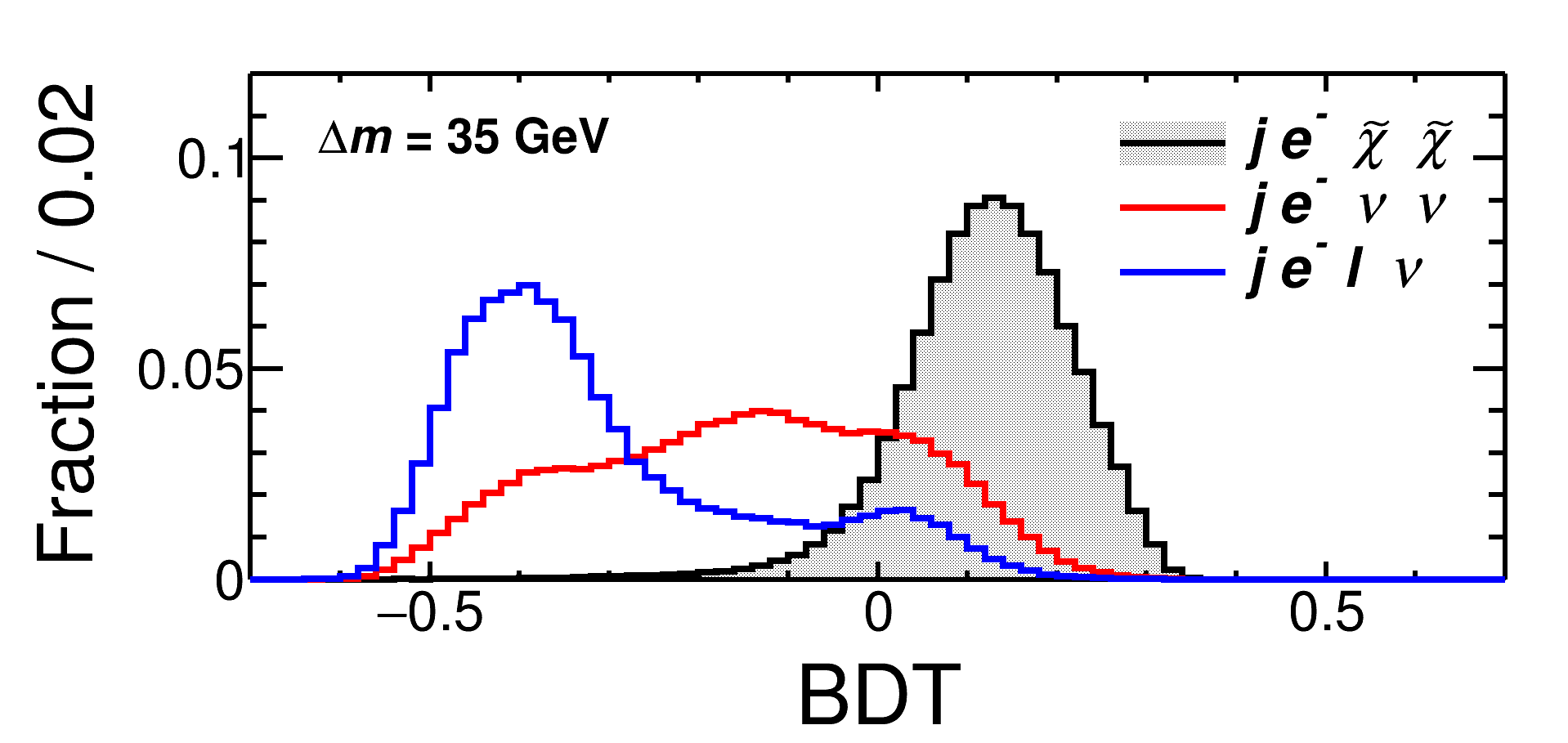}
\includegraphics[width=4cm,height=3cm]{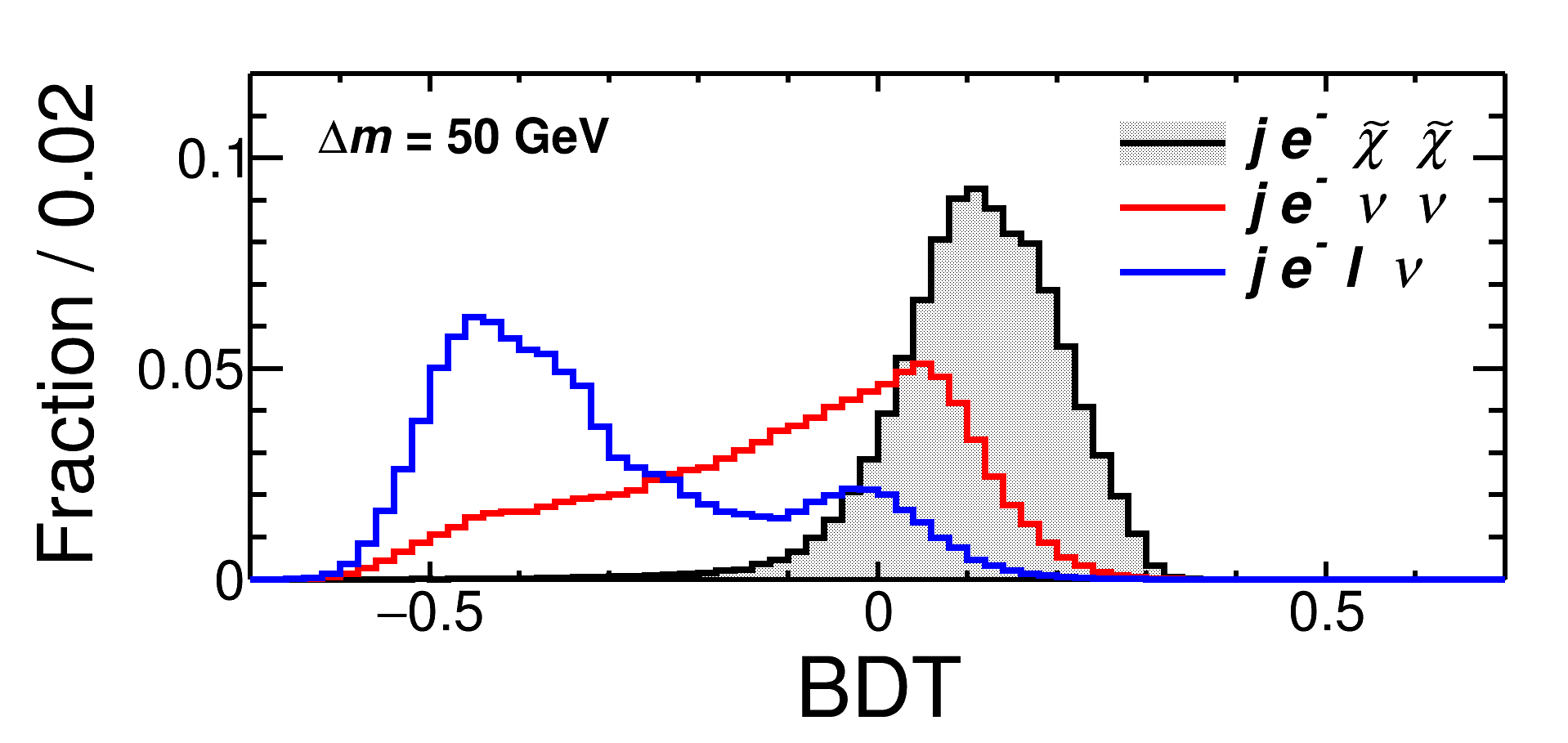}
\includegraphics[width=4cm,height=3cm]{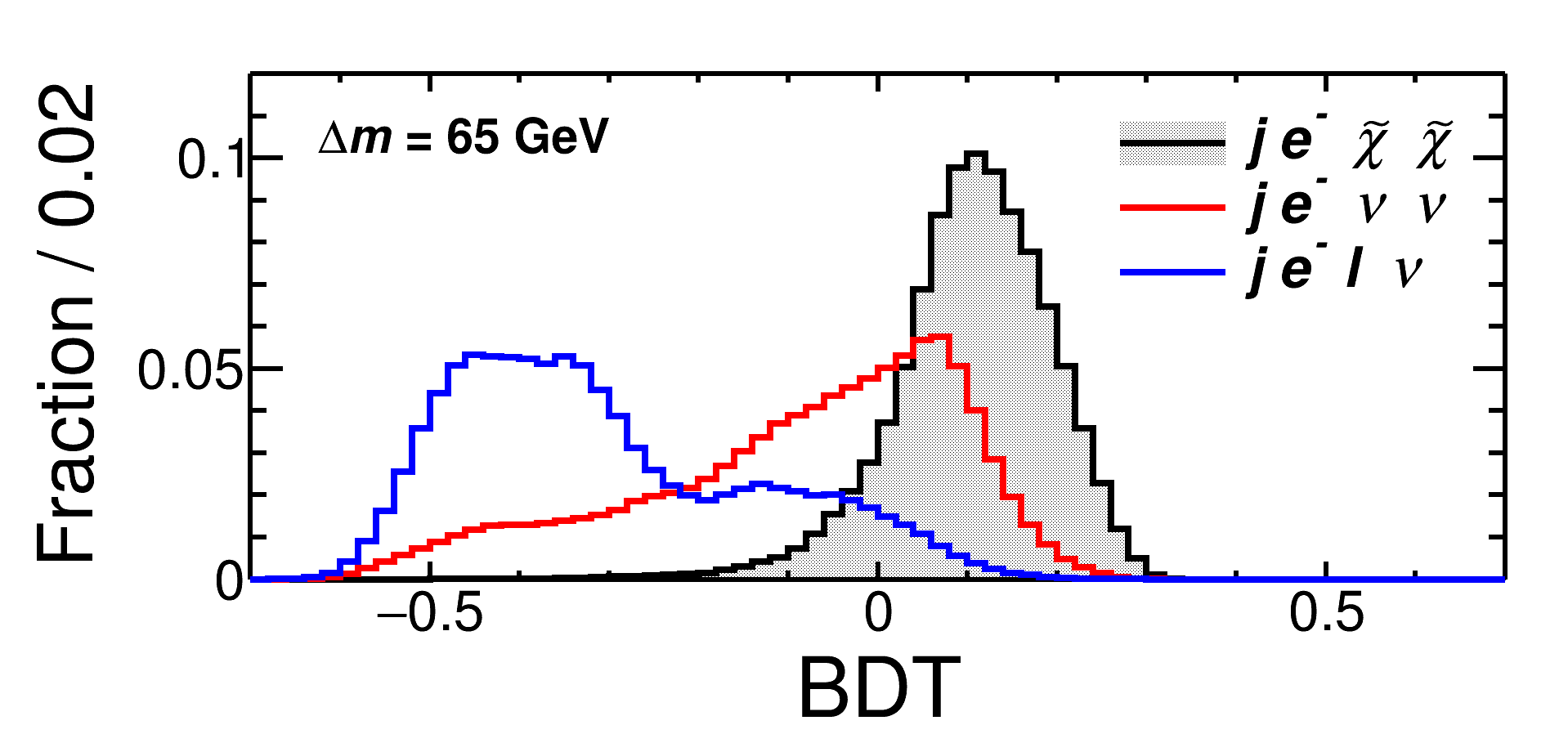}
\caption{Distributions of the BDT response when fixing $m_{\chargino1, \neutralino2} = $ 400 GeV and varying the mass difference $\Delta m = m_{\slepton} - m_{\chargino1, \neutralino2}$ in the compressed-slepton scenario at the FCC-eh.
}
\label{fig:BDT_FCCeh_compressedSlep_varySlep}
\end{figure}

The effect of varying the assumption on $\Delta m$ is evaluated. 
A range between 5 to 65 GeV is considered, with  $m_{\chargino1, \neutralino2}$ being fixed at 400 GeV~\footnote{When $\Delta m <$ 9 GeV, $\tan\beta$ is changed from 30 to smaller value such that the LSP can still be $\neutralino1$, rather than $\sneutrino$.}.
The distributions of the BDT response at the FCC-eh for different $\Delta m$ assumptions are presented in Fig.~\ref{fig:BDT_FCCeh_compressedSlep_varySlep}.
Although the same background is analyzed, the BDT distributions still change for different $\Delta m$ assumptions, because the signal kinematics varies with $\Delta m$ values.
As $\Delta m$ increases,
the kinematics of the 2-neutrino background $j\, e^-\, \nu \nu$ becomes similar to that of the signal process $j\, e^-\, \tilde{\chi} \tilde{\chi}$, which renders its BDT distribution largely overlapping with that of the signal. When $\Delta m > $ 35 GeV, part of the 1-neutrino background $j\, e^-\, \ell \nu$ events also begins to mimic the signal.
Because of clearer distinctions between the BDT distributions of the signal and background, the BDT is most effective in rejecting the SM background if $\Delta m$ between 10 and 40~GeV. 

\begin{figure}[h]
\includegraphics[width=7.5cm,height=4.5cm]{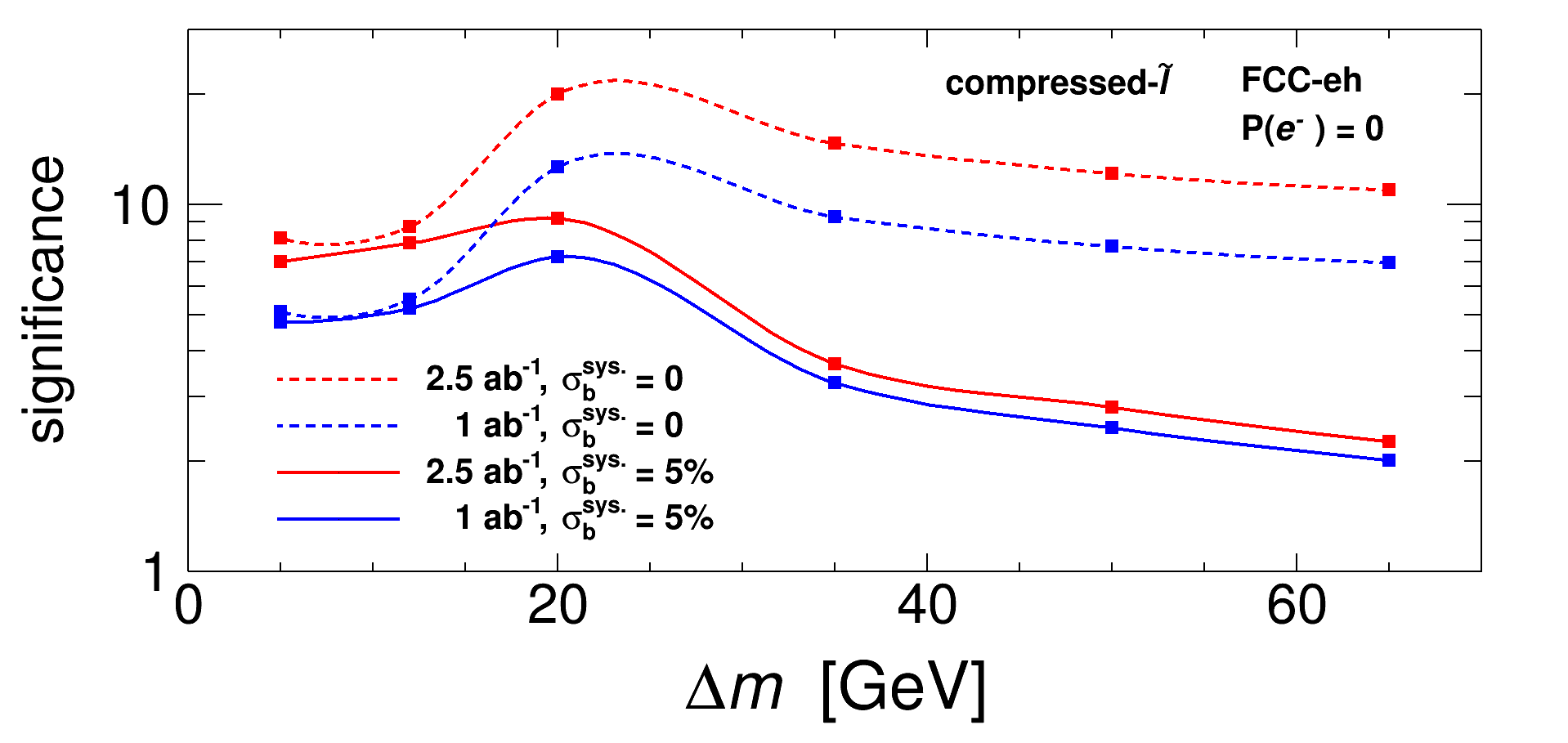}
\caption{Significance shown as a function of the mass difference $\Delta m = m_{\slepton} - m_{\chargino1, \neutralino2}$ when fixing electroweakino masses,  $m_{\chargino1, \neutralino2} = $ 400 GeV, in the compressed-slepton scenario at the FCC-eh.}
\label{fig:sgf_FCCeh_compressedSlep_varySlep}
\end{figure}

The significances assuming either 5\% or 0\% systematic uncertainties on the background as a function of the mass differences $\Delta m$ are reported in  Fig.~\ref{fig:sgf_FCCeh_compressedSlep_varySlep} for luminosities of 1 and 2.5  $\iab$.
It is found that the $\Delta m$ = 20 GeV can be tested with maximum efficiency.
This is because, as shown in Fig.~\ref{fig:BDT_FCCeh_compressedSlep_varySlep}, for small $\Delta m$, the BDT distributions present a better separation between signal and background.
However, when $\Delta m <<$ 20 GeV, the electron from the slepton and sneutrino decays become too soft to pass the pre-selections, $p_T(e^-) > $ 10 GeV, suppressing the signal number of events.

It is worth noting that, although the significances for $\Delta m =$ 12 and 5 GeV are slightly lower than $\Delta m =$ 20 GeV, when 5\% systematic uncertainty on background is included, they are still larger than the benchmark case with $\Delta m =$ 35 GeV. However, when no systematic uncertainty on background is considered, $\Delta m =$ 35 GeV gives larger significance compared to $\Delta m =$ 12 and 5 GeV. 
Overall, significances are quite sensitive to the assumption of systematic uncertainty when $\Delta m >$ 20 GeV. Therefore, to enhance the discovery potential to the SUSY electroweak sector, it will be important to control the level of systematic uncertainties on the background.

\section{Decoupled-Slepton Scenario}
\label{sec:reuslts_decoupledSlep}

\noindent
In the decoupled-slepton scenario, where the sleptons are assumed to have multi-TeV mass, the signal events are mainly arising from the VBF processes of charginos and neutralinos, $p e^- \to j e^-\tilde\chi\tilde\chi$.
As this is a four-body production process, the signal cross sections are considerably lower than in the case of compressed-slepton scenario (cf.~Fig.~\ref{fig:crs}). At the LHeC they are of the order of 1 fb or lower, hence the decoupled-slepton scenario is only considered for the FCC-$eh$ case.

\begin{figure}[h]
\includegraphics[width=7.5cm,height=4.5cm]{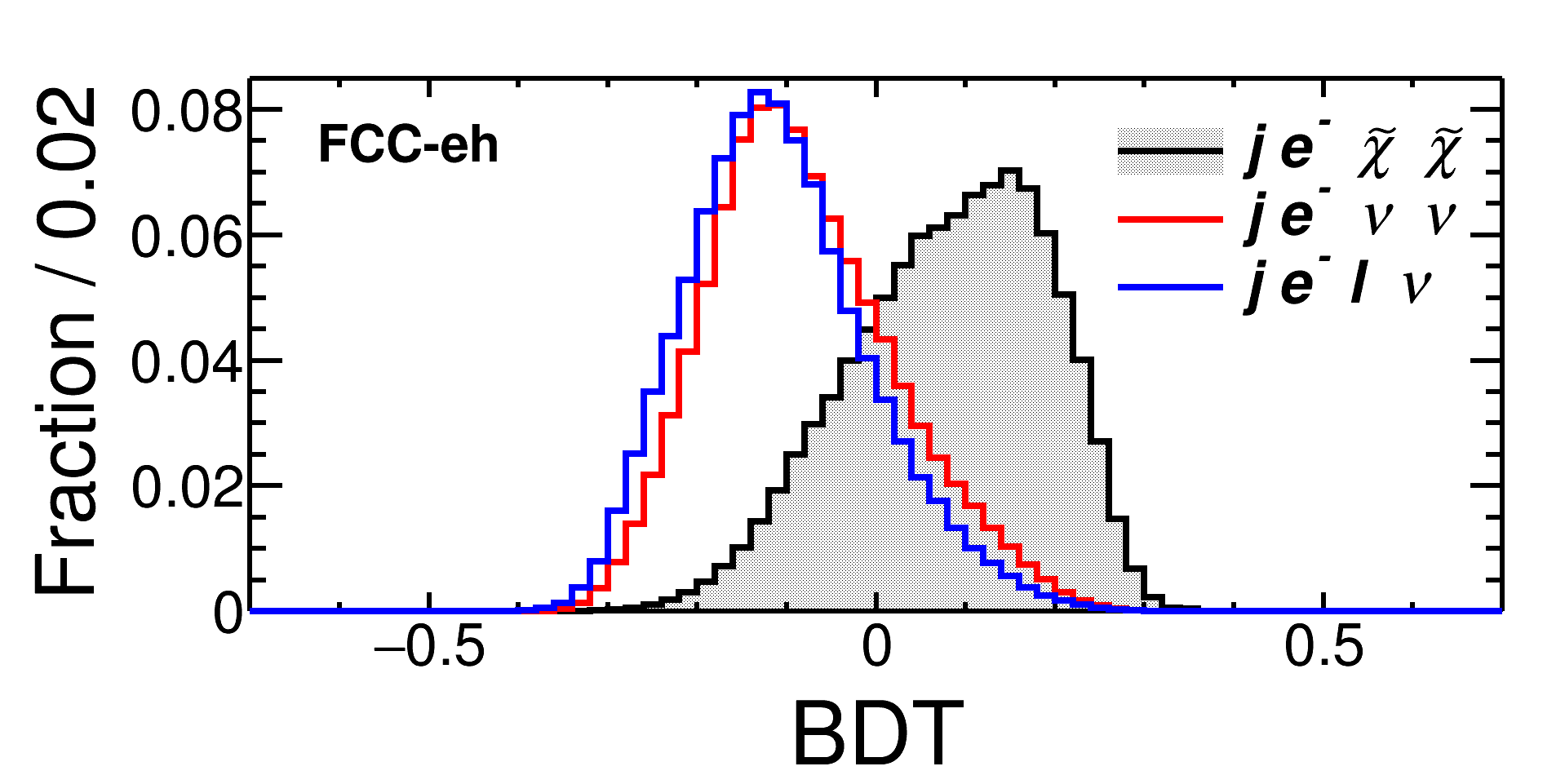}
\caption{Distributions of BDT response at the FCC-eh
for signal $j\, e^-\, \tilde{\chi} \tilde{\chi}$ with $m_{\chargino1, \neutralino2}$ = 250 GeV (black with filled area) in the decoupled-slepton scenario, and SM background processes of $j\, e^-\, \nu \nu$ (red) and $j\, e^-\, \ell \nu$ (blue) after applying the pre-selections. }
\label{fig:BDT_FCCeh_decoupledSlep_m250}
\end{figure}

As in the previous section, the kinematic distributions of input observables for events passing the pre-selections are shown in Fig.~\ref{fig:obs_FCCeh_decoupledSlep_m250} in the Appendix~\ref{app:Obs}, where the signal events are from the process $p e^- \to j e^-\tilde\chi\tilde\chi$ with $m_{\chargino1, \neutralino2}=250\GeV$ and the SM backgrounds are from of $j\, e^-\, \nu \nu$ and $j\, e^-\, \ell \nu$.
The corresponding BDT-inputs distributions are presented in Fig.~\ref{fig:BDT_FCCeh_decoupledSlep_m250}.
Table~\ref{tab:FCCeh_decoupledSlep_m250} shows the number of events after each   selection is applied sequentially corresponding to 1 $\iab$. In the last row,  significances calculated assuming 5\% systematic uncertainties for the backgrounds are also included.

\begin{table}[h]
\begin{tabular}{c|c|cc}
\hline
\hline
FCC-eh [1 $\iab$] & Signal & \multicolumn{2}{c}{Background} \\
\hline
$m_{\chargino1, \neutralino2}$ [GeV]&  250 &\multirow{2}{*}{$j\, e^-\, \nu \nu$} &\multirow{2}{*}{$j\, e^-\, \ell \nu$} \\
\,\,\,\,\,\,\,\,\,\,\,$m_{\slepton}\,$ [GeV]  &    - &  & \\
\hline
initial            &  909 & $1.08 \times 10^6$ & $7.96 \times 10^6$ \\
Pre-selection      &  399 & $3.87 \times 10^5$ & $5.71 \times 10^5$ \\
${\rm BDT}>0.251$  & 14 &                326 &                357 \\
\hline
$\sigma_{\rm stat+syst}$ & 0.3 & &  \\
\hline
\hline
\end{tabular}
\caption{Number of events after selections applied sequentially on the signal $j\, e^-\, \tilde{\chi} \tilde{\chi}$ with $m_{\chargino1, \neutralino2}$ = 250 GeV in the decoupled-slepton scenario, and for the SM background processes $j\, e^-\, \nu \nu$ and $j\, e^-\, \ell \nu$. The numbers correspond to an integrated luminosity of $1~\mathrm{ab}^{-1}$ at the FCC-eh with unpolarized electron beam. The significances, calculated including 5\% systematic uncertainties on the background, are presented in the last row.}
\label{tab:FCCeh_decoupledSlep_m250}
\end{table}

\begin{figure}[h]
\includegraphics[width=7.5cm,height=4.5cm]{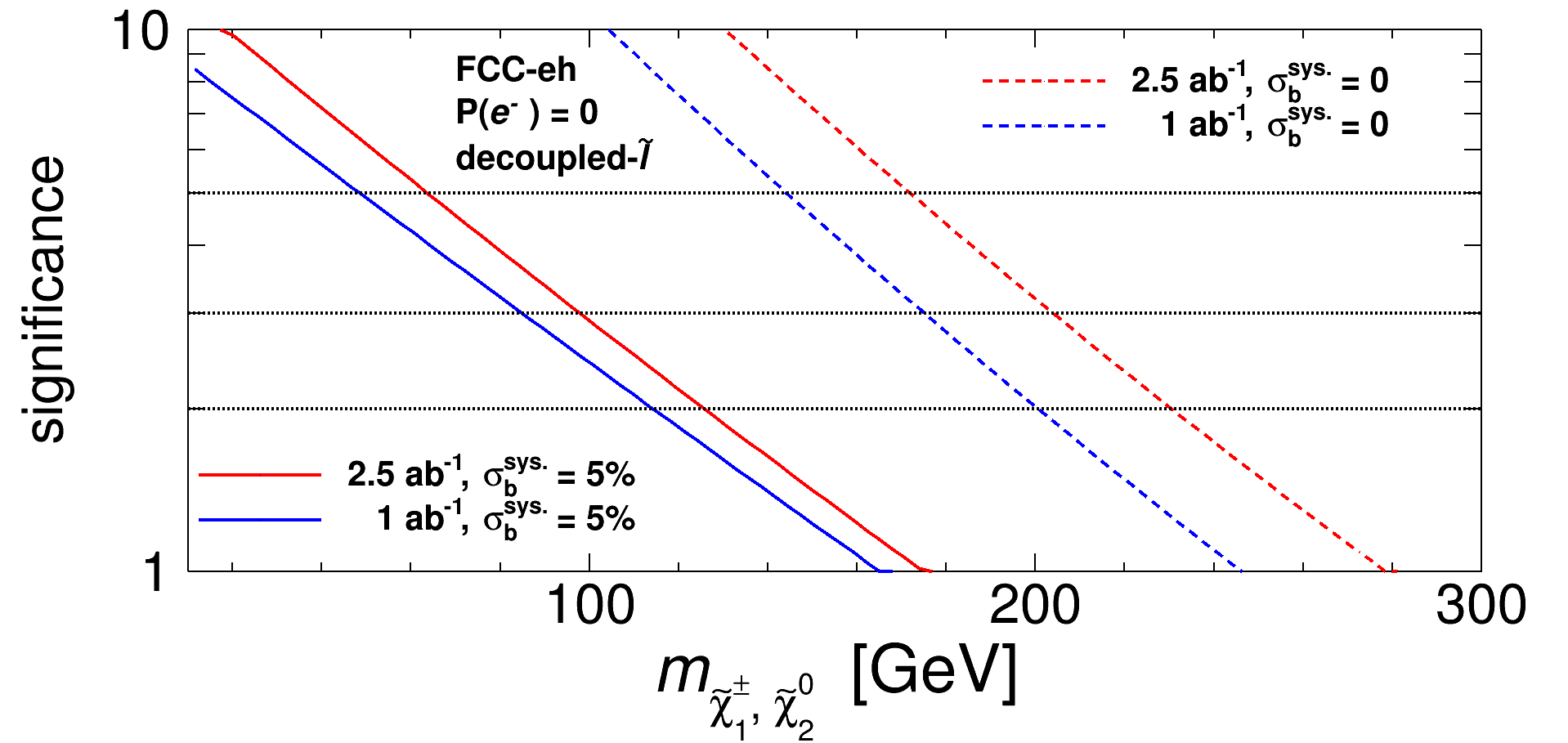}
\caption{Significances as a function of the masses of $\chargino1$ and $\neutralino2$ for the decoupled-slepton scenario at the FCC-$eh$ with unpolarized beams and integrated luminosities of 1 $\iab$ and 2.5 $\iab$. For dashed (solid) curve, a systematic uncertainty of 0\% (5\%) on the background is considered.}
\label{fig:sgf_decoupledSlep}
\end{figure}

Figure~\ref{fig:sgf_decoupledSlep} shows the significance as a function of $m_{\chargino1,\neutralino2}$.
When considering 0\% (5\%) systematic uncertainty on the background, 
the 2-$\sigma$ limits on the $\chargino1$ and $\neutralino2$ masses are 230 (125) GeV for 2.5 $\iab$ luminosity.
Because of the small signal production, the discovery power for the decoupled-slepton scenario is limited at the future $ep$ colliders.

\section{Effects of Beam Polarization}
\label{sec:beamPol}

\noindent
The possibility of having a polarized electron beam at the LHeC and FCC-eh colliders could potentially enhance the sensitivity to electroweakinos.  A high degree of longitudinal polarization could be envisaged ($|{\rm P} (e^{-})| = 80$\%).
When applying an electron beam polarization of -80\% (+80\%) at the FCC-eh, the production cross sections of SM background processes increase (decrease) by about 55\% (20\%), compared to the rate in case of unpolarized electron beam. 
On the other hand, in the case of the compressed-slepton scenario with benchmark $m_{\chargino1, \neutralino2}$ = 400 GeV and $m_{\slepton}$ = 435 GeV, the signal production cross section increases (decreases) by about 80\%.
In the decoupled-slepton scenario with $m_{\chargino1, \neutralino2}$ = 250 GeV, the signal production cross section increases (decreases) by about 40\%. 
The changes in the cross sections are due to the electroweak nature of the interactions involved and the assumptions made: the sleptons considered are superpartners of left-handed electrons, and the gauge bosons and Wino-like electroweakinos couple more to left-handed electrons or sleptons.
A negative polarization can be therefore advantageous for such searches of electroweak SUSY particles, whilst a positive polarization might reduce the sensitivity, although a new optimization of the analysis would be needed in this case. 

Compared with the unpolarized electron beam with 1 $\iab$ luminosity, the -80\% electron beam polarization increases the significances for the benchmark mass hypothesis in compressed-slepton scenarios by about 40\% (15\%), from 9 to 13 (3.3 to 3.8) when including 0\% (5\%) systematic uncertainty on the background. In case of decoupled-slepton scenarios, the -80\% electron beam polarization increases the significances by about 30\% (10\%), from 1.0 to 1.3 (0.32 to 0.35) when including 0\% (5\%) systematic uncertainty on the background. 
Hence, thanks to the much larger signal production, our preliminary analysis indicates that the -80\% electron beam polarization could lead to a stronger discovery potential for the SUSY electroweak sector especially for the compressed-slepton scenario with small systematic uncertainty.
A more accurate simulation and estimate of acceptances and yields in case of polarized electron beams is needed considering several mass hierarchy hypotheses, which is beyond the scope of this study and we leave it for future studies.

\section{Conclusions and Discussions}
\label{sec:Summary}

\noindent
The LHC experiments have produced impressive constraints on the masses of the SUSY coloured sector. 
Because of the low direct production cross sections for neutralinos, charginos and sleptons, the LHC constraints on their masses are limited, particularly in the compressed scenarios where the experimental analysis is challenging, due to the presence of soft momenta decay products.
In this article, we have reported a search strategy for the SUSY electroweak sector and its discovery potential at the future electron-proton colliders, LHeC and FCC-eh.
Our study shows that the search for the SUSY electroweak sector in the compressed scenarios at the $ep$ colliders could be complementary to the analyses at the $pp$ colliders, mainly due to the cleaner environment and smaller SM background at $ep$ colliders.

We focus on the compressed scenarios where the LSP $\neutralino1$ is Bino-like, $\chargino1$ and $\neutralino2$ are Wino-like with almost degenerate masses, and the mass difference between $\neutralino1$ and $\chargino1$ is small. 
The signal is produced via the process ``$p\, e^- \to j\, e^-\, \tilde{\chi} \tilde{\chi}$", where $\tilde{\chi}=\neutralino1$, $\chargino1$ or $\neutralino2$. LO cross sections are considered for the SUSY signal models, hence results are conservative.  The kinematic observables are input to the TMVA package to perform a multivariate analysis at the detector level.

In the compressed-slepton scenario, the case where the left-handed slepton $\sleptonL$ and sneutrino $\sneutrino$ are slightly heavier than $\chargino1$ or $\neutralino2$ is considered.
When fixing the mass difference $\Delta m = m_{\slepton} - m_{\chargino1, \neutralino2} = $ 35 GeV and 
assuming no systematic uncertainty on the background, our analysis indicates that the 2 (5)-$\sigma$ limits on the $\chargino1$, $\neutralino2$ mass are 616 (517) GeV for 2.5 $\iab$ luminosity at the FCC-eh, and 266 (227) GeV for 1 $\iab$ luminosity at the LHeC, respectively. 
The effects of varying $\Delta m$ are investigated: fixing $m_{\chargino1, \neutralino2}$ to be 400 GeV, we find that at the FCC-eh the significance is maximal when $\Delta m$ is around 20 GeV.

In the decoupled-slepton scenarios where only $\neutralino1$, $\chargino1$ and $\neutralino2$ are light and other SUSY particles are heavy and decoupled, 
the 2-$\sigma$ limits obtained on the $\chargino1$, $\neutralino2$ mass are 230 GeV for 2.5 $\iab$ luminosity at the FCC-eh when 
assuming no
systematic uncertainty on the background. Large systematic uncertainties on the SM background processes can substantially affect the sensitivity, hence good control of experimental and theoretical sources of uncertainties is very important. 
Finally, it is found that the possibility of having a negatively polarized electron beam (${\rm P} (e^{-})= 80$\%) could potentially extend the sensitivity to electroweakinos by up to 40\%. 

Searches at $pp$ colliders usually target either hard or very soft leptons from the decays of the sleptons, charginos or neutralinos, which corresponds to either large or very small $\Delta m(\slepton, \neutralino1)$ and $\Delta m(\chargino1/\neutralino2, \neutralino1)$. Given the difficulty to probe intermediate regions with $\Delta m(\slepton, \neutralino1) \sim 35$ GeV and $\Delta m(\chargino1/\neutralino2, \neutralino1) \sim 20 - 50$ GeV, these scenarios may still be elusive even after the HL-LHC.
Therefore, measurements at $ep$ colliders may prove to offer complementary or additional reaches, in particular for the compressed scenarios.

\begin{acknowledgments}
\noindent
We thank Oliver Fisher, Satoshi~Kawaguchi, Max~Klein, Uta~Klein, and Peter~Kostka for helpful communication. 
We appreciate comments from members in the LHeC / FCC-eh Higgs and BSM physics study groups.
K.W.\ thanks Christophe~Grojean and Cai-dian~L\"u for their supports.
G.A.\ is supported by National Science and Engineering Research Council (Canada).
The work of S.I.\ is partially supported by the MIUR-PRIN project 2015P5SBHT 003 ``Search for the Fundamental Laws and Constituents''.
K.W.\ is supported by the Excellent Young Talents Program of the Wuhan University of Technology, the National Natural Science Foundation of China under grant no.~11905162, and the CEPC theory grant (2019--2020) of IHEP, CAS.
\end{acknowledgments}



\appendix

\section{Distributions of Input Observables}
\label{app:Obs}

\begin{figure}[H] 
\subfigure{
\includegraphics[width=4cm,height=3cm]{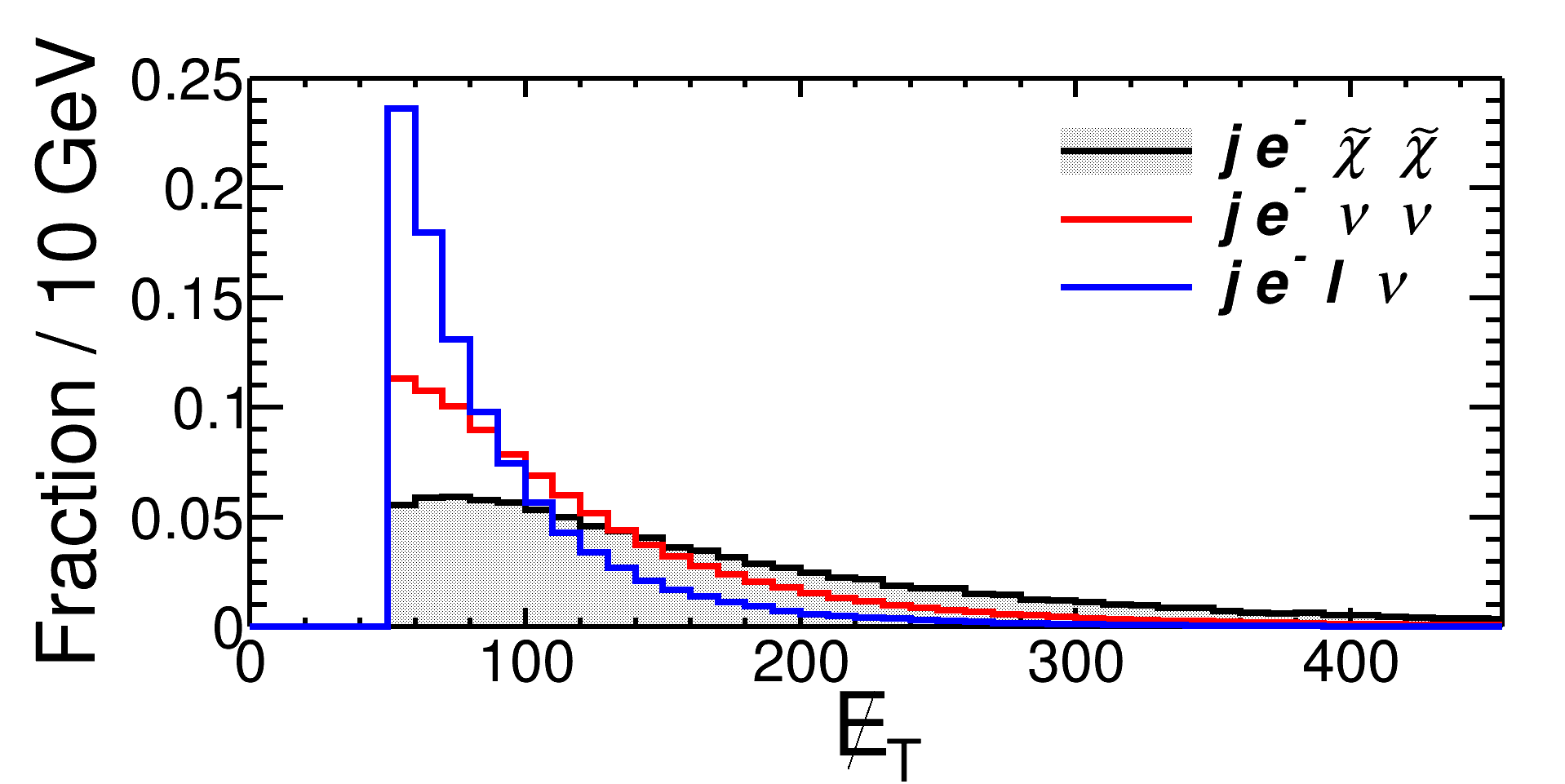}
\includegraphics[width=4cm,height=3cm]{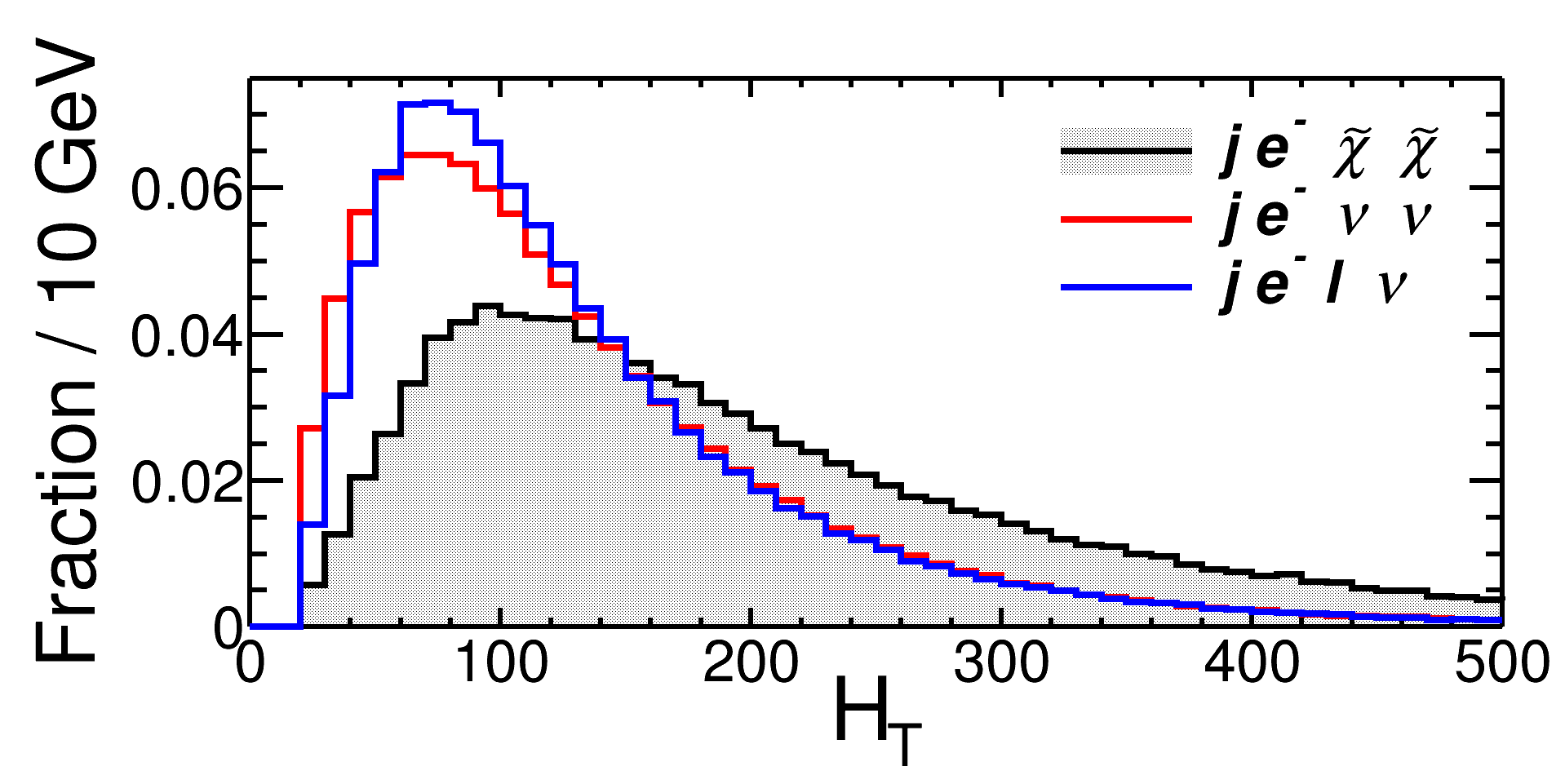}
}
\end{figure}
\addtocounter{figure}{-1}
\vspace{-1.1cm}
\begin{figure}[H] 
\addtocounter{figure}{1}
\subfigure{
\includegraphics[width=4cm,height=3cm]{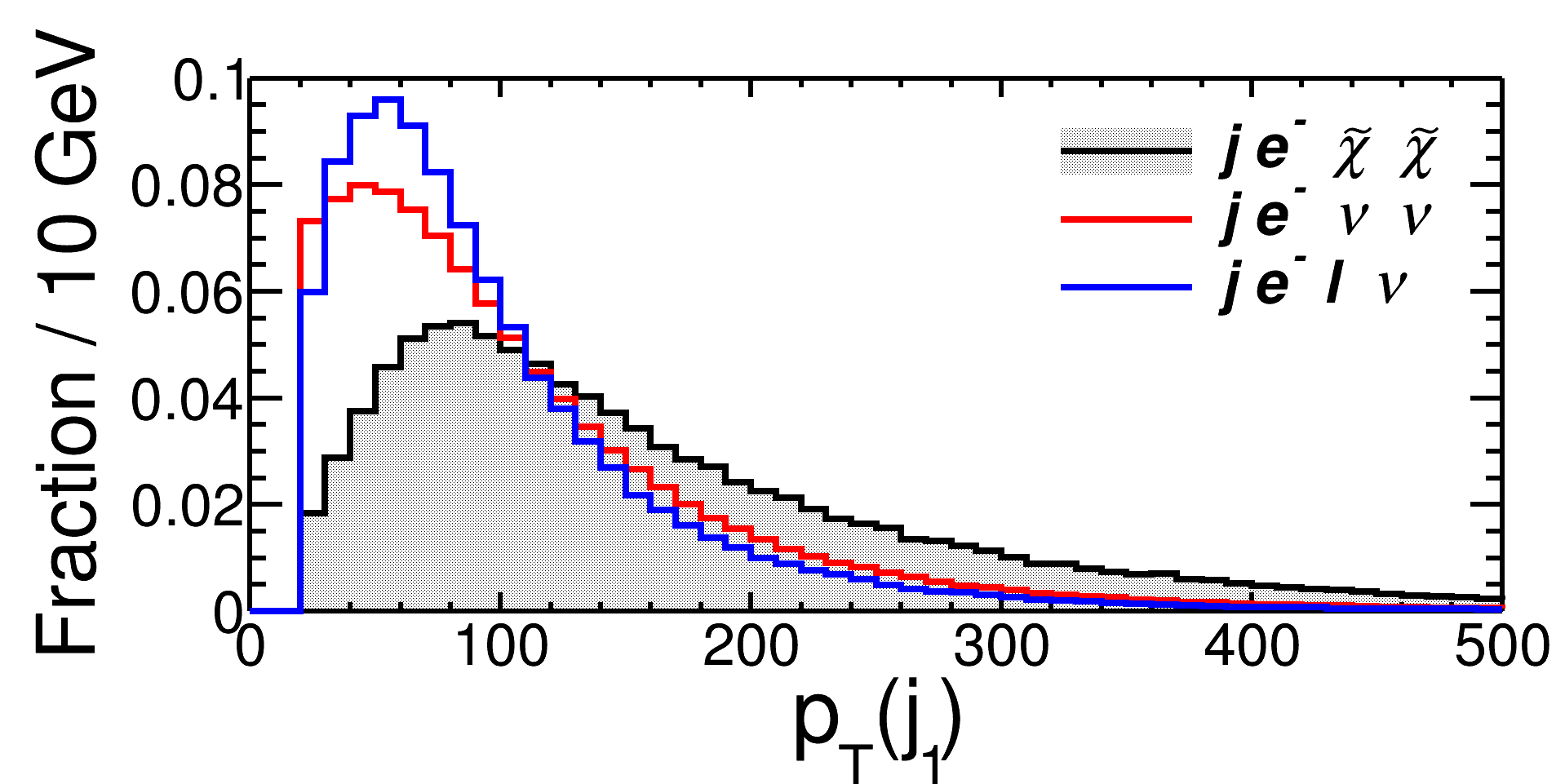}
\includegraphics[width=4cm,height=3cm]{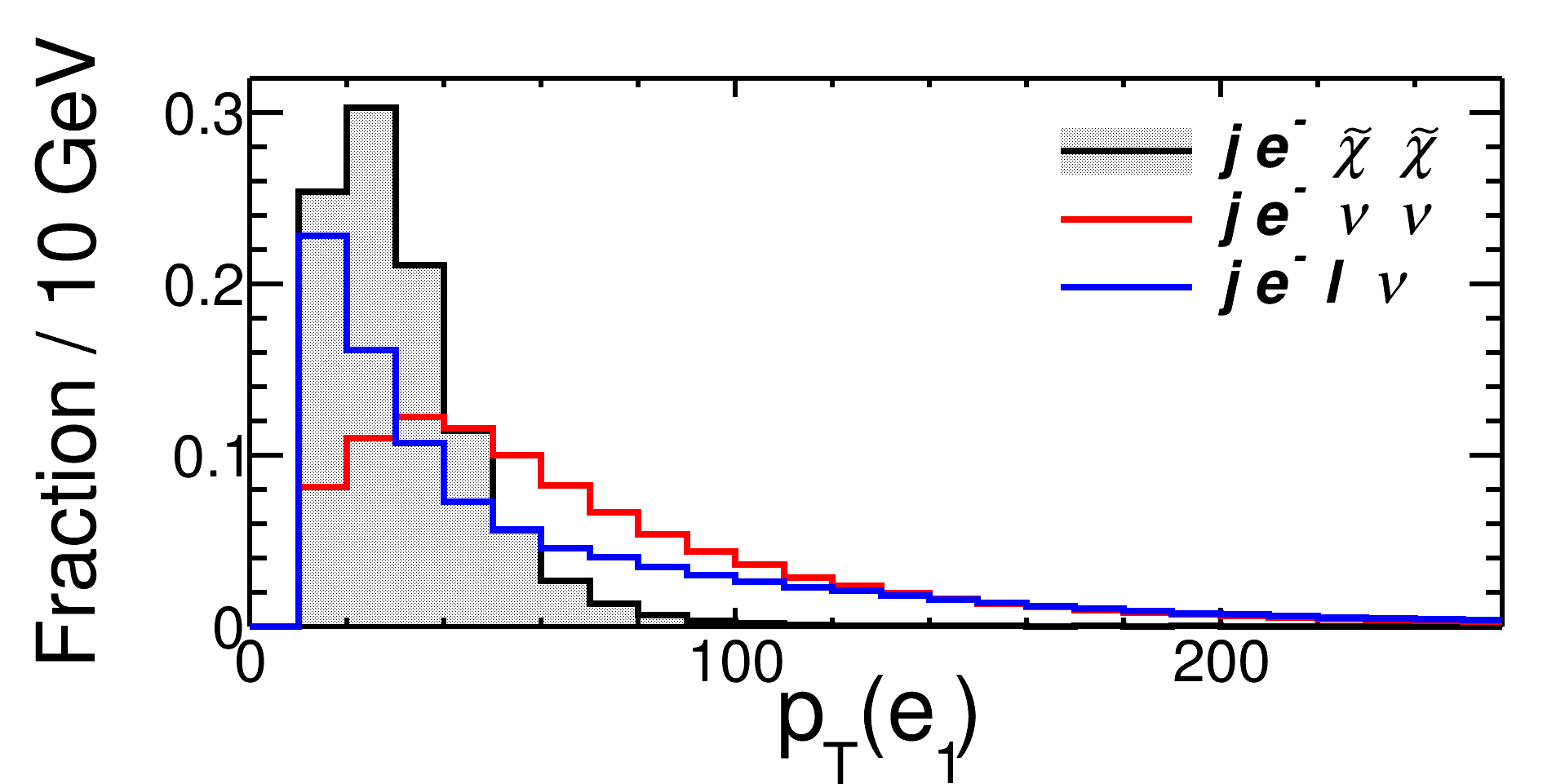}
}
\end{figure}
\vspace{-1.1cm}
\begin{figure}[H] 
\addtocounter{figure}{-1}
\subfigure{
\includegraphics[width=4cm,height=3cm]{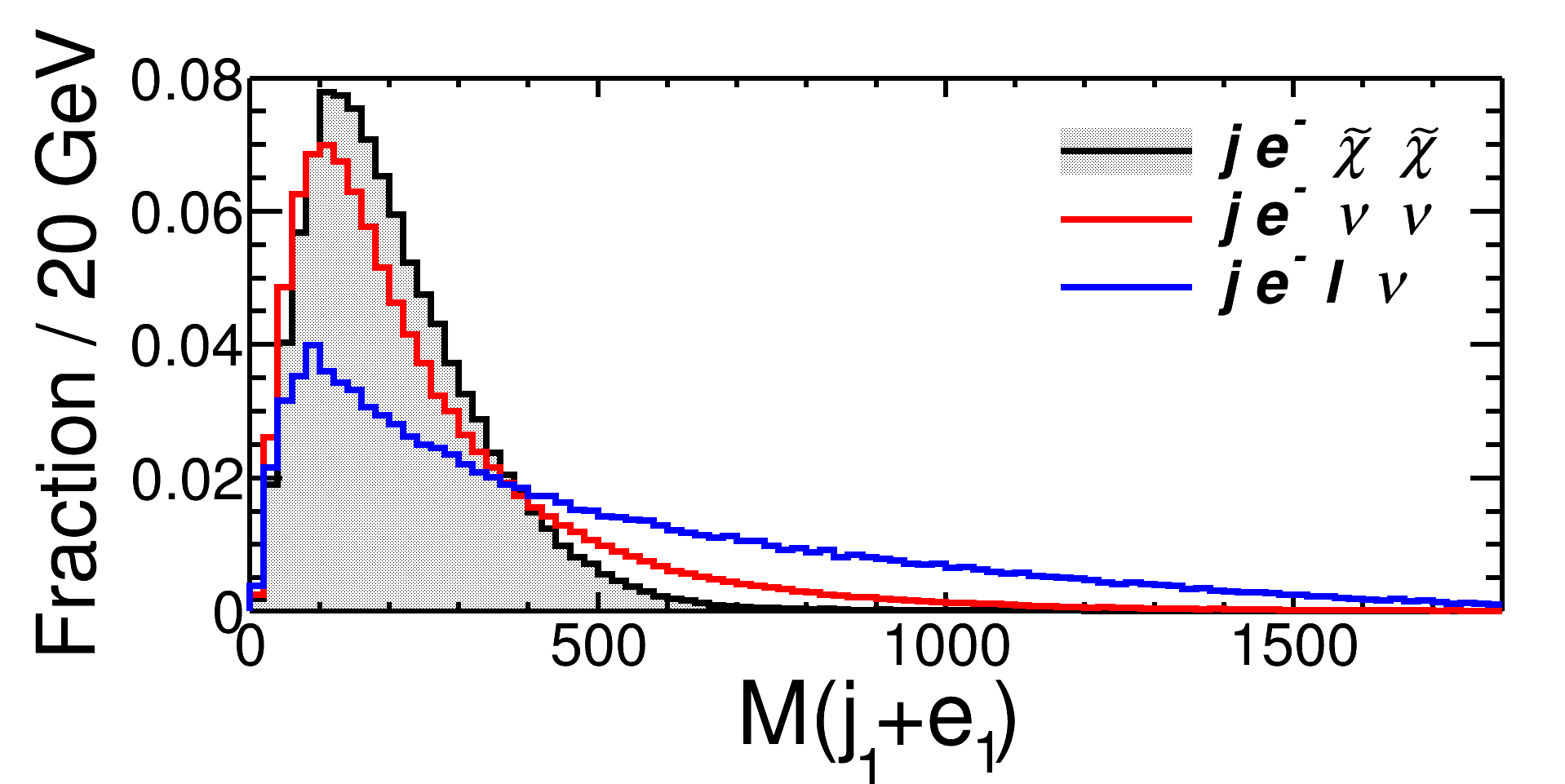}
\includegraphics[width=4cm,height=3cm]{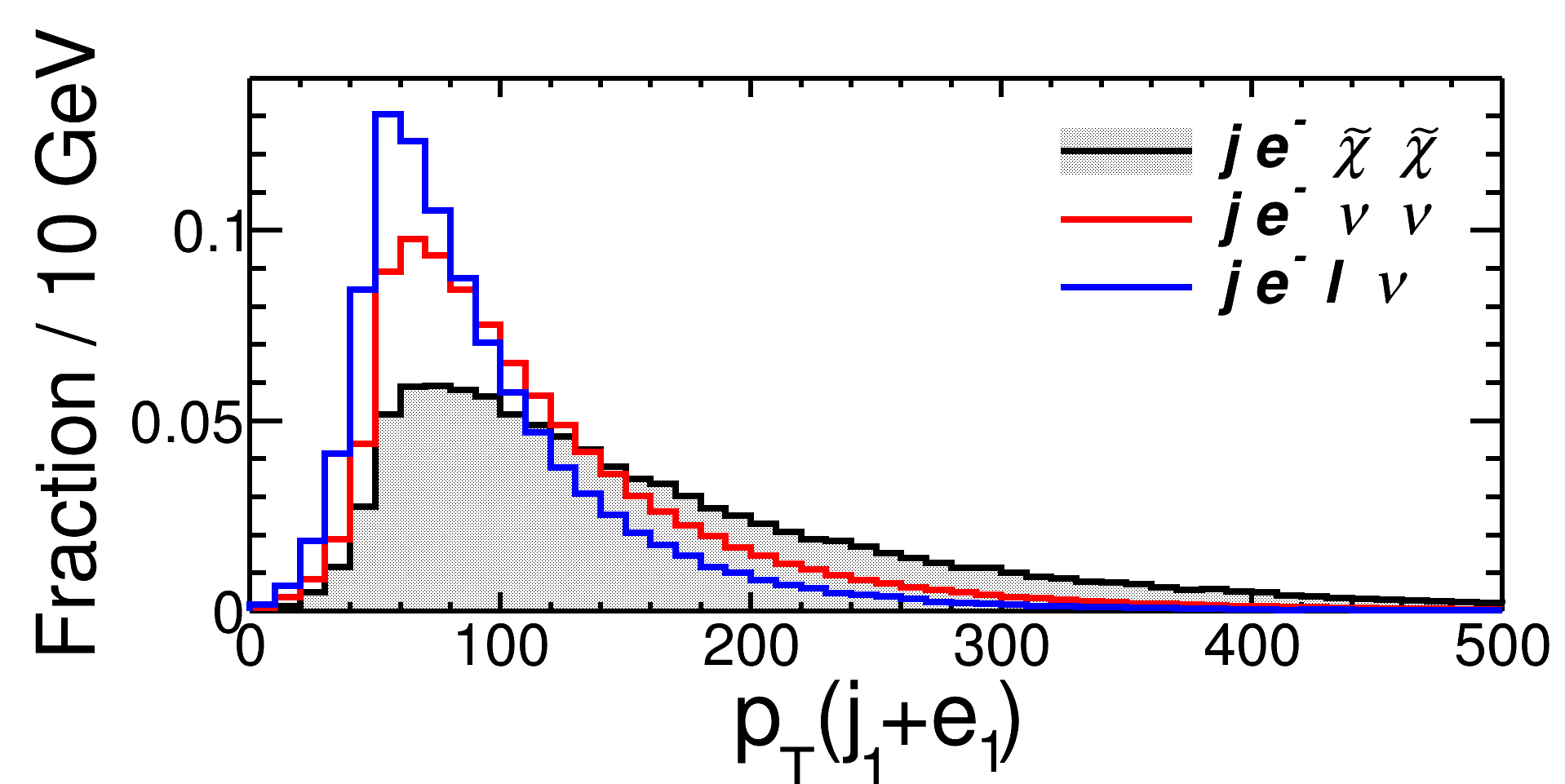}
}
\end{figure}
\vspace{-1.1cm}
\begin{figure}[H] 
\addtocounter{figure}{1}
\subfigure{
\includegraphics[width=4cm,height=3cm]{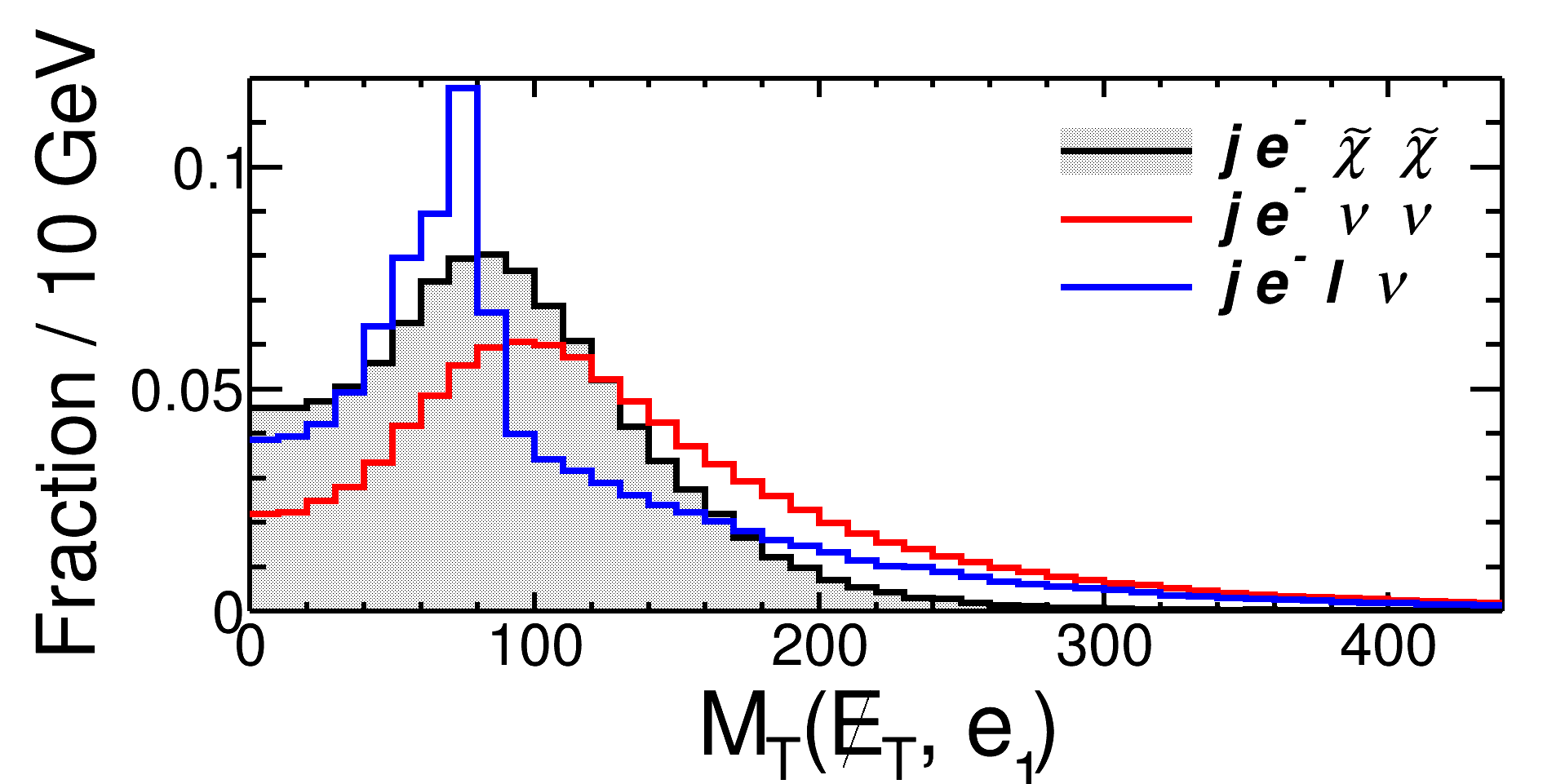}
\includegraphics[width=4cm,height=3cm]{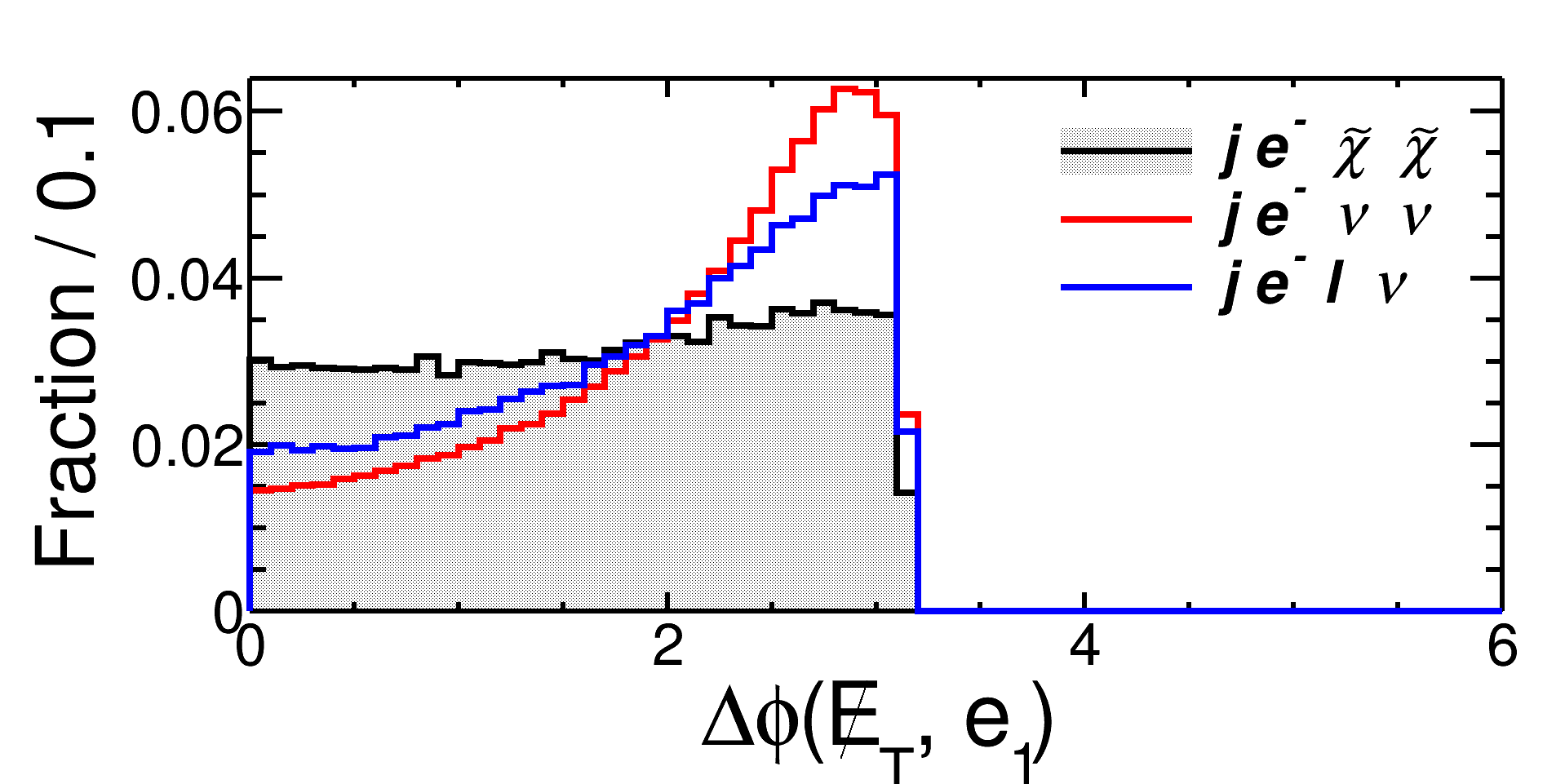}
}
\end{figure}
\vspace{-1.1cm}
\begin{figure}[H] 
\addtocounter{figure}{-1}
\subfigure{
\includegraphics[width=4cm,height=3cm]{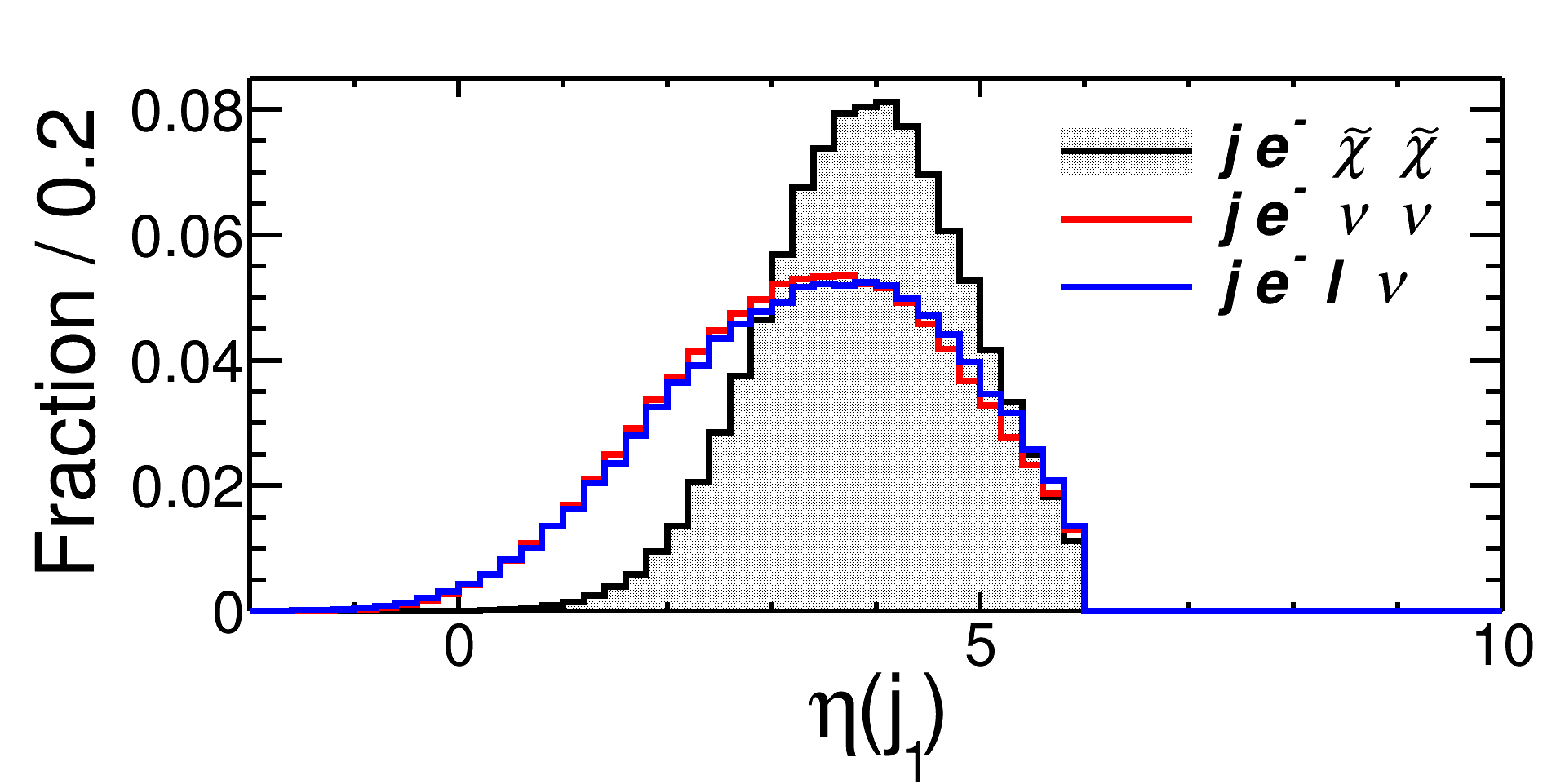}
\includegraphics[width=4cm,height=3cm]{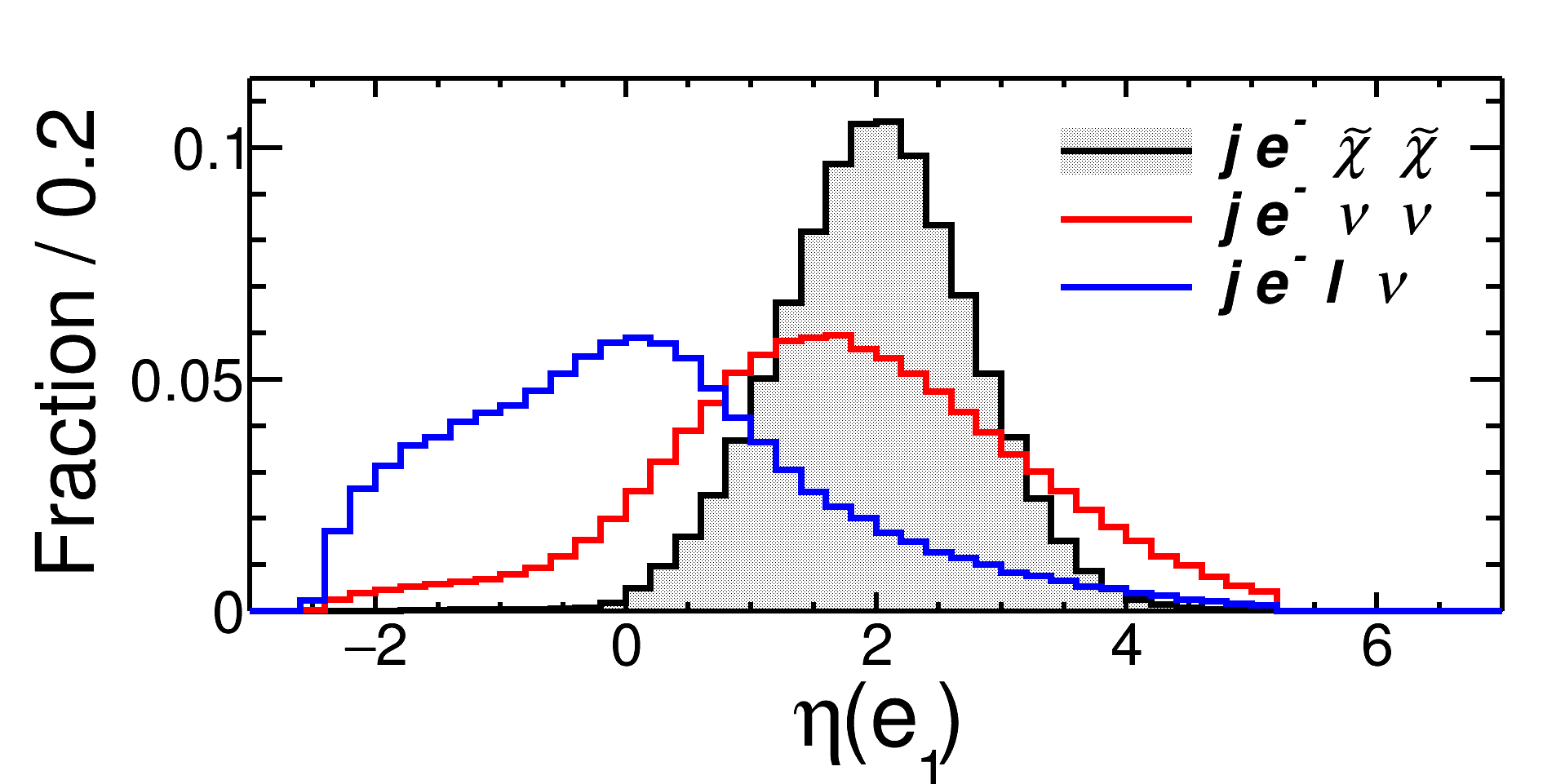}
}
\end{figure}
\vspace{-1.1cm}
\begin{figure}[H] 
\addtocounter{figure}{1}
\subfigure{
\includegraphics[width=4cm,height=3cm]{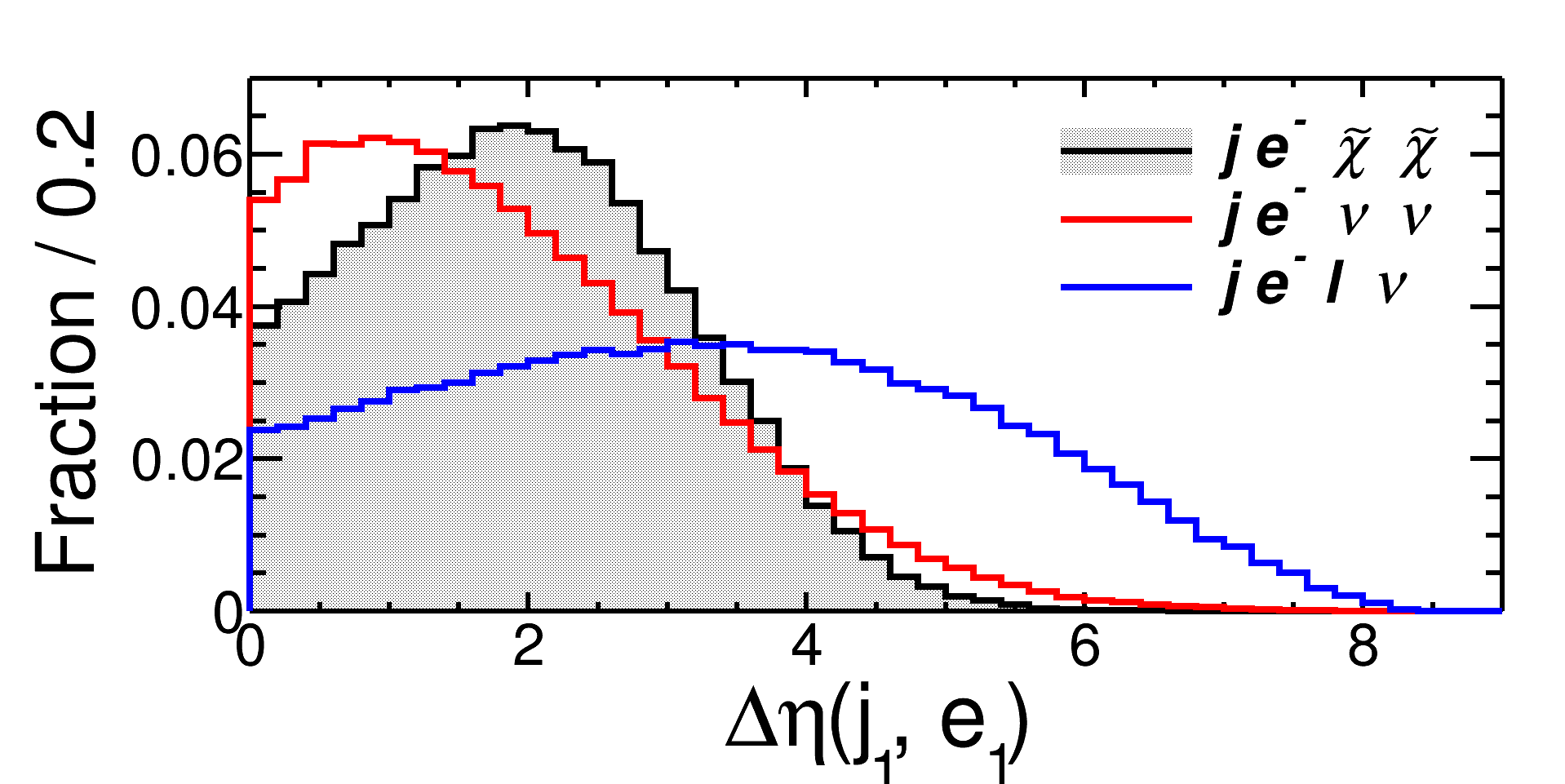}
\includegraphics[width=4cm,height=3cm]{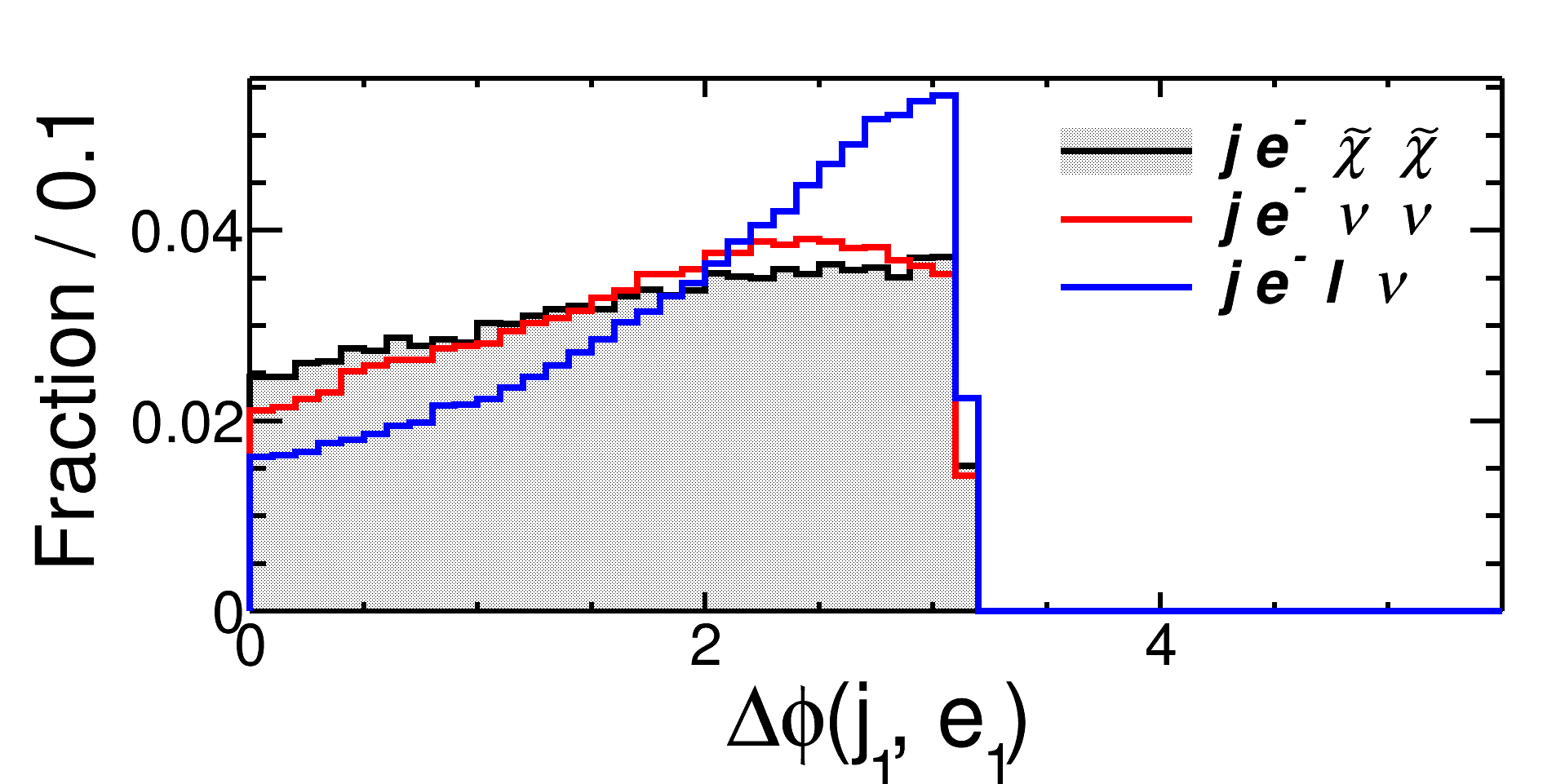}
}
\caption{Kinematical distributions of some input observables  for signal $j\, e^-\, \tilde{\chi} \tilde{\chi}$ with $m_{\chargino1, \neutralino2}$ = 400 GeV (black with filled area) in the compressed-slepton scenario, and for the SM background of the $j\, e^-\, \nu \nu$ (red) and $j\, e^-\, \ell \nu$ (blue) processes after applying the pre-selection cuts at the FCC-eh with an unpolarized electron beam. }
\label{fig:obs_FCCeh_compressedSlep_m400}
\end{figure}

\begin{figure}[H] 
\subfigure{
\includegraphics[width=4cm,height=3cm]{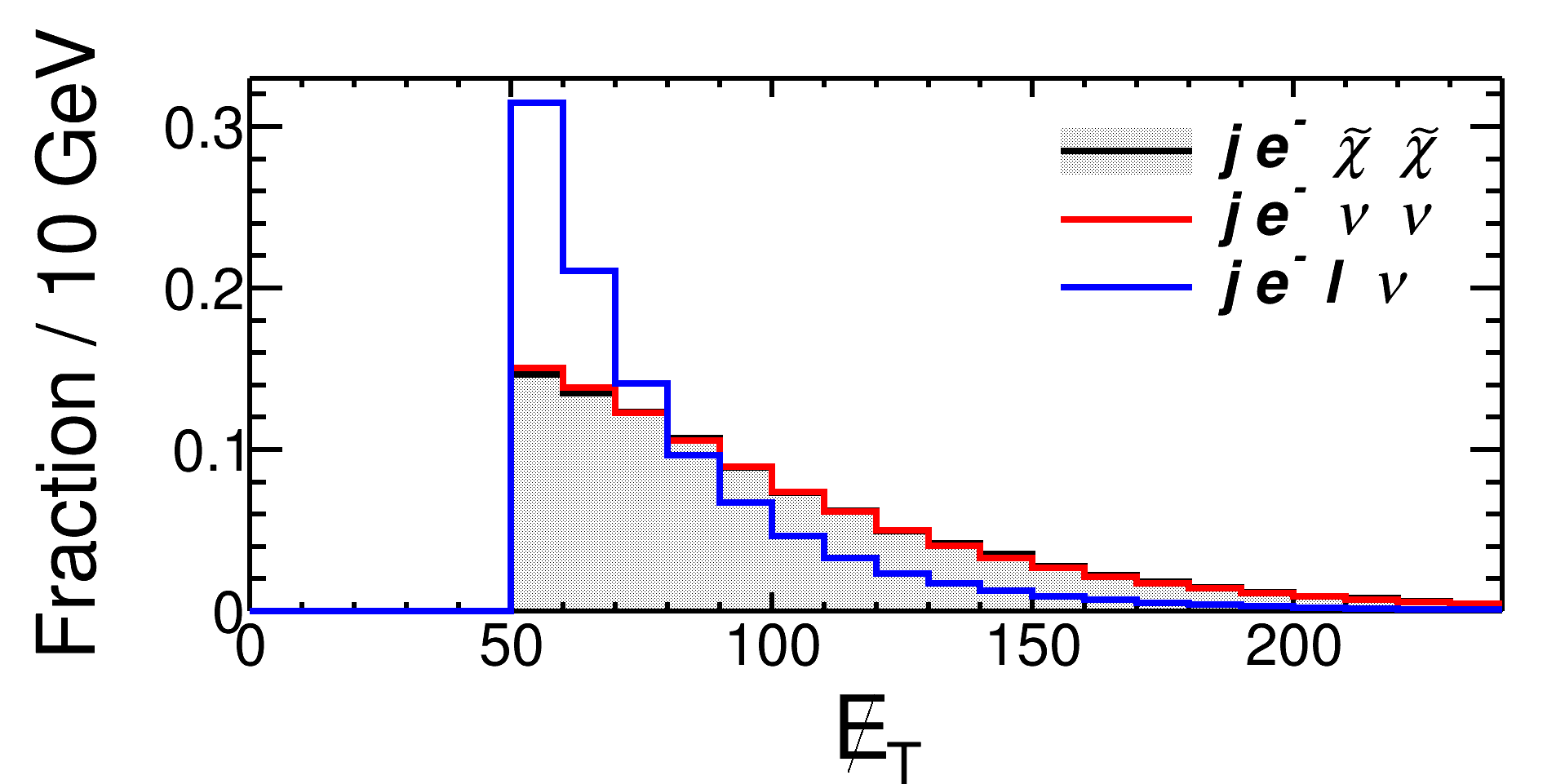}
\includegraphics[width=4cm,height=3cm]{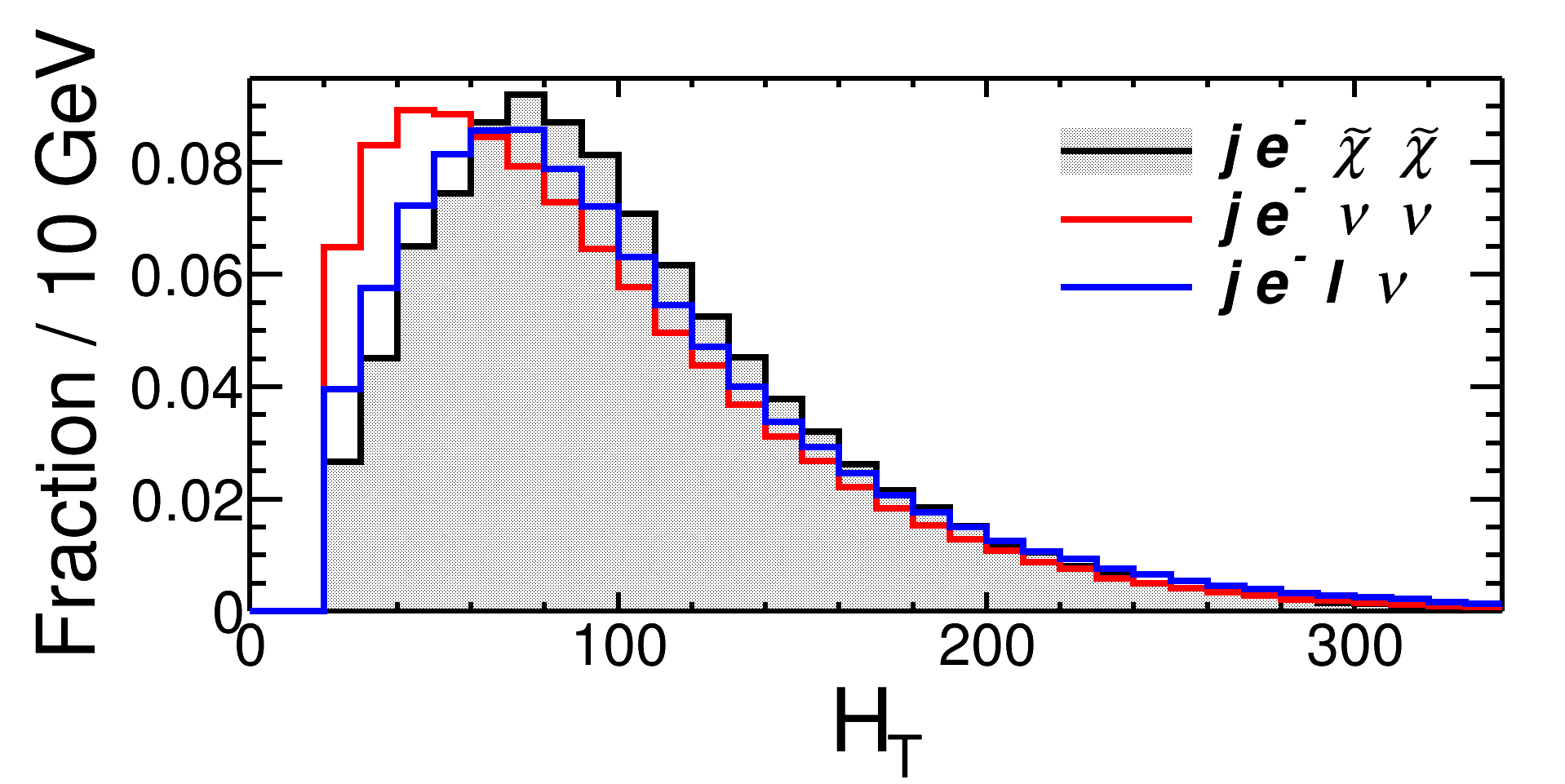}
}
\end{figure}
\addtocounter{figure}{-1}
\vspace{-1.1cm}
\begin{figure}[H] 
\addtocounter{figure}{1}
\subfigure{
\includegraphics[width=4cm,height=3cm]{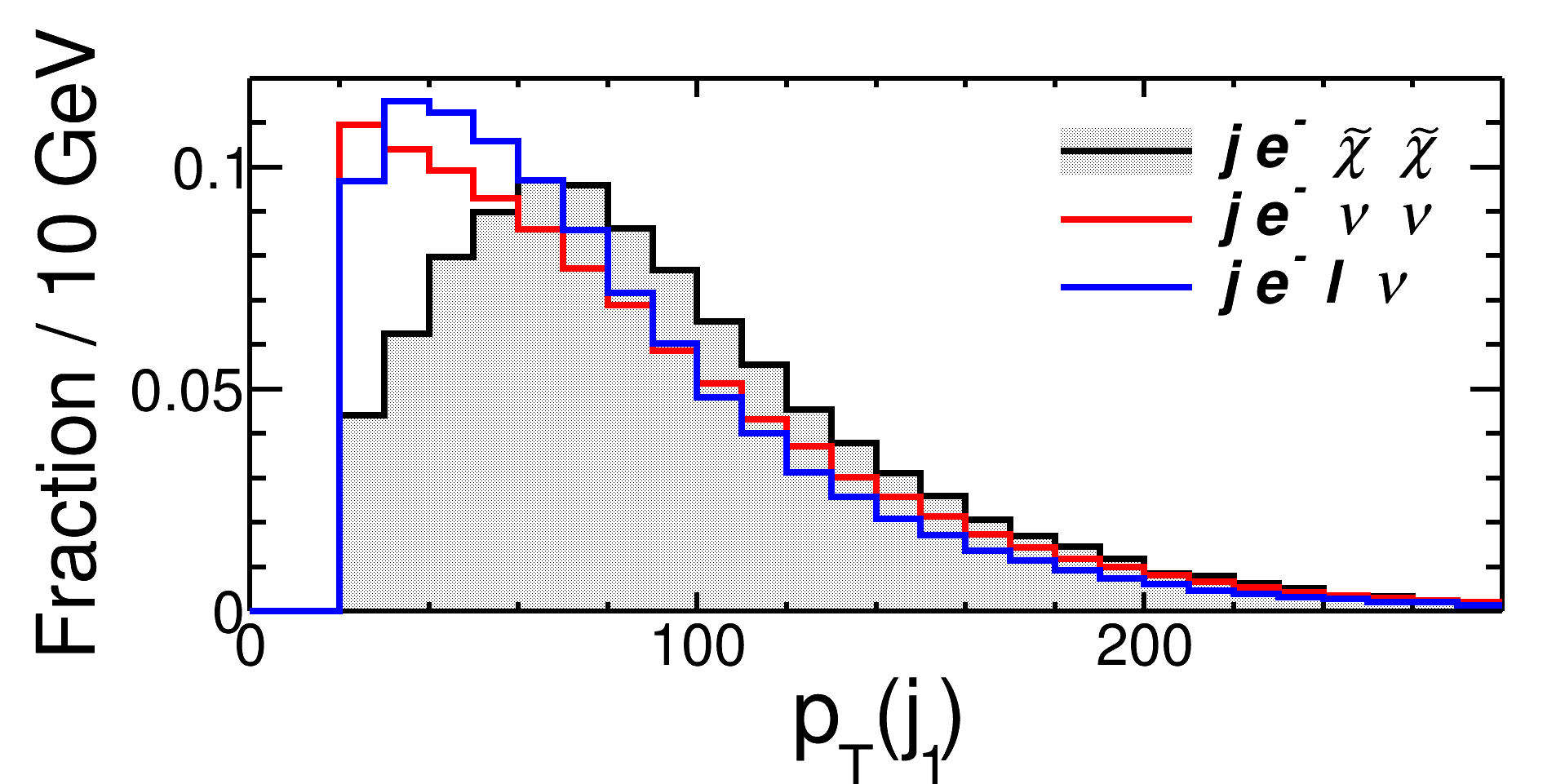}
\includegraphics[width=4cm,height=3cm]{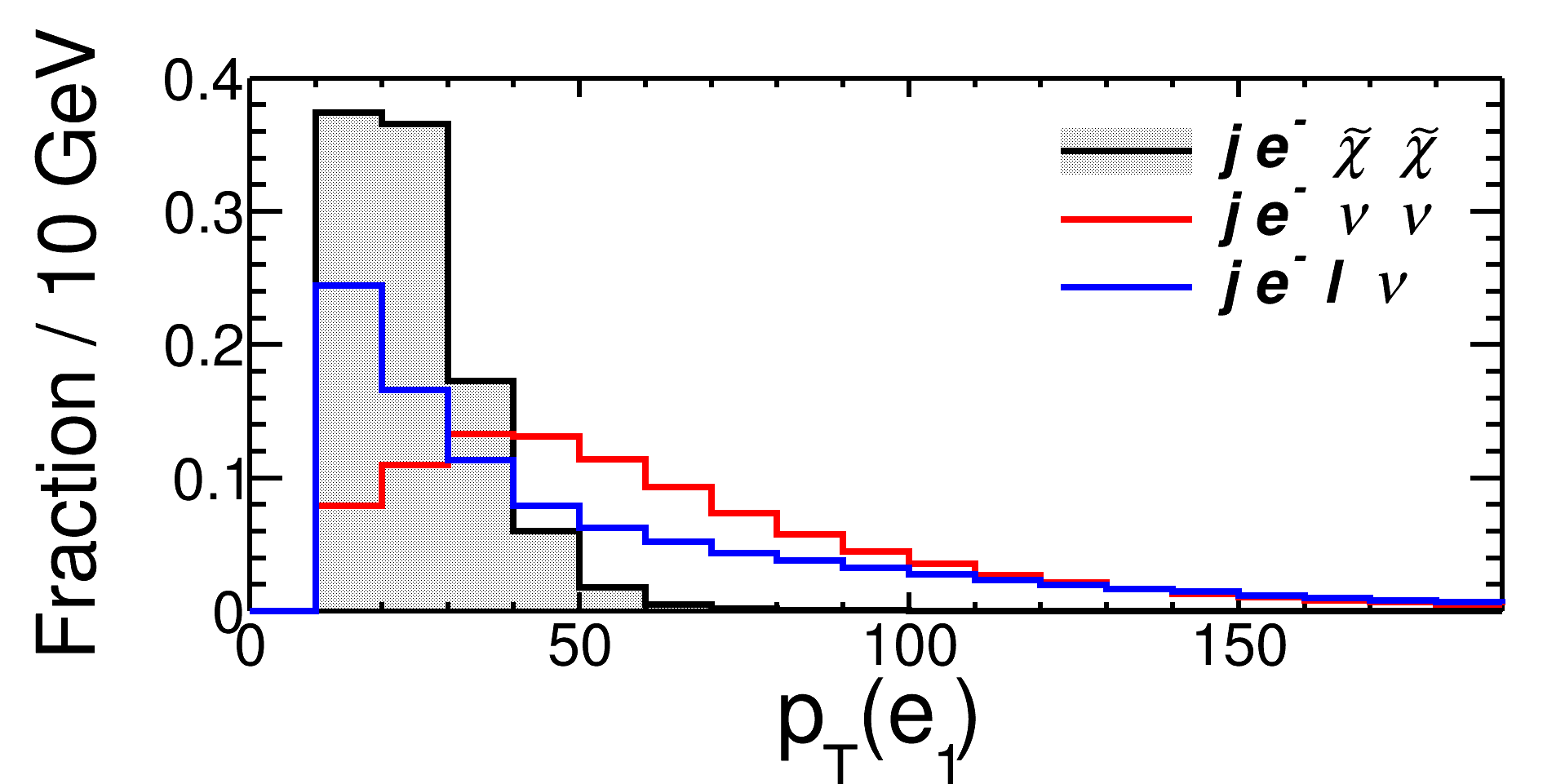}
}
\end{figure}
\vspace{-1.1cm}
\begin{figure}[H] 
\addtocounter{figure}{-1}
\subfigure{
\includegraphics[width=4cm,height=3cm]{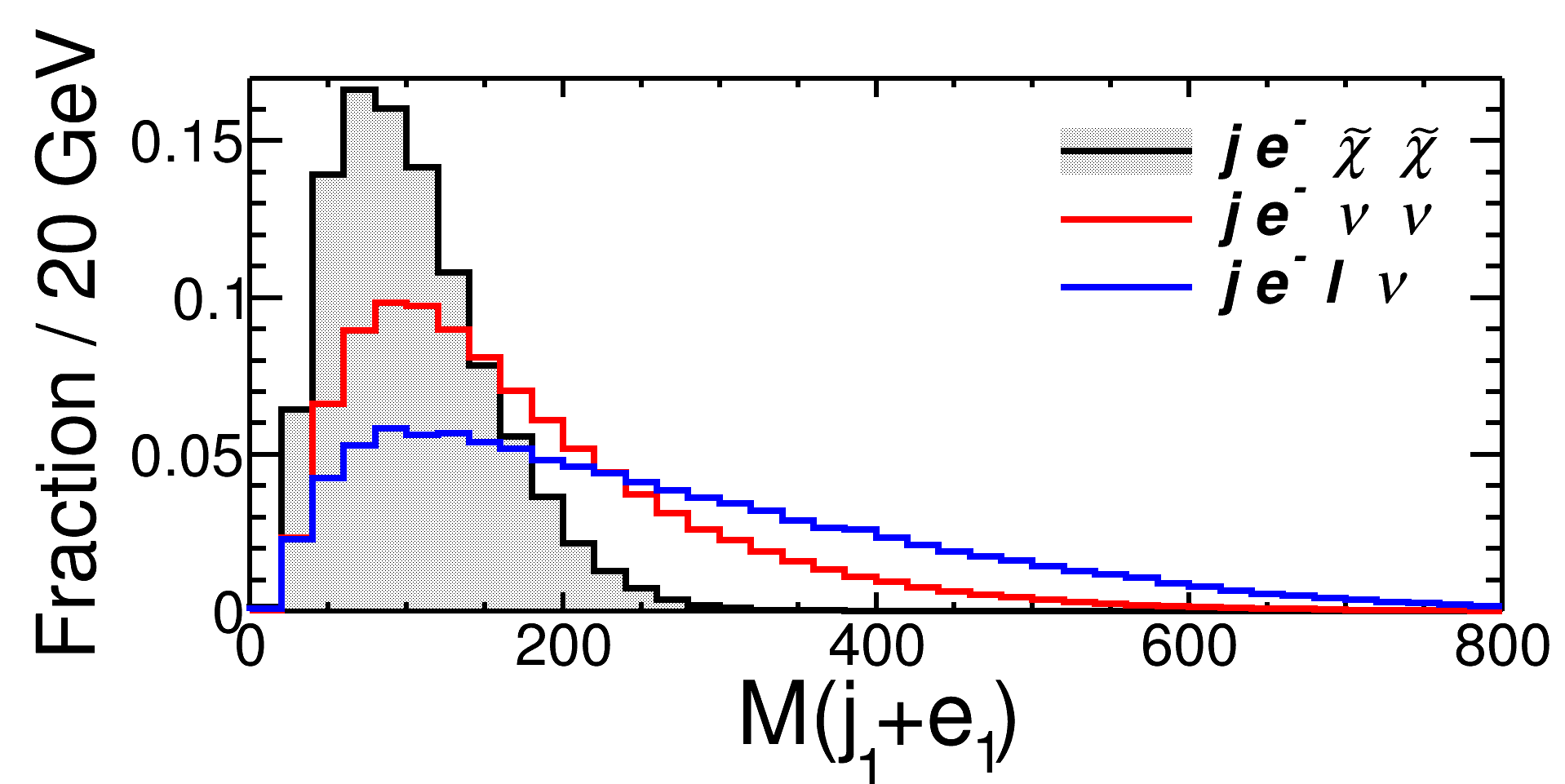}
\includegraphics[width=4cm,height=3cm]{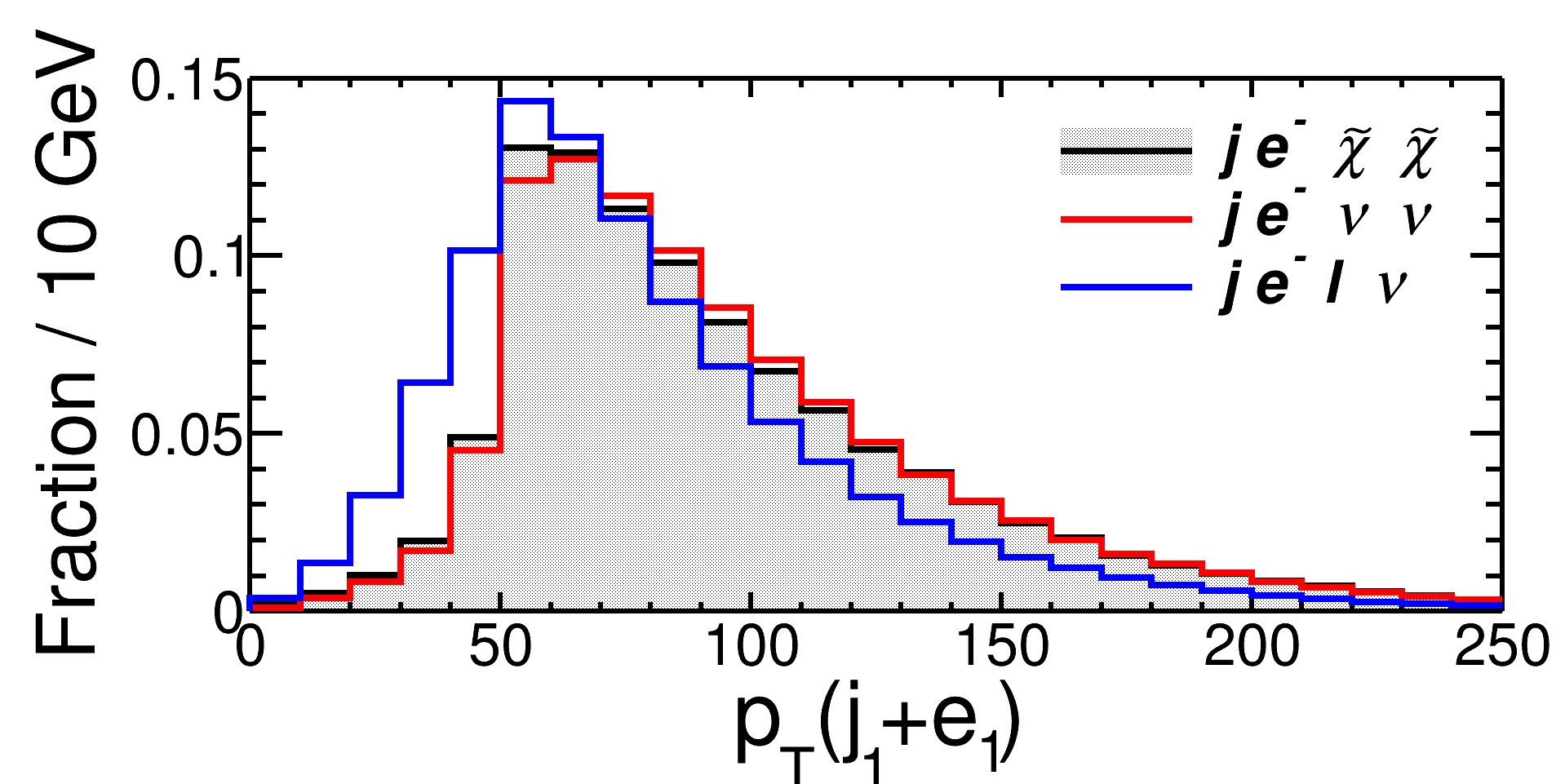}
}
\end{figure}
\vspace{-1.1cm}
\begin{figure}[H] 
\addtocounter{figure}{1}
\subfigure{
\includegraphics[width=4cm,height=3cm]{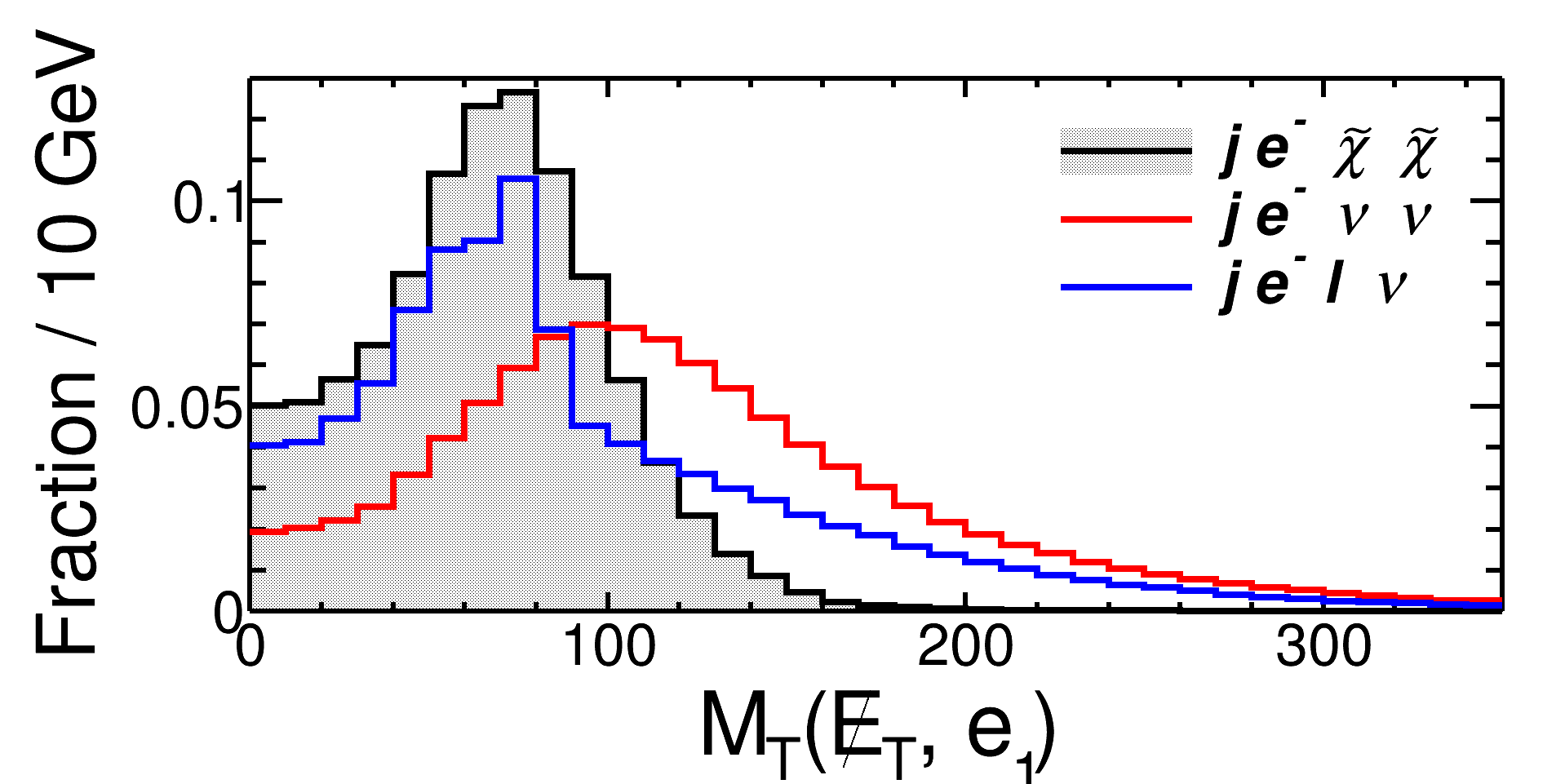}
\includegraphics[width=4cm,height=3cm]{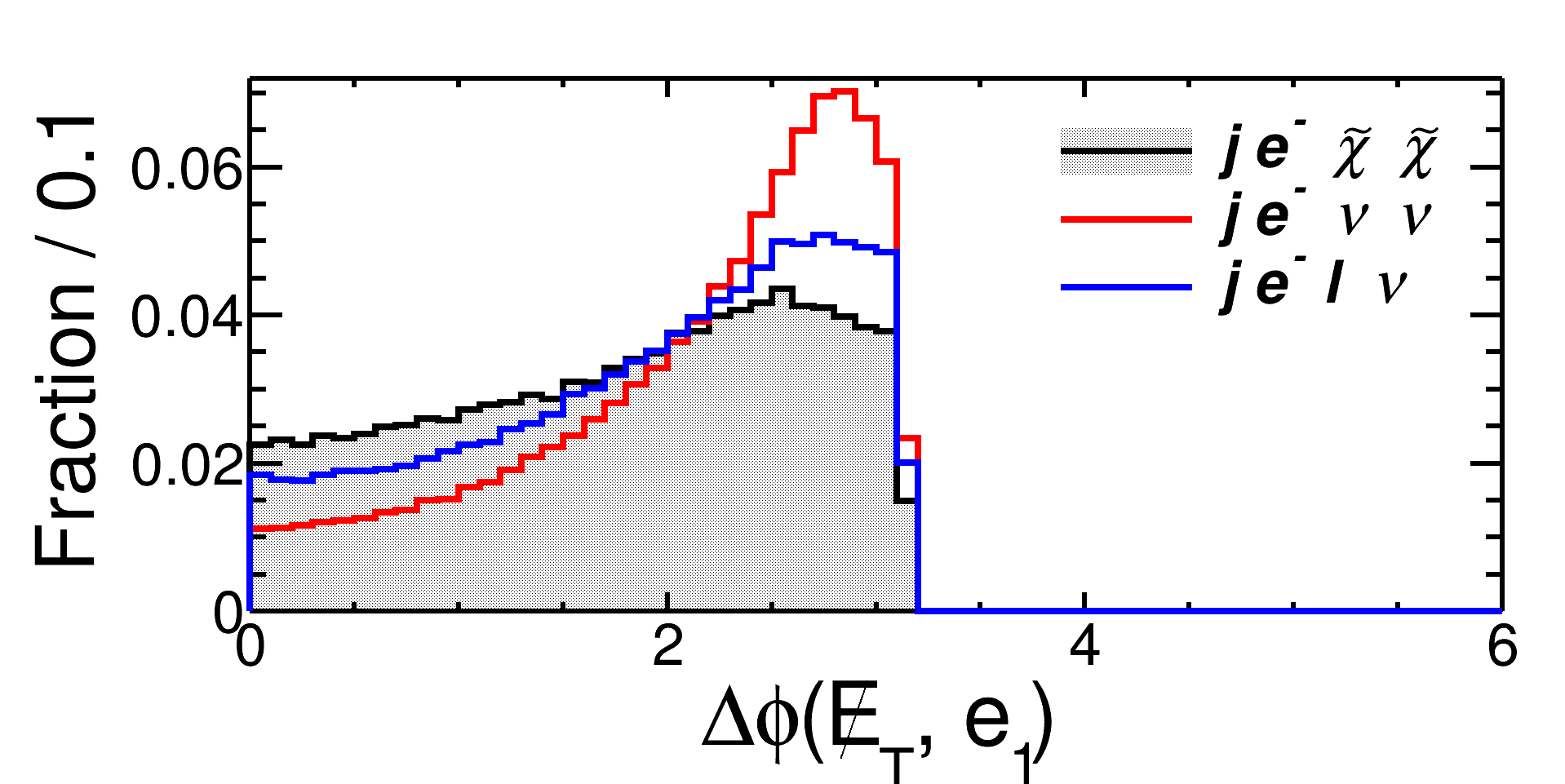}
}
\end{figure}
\vspace{-1.1cm}
\begin{figure}[H] 
\addtocounter{figure}{-1}
\subfigure{
\includegraphics[width=4cm,height=3cm]{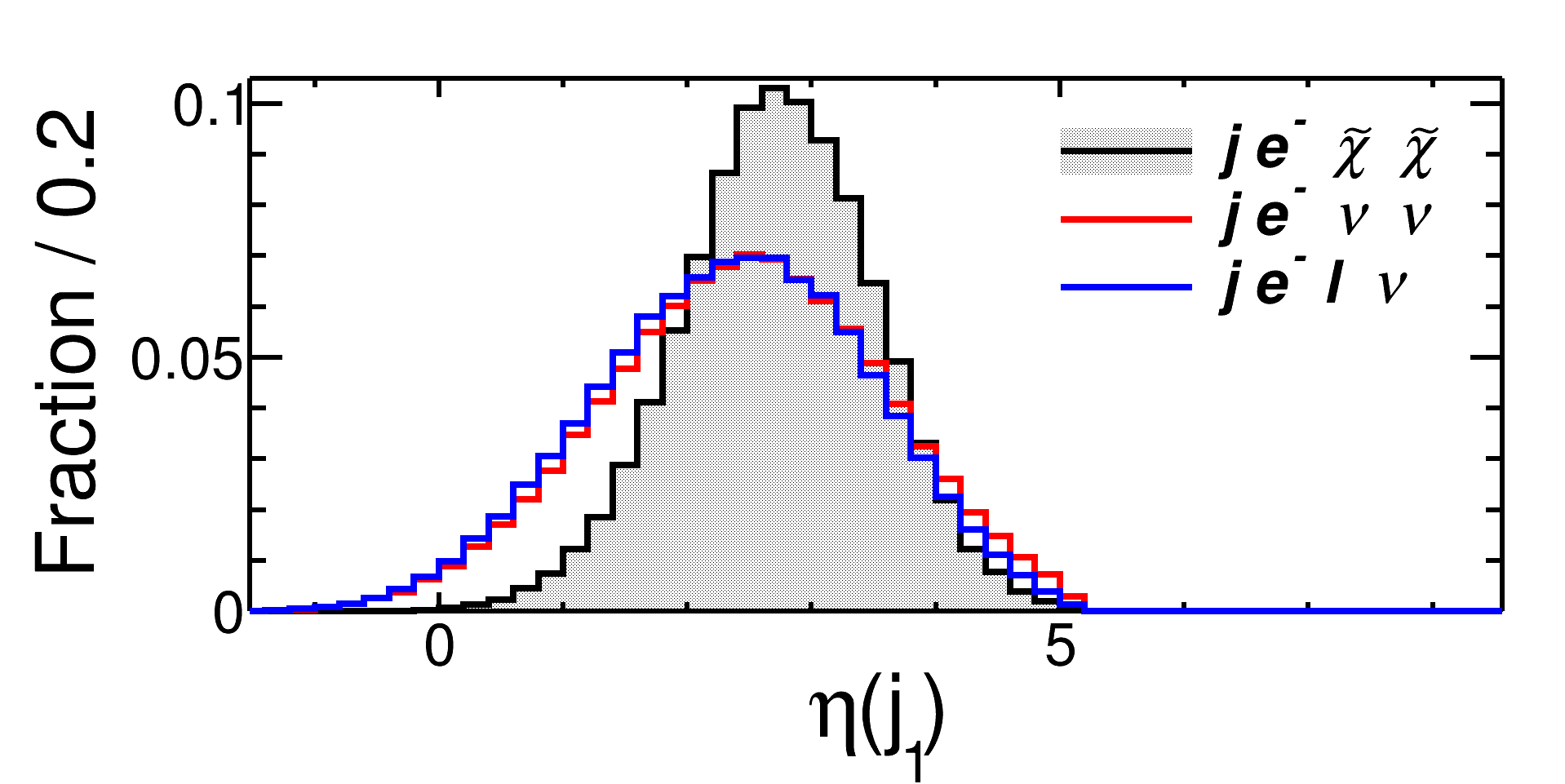}
\includegraphics[width=4cm,height=3cm]{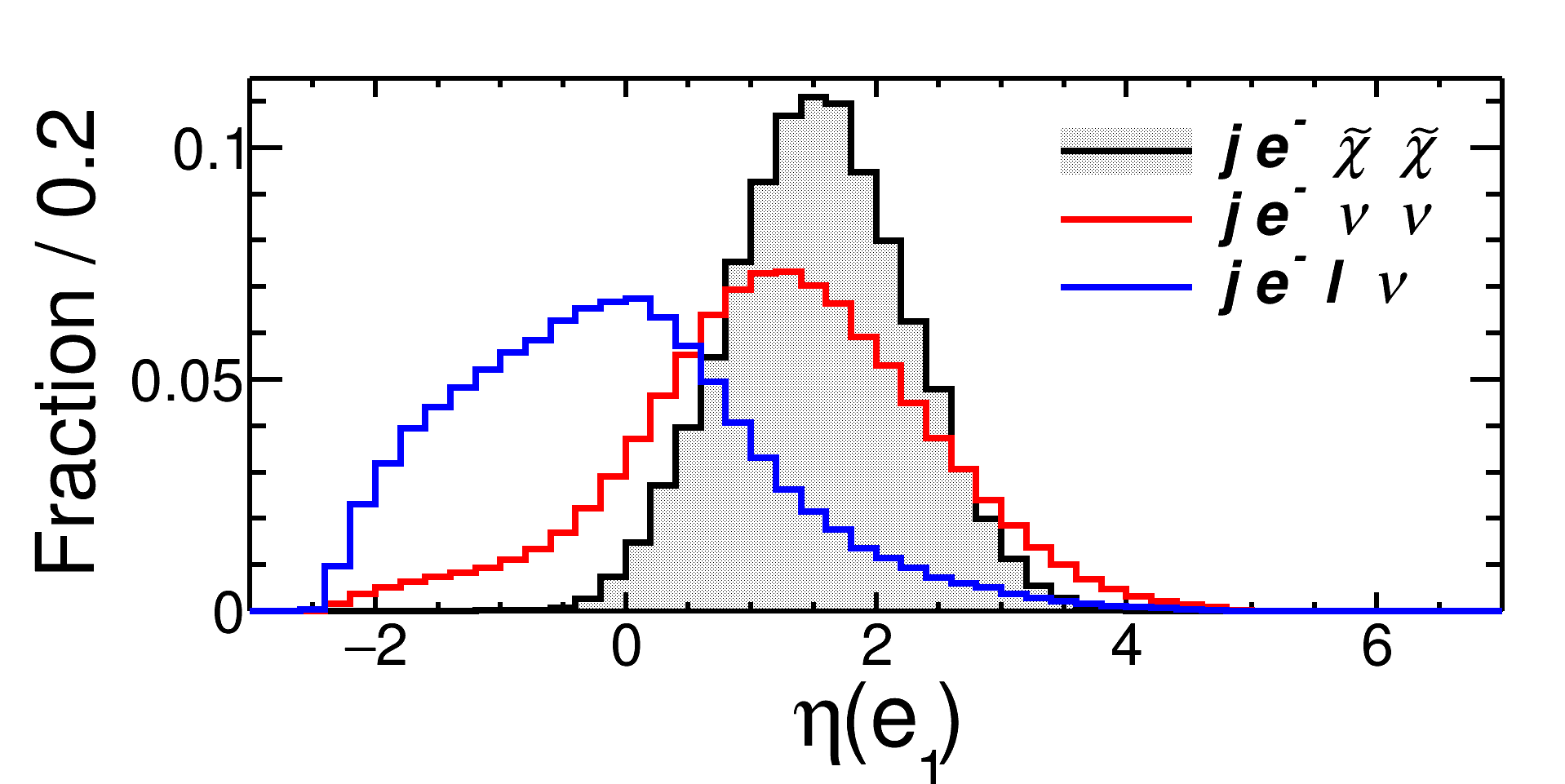}
}
\end{figure}
\vspace{-1.1cm}
\begin{figure}[H] 
\addtocounter{figure}{1}
\subfigure{
\includegraphics[width=4cm,height=3cm]{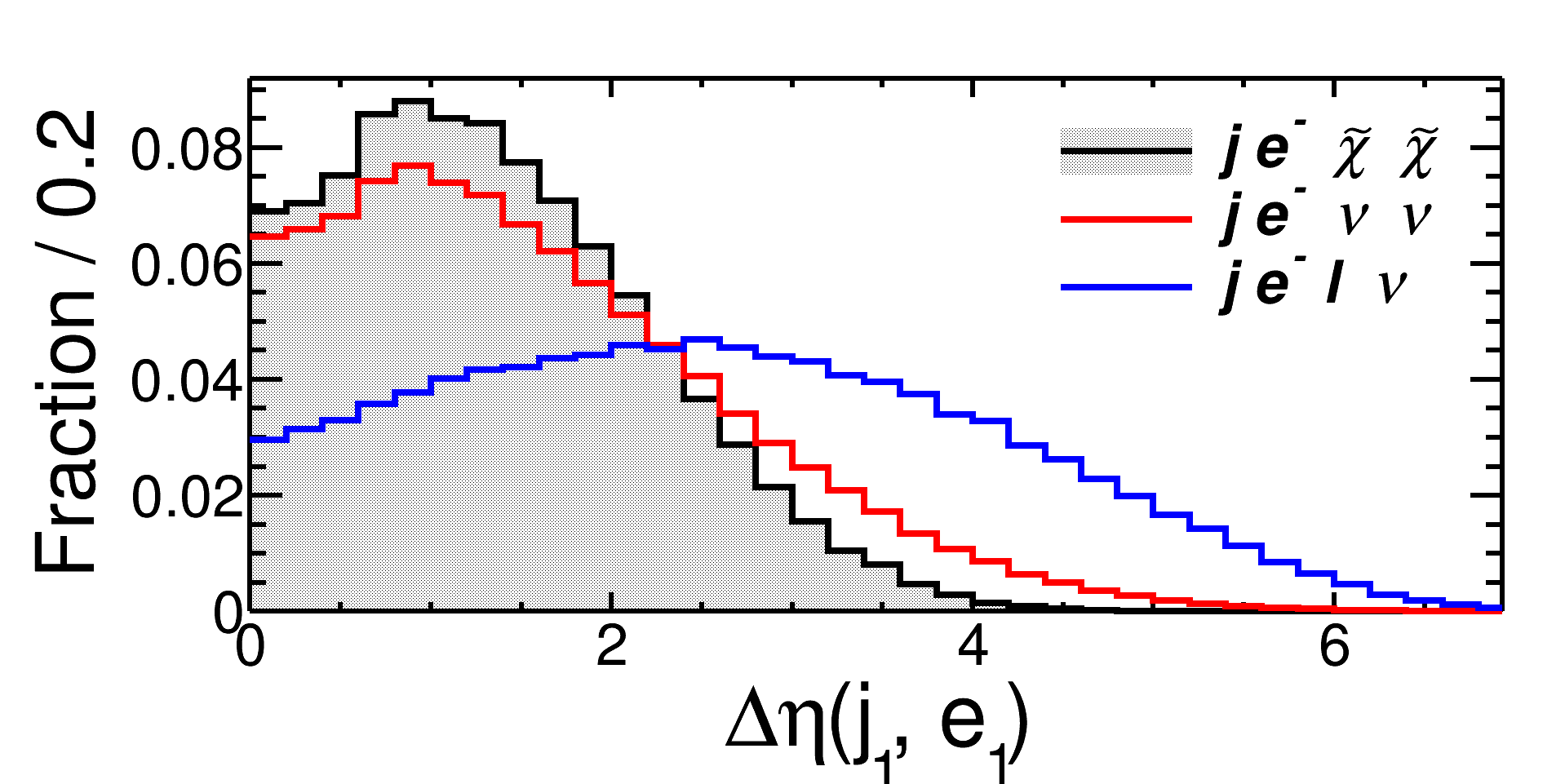}
\includegraphics[width=4cm,height=3cm]{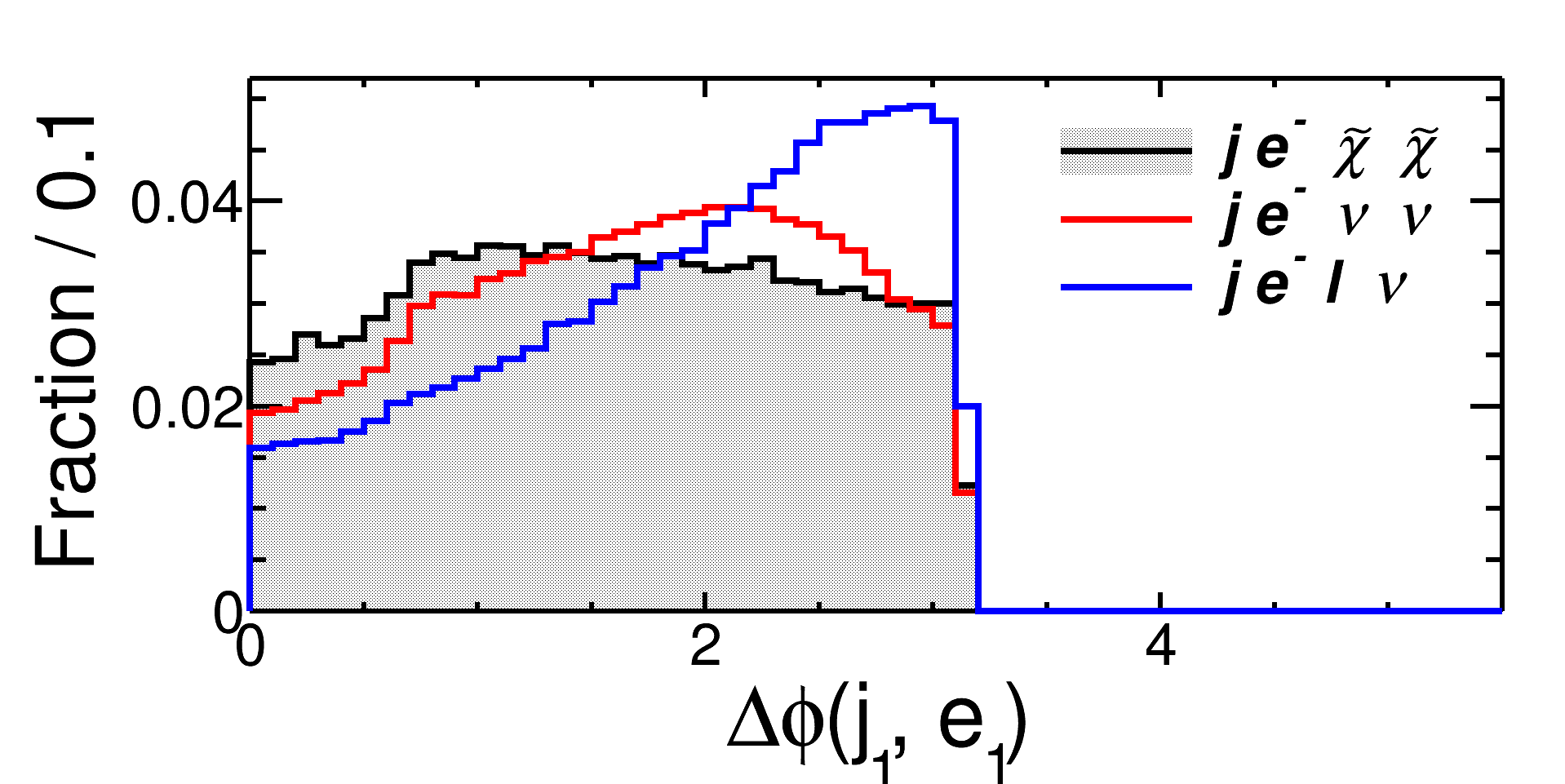}
}
\caption{Kinematical distributions of some input observables  for signal $j\, e^-\, \tilde{\chi} \tilde{\chi}$ with $m_{\chargino1, \neutralino2}$ = 250 GeV (black with filled area) in the compressed-slepton scenario, and for the SM background of the $j\, e^-\, \nu \nu$ (red) and $j\, e^-\, \ell \nu$ (blue) processes after applying the pre-selection cuts at the LHeC with an unpolarized electron beam. }
\label{fig:obs_LHeC_compressedSlep_m250}
\end{figure}

\begin{figure}[H] 
\subfigure{
\includegraphics[width=4cm,height=3cm]{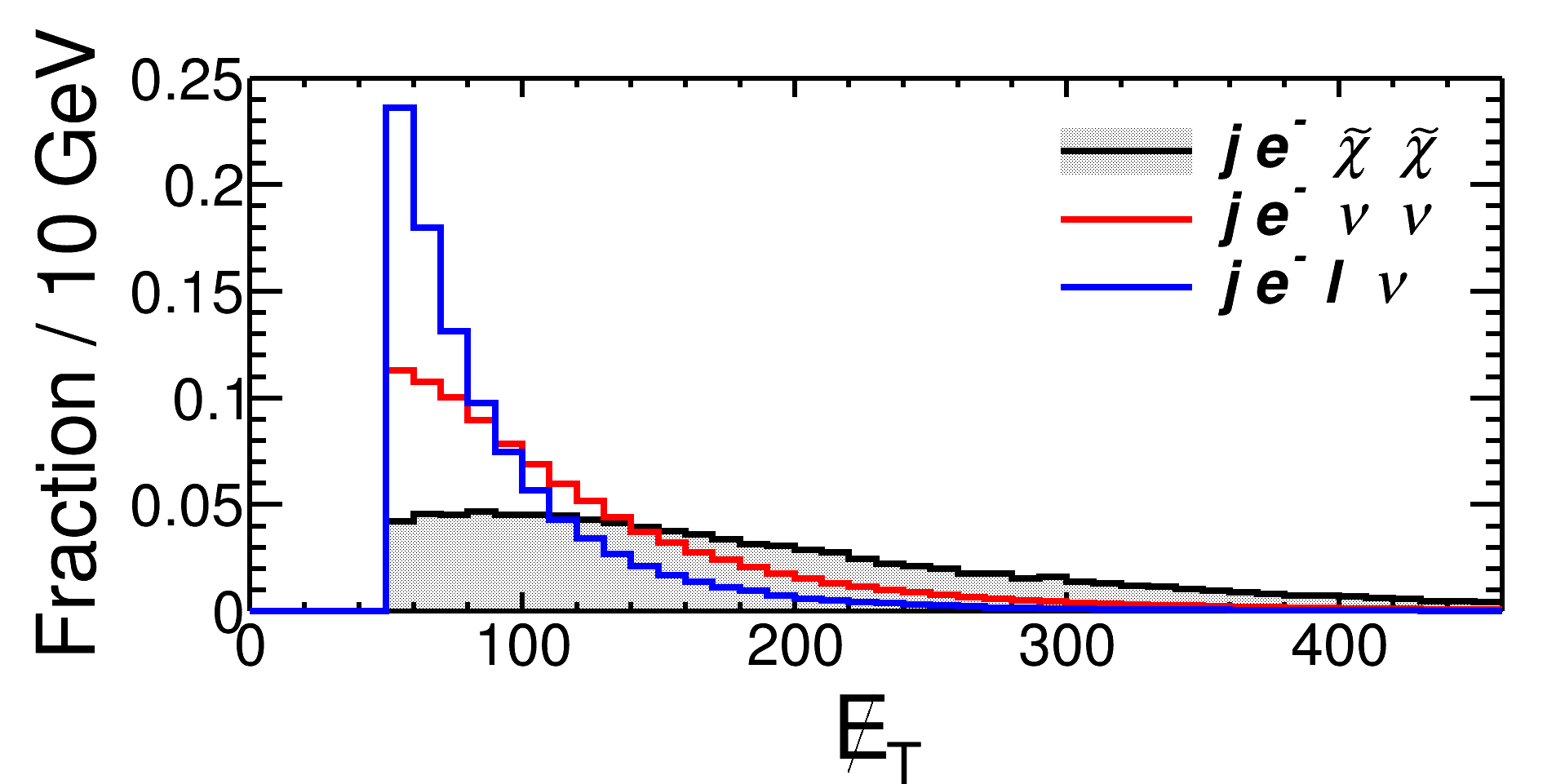}
\includegraphics[width=4cm,height=3cm]{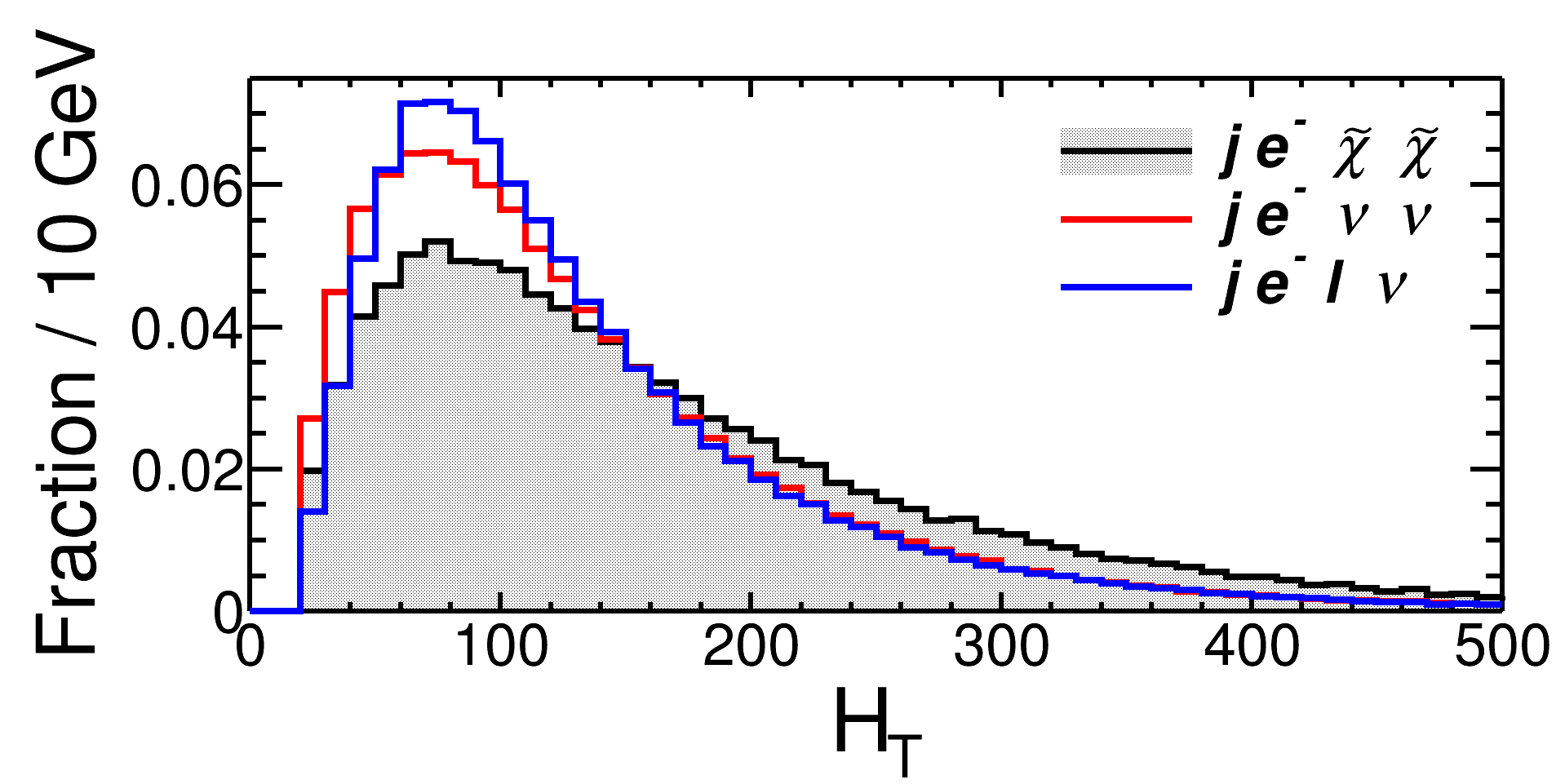}
}
\end{figure}
\addtocounter{figure}{-1}
\vspace{-1.1cm}
\begin{figure}[H] 
\addtocounter{figure}{1}
\subfigure{
\includegraphics[width=4cm,height=3cm]{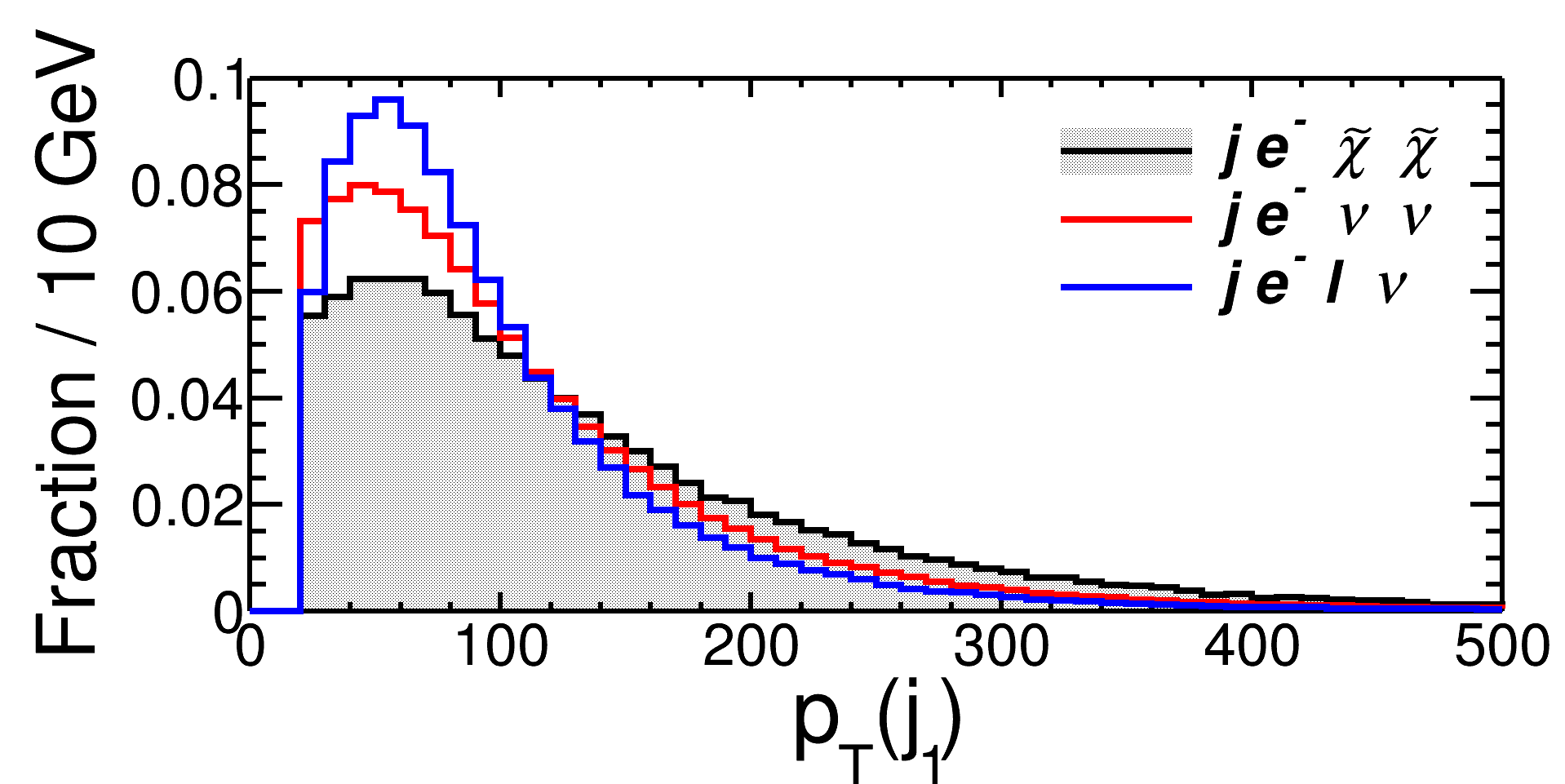}
\includegraphics[width=4cm,height=3cm]{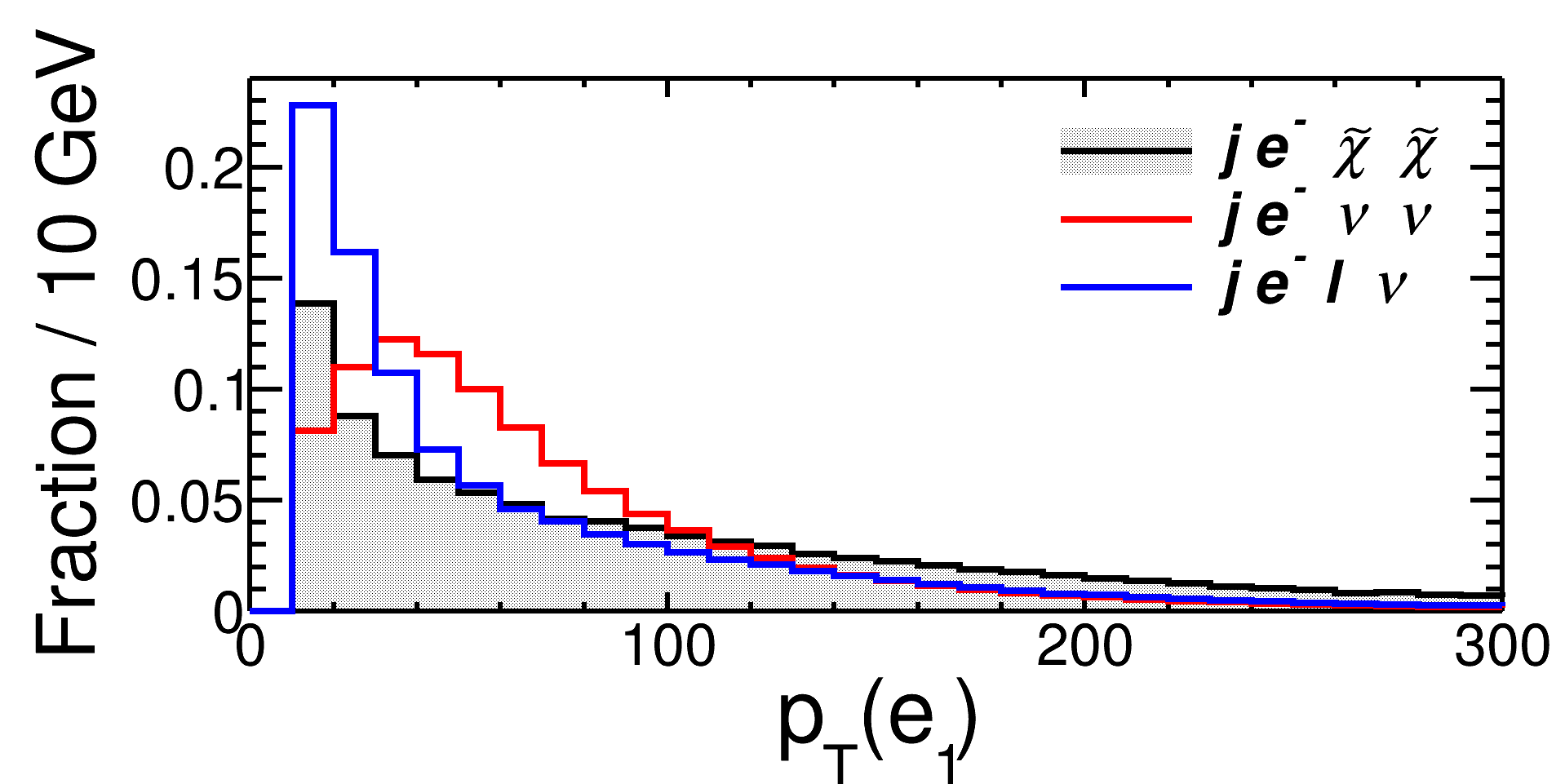}
}
\end{figure}
\vspace{-1.1cm}
\begin{figure}[H] 
\addtocounter{figure}{-1}
\subfigure{
\includegraphics[width=4cm,height=3cm]{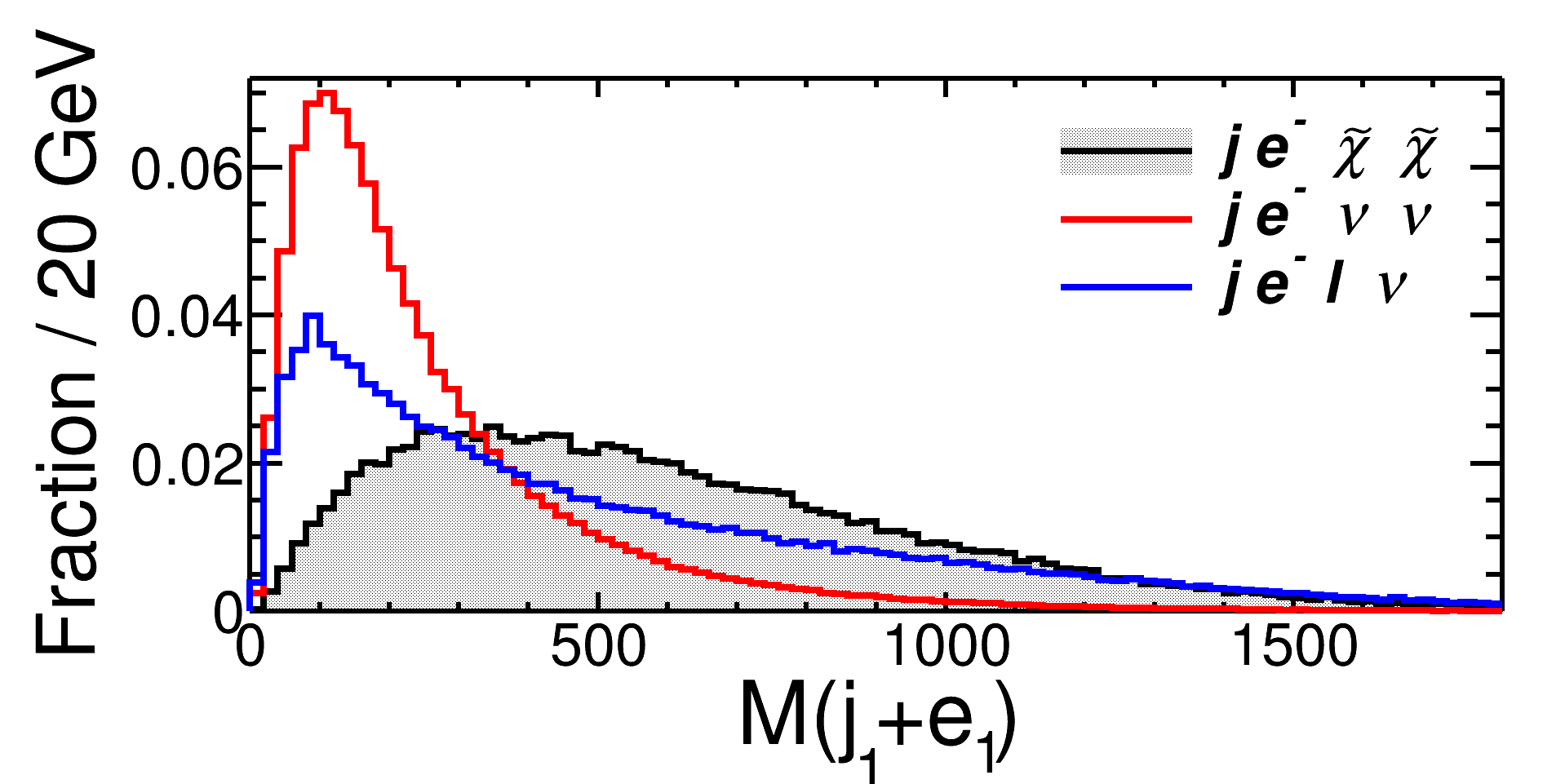}
\includegraphics[width=4cm,height=3cm]{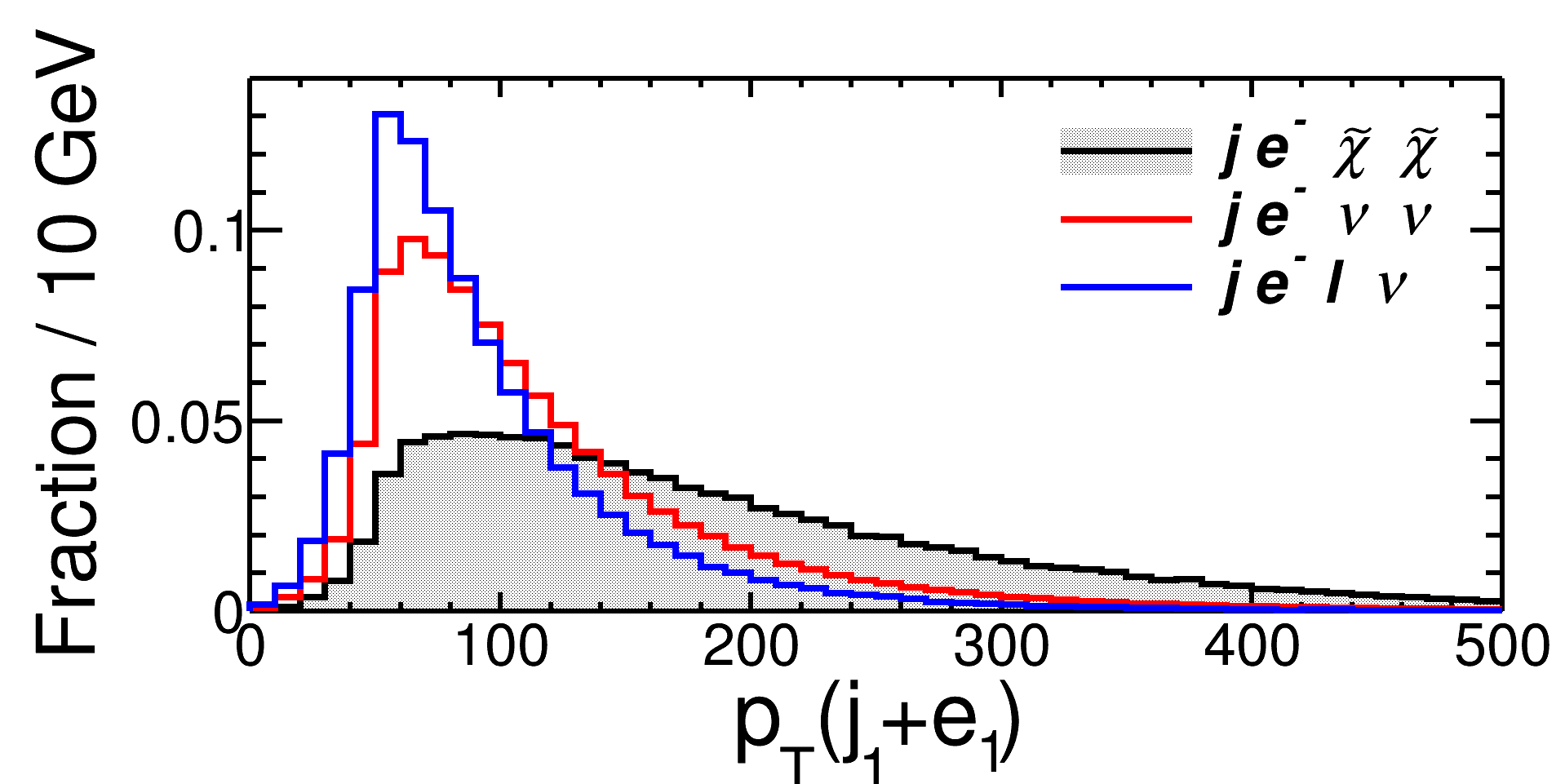}
}
\end{figure}
\vspace{-1.1cm}
\begin{figure}[H] 
\addtocounter{figure}{1}
\subfigure{
\includegraphics[width=4cm,height=3cm]{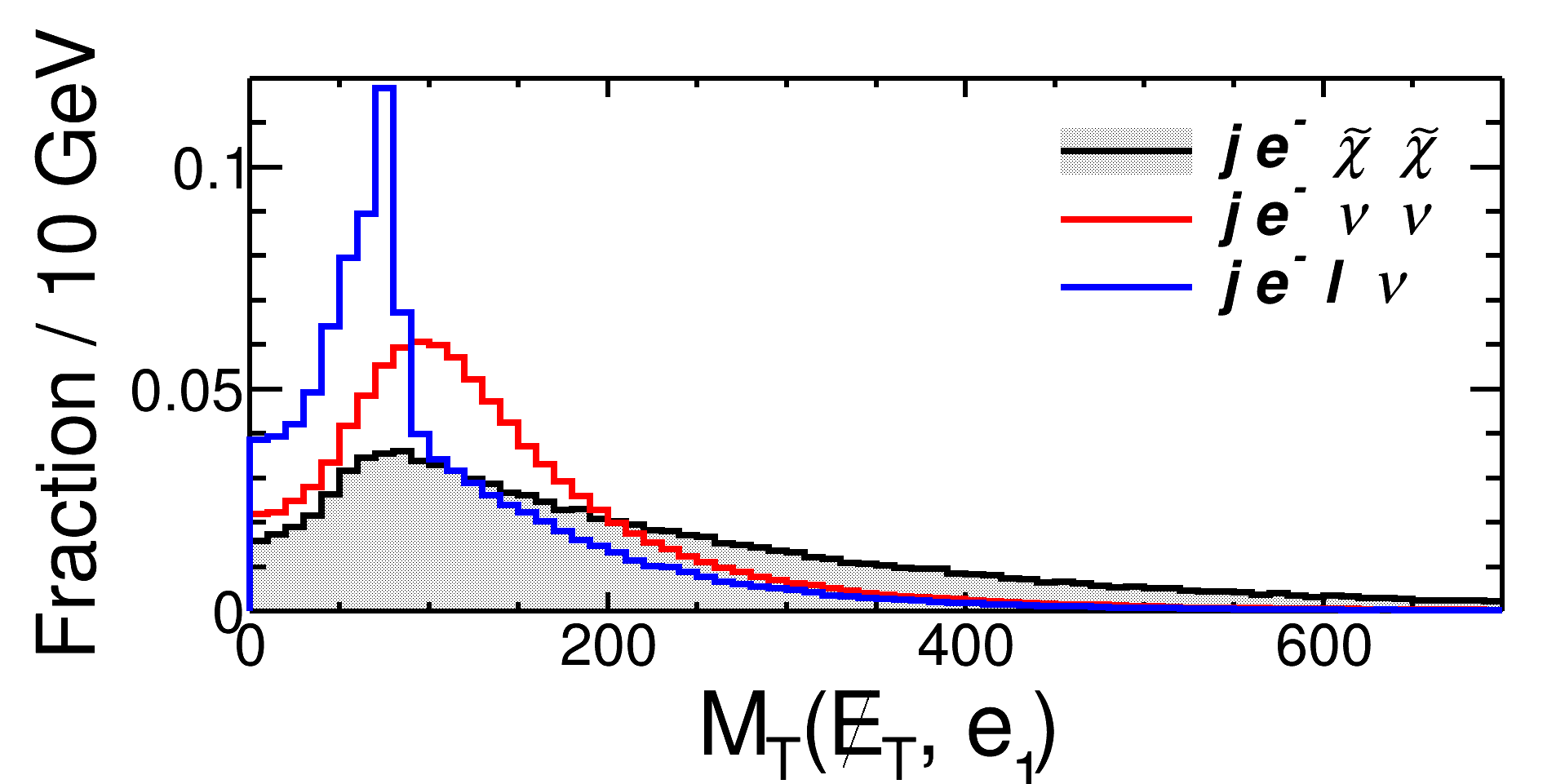}
\includegraphics[width=4cm,height=3cm]{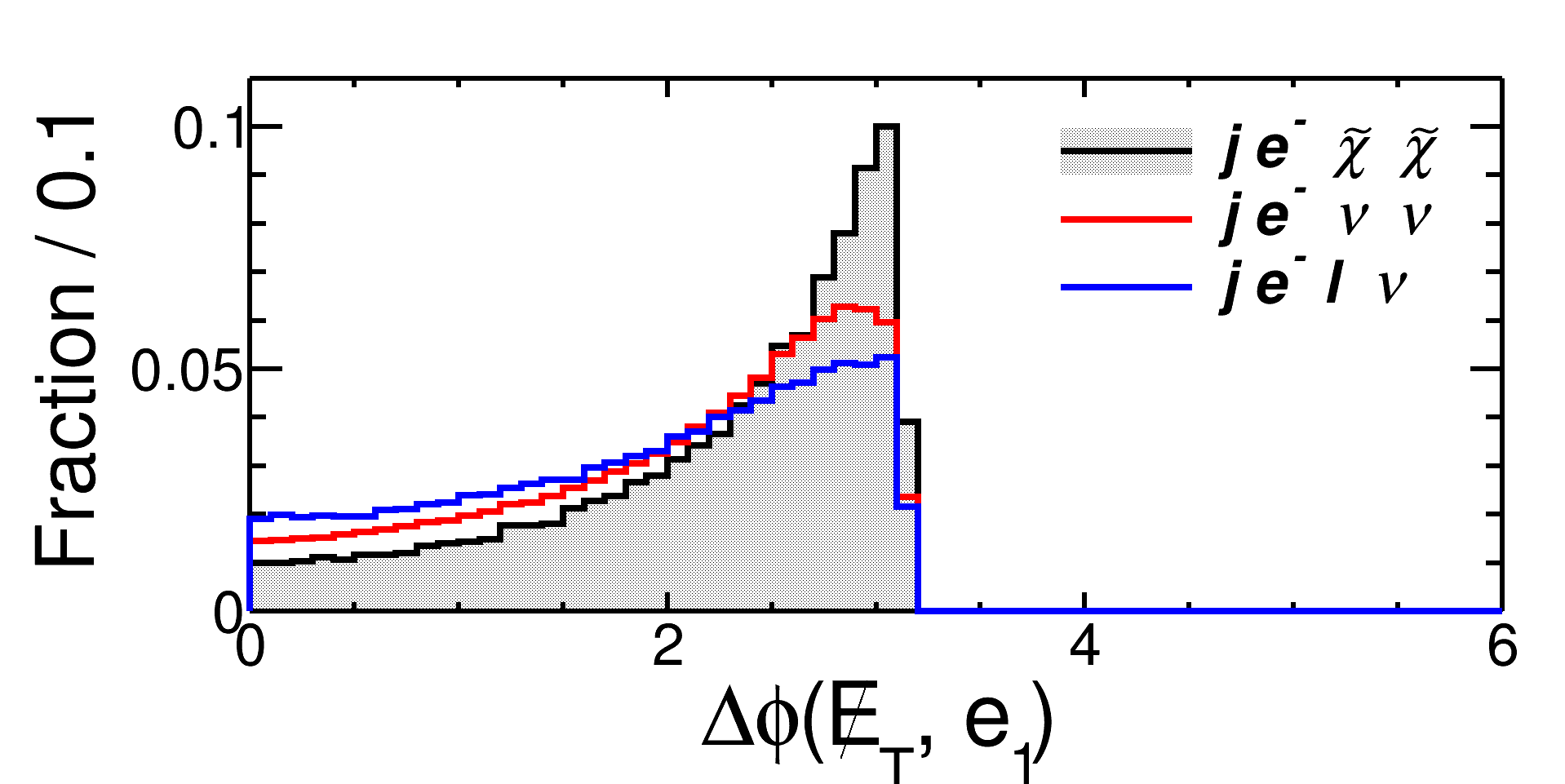}
}
\end{figure}
\vspace{-1.1cm}
\begin{figure}[H] 
\addtocounter{figure}{-1}
\subfigure{
\includegraphics[width=4cm,height=3cm]{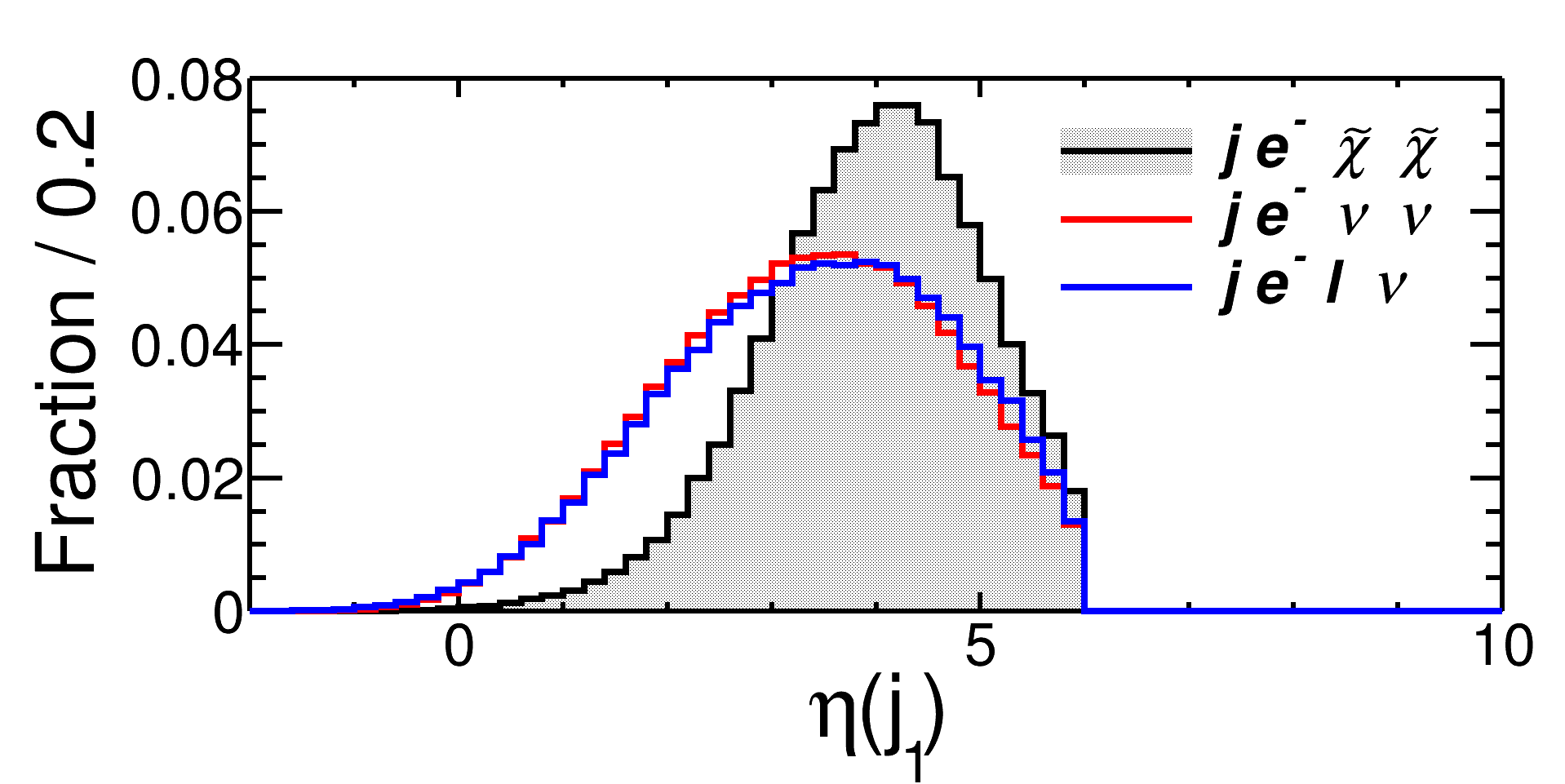}
\includegraphics[width=4cm,height=3cm]{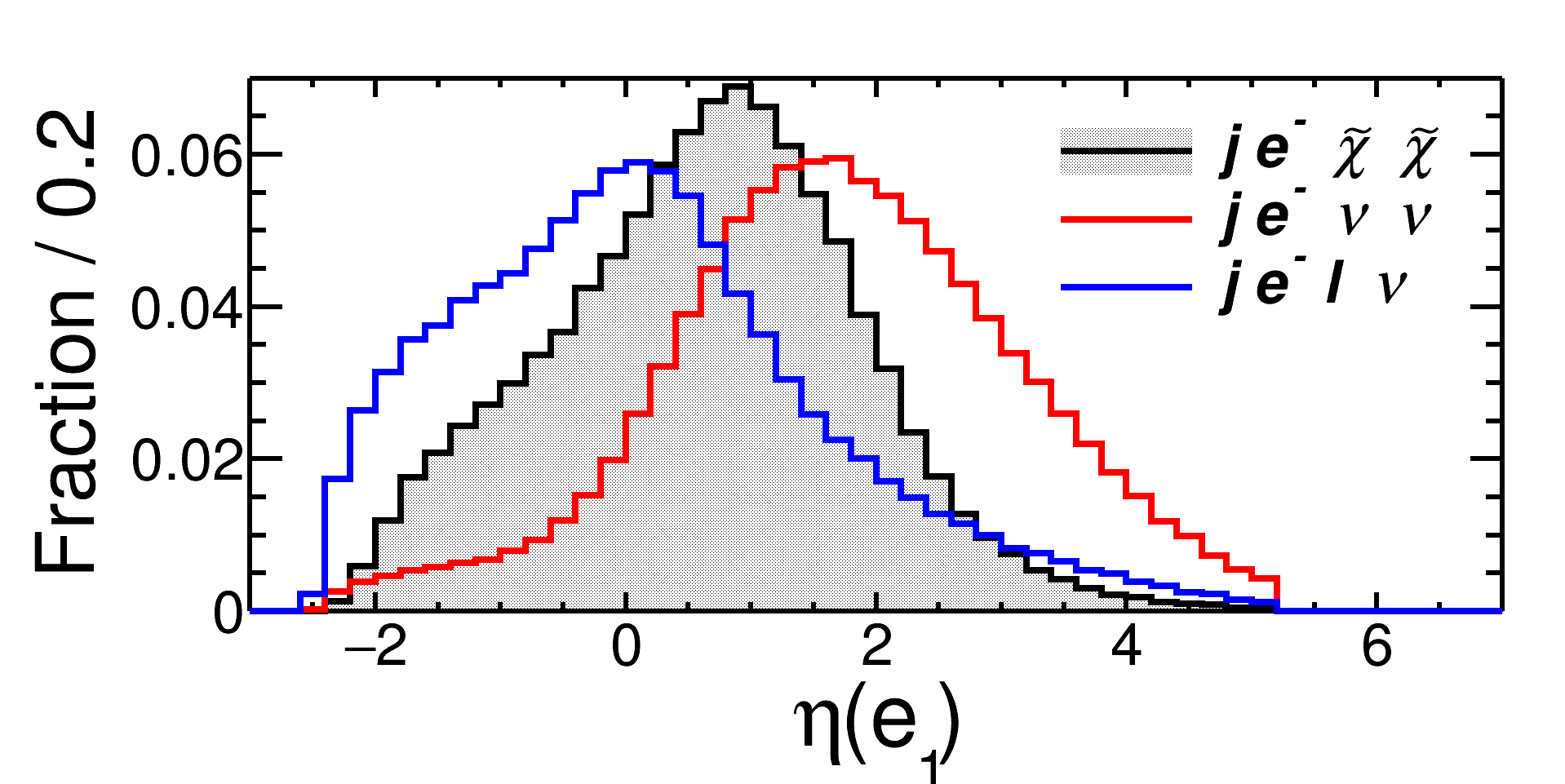}
}
\end{figure}
\vspace{-1.1cm}
\begin{figure}[H] 
\addtocounter{figure}{1}
\subfigure{
\includegraphics[width=4cm,height=3cm]{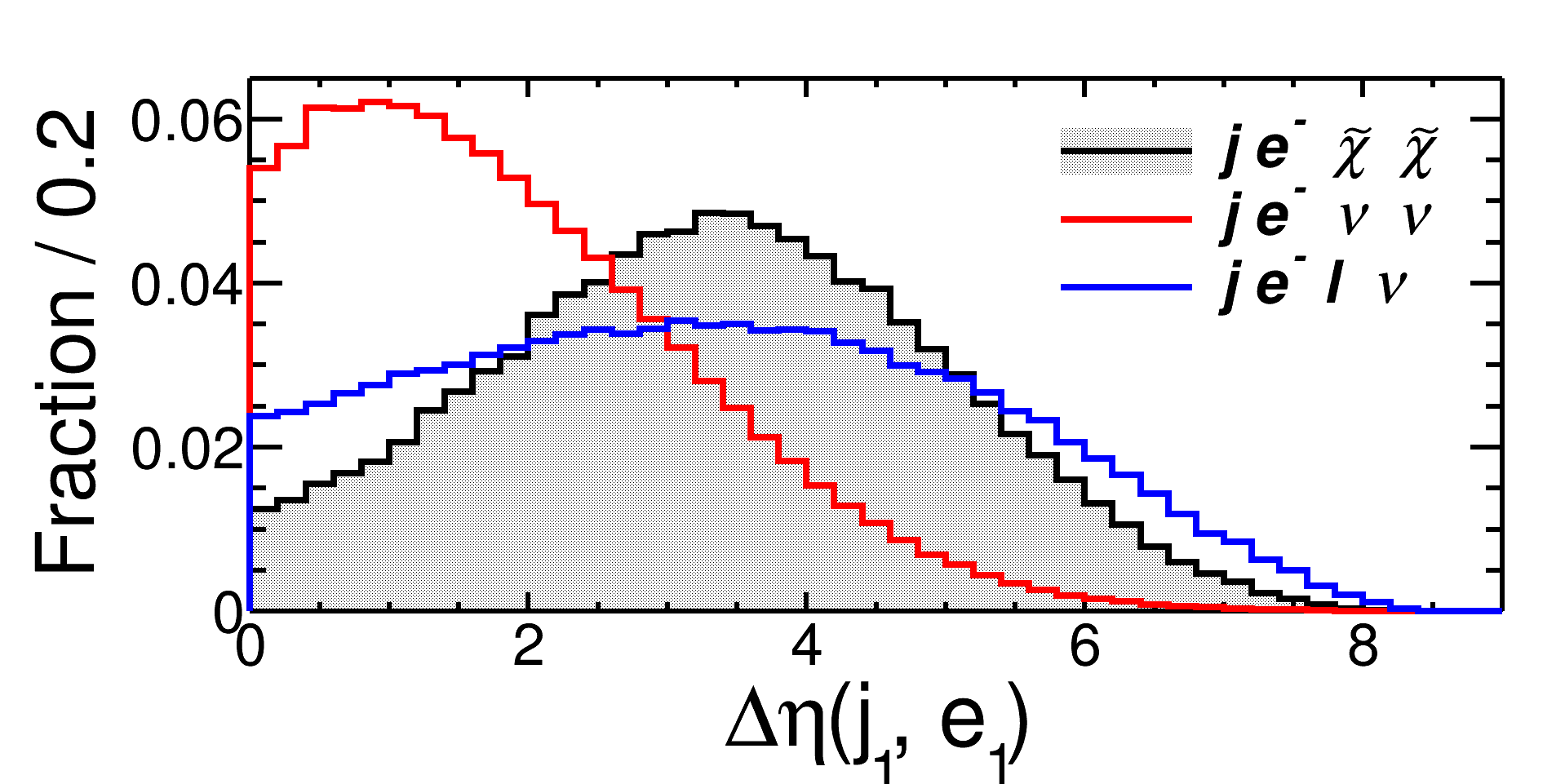}
\includegraphics[width=4cm,height=3cm]{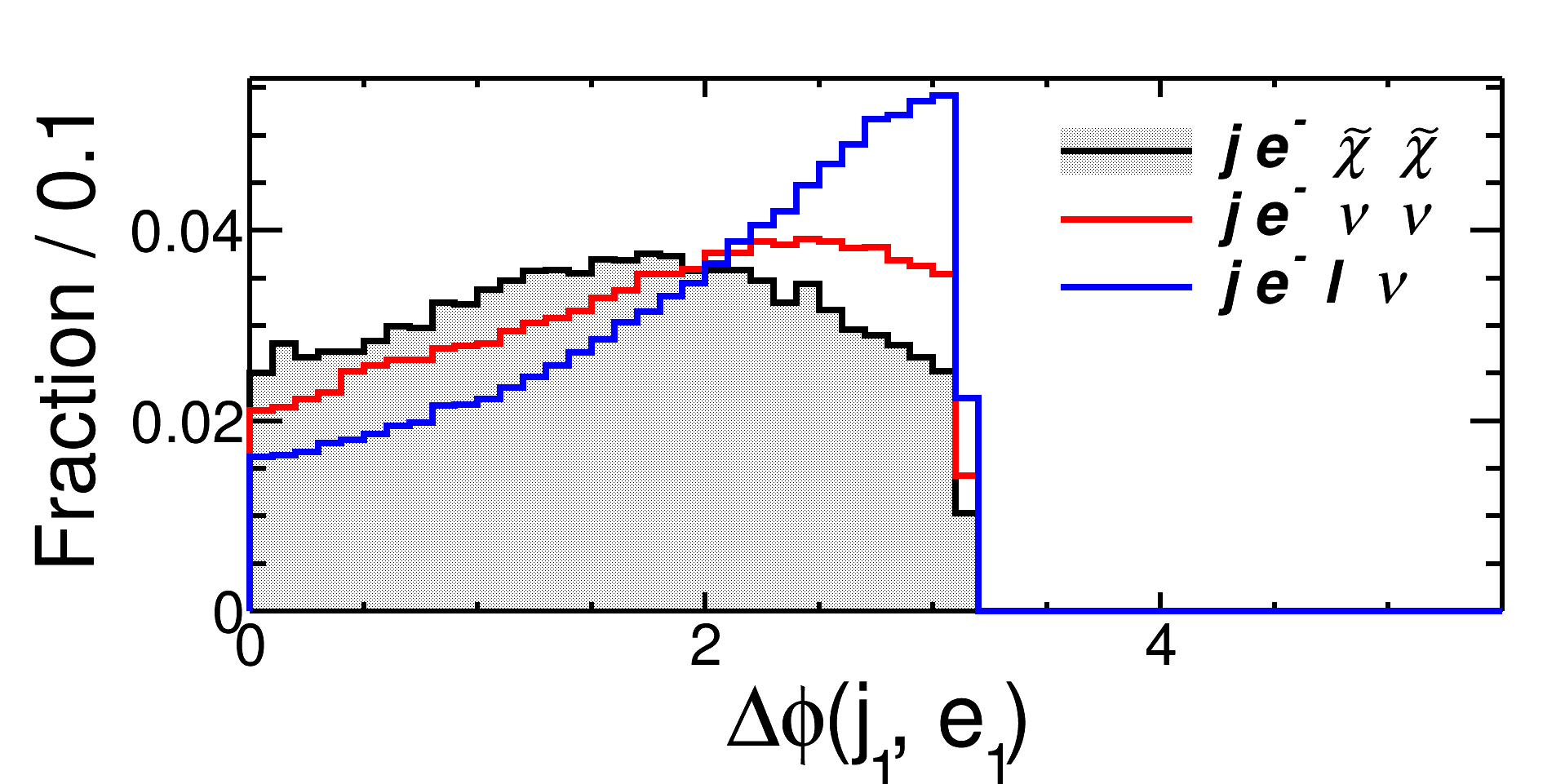}
}
\caption{Kinematical distributions of some input observables  for signal $j\, e^-\, \tilde{\chi} \tilde{\chi}$ with $m_{\chargino1, \neutralino2}$ = 250 GeV (black with filled area) in the decoupled-slepton scenario, and for the SM background of the $j\, e^-\, \nu \nu$ (red) and $j\, e^-\, \ell \nu$ (blue) processes after applying the pre-selection cuts at the FCC-eh with an unpolarized electron beam. }
\label{fig:obs_FCCeh_decoupledSlep_m250}
\end{figure}


\bibliography{Refs}
\bibliographystyle{h-physrev5}

\end{document}